      [1999/12/01 v1.4c Il Nuovo Cimento]
\documentclass{cimento}

\usepackage{amssymb,amsmath,latexsym,mathrsfs}
\usepackage{graphicx}
\usepackage{epsfig}
\usepackage{soul}
\usepackage[caption=false]{subfig}
\usepackage{color}
 \usepackage{url}

\def\be{\begin{eqnarray}}
\def\ee{\end{eqnarray}}
\def\beq{\begin{equation}}
\def\eeq{\end{equation}}
\def\HH{${\rm {H_2}}\,\,$}

\def\gtsima{$\; \buildrel \over \sim \;$}
\def\ltsima{$\; \buildrel < \over \sim \;$}

\def\gsim{\lower.5ex\hbox{\gtsima}}
\def\lsim{\lower.5ex\hbox{\ltsima}}
\def\simgt{\lower.5ex\hbox{\gtsima}}
\def\simlt{\lower.5ex\hbox{\ltsima}}
\def\Lya{Ly${\alpha}\,\,$}

\def\nhi{{N_{\rm HI}}}
\def\gta{\, {}^>_\sim \,}

\def\HI{\hbox{H~$\scriptstyle\rm I$}}
\def\HeIs{{\rm He\, \scriptstyle I}}
\def\HeIIs{{\rm He\, \scriptstyle II}}
\def\HeIIIs{{\rm He\, \scriptstyle III}}
\def\HeI{\hbox{He~$\scriptstyle\rm I$}}
\def\HeII{\hbox{He~$\scriptstyle\rm II$}}
\def\HeIII{\hbox{He~$\scriptstyle\rm III$}}

\def\HIs{{\rm H\, \scriptstyle I}}
\def\HIIs{{\rm H\, \scriptstyle II}}

\def\Lya{Ly$\alpha\ $}
\def\Lyb{Ly$\beta\ $}

\def\gtsima{$\; \buildrel > \over \sim \;$}
\def\gsim{\lower.5ex\hbox{\gtsima}}
\newcommand{\HII}{\rm H\, {\scriptstyle II} }
\newcommand{\apj}{ApJ}
\newcommand{\aj}{AJ}
\newcommand{\mnras}{MNRAS}
\newcommand{\apjl}{ApJL}

\def\beq{\begin{equation}}
\def\eeq{\end{equation}}

\def\beqa{\begin{eqnarray}}
\def\eeqa{\end{eqnarray}}

\def\nb{\bar{n}}

\def\Ng{N_\gamma}

\def\HI{\rm H\,I}

\def\Ni{N_{\rm ion}}

\newcommand{\bear}{\begin{eqnarray}}
\newcommand{\ear}{\end{eqnarray}}
\newcommand{\nline}{\nonumber \\}
\newcommand{\f}{\frac}
\newcommand{\de}{{\rm d}}
\newcommand{\del}{\partial}

\def\OVI{\hbox{O~$\scriptstyle\rm VI\ $}}

\def\lyc{Ly$\gamma$ }

\title{Reionization of the Intergalactic Medium}
\author{Andrea Ferrara\from{ins:x}\ETC,
Stefania Pandolfi\from{ins:y}\\}
\instlist{\inst{ins:x} Scuola Normale Superiore, Pisa, Italy
  \inst{ins:y} Dark Cosmology Centre, Niels Bohr Institute, University of Copenhagen}
\PACSes{\PACSit{00.00}{}
\PACSit{---.---}{\ldots}}
\begin{document}

\maketitle

\begin{abstract}
After recombination the cosmic gas was left in a cold and neutral state. However, as the first stars and black holes formed within early galactic systems, their UV and X-ray radiation induced a gradual phase transition of the intergalactic gas into the warm and ionized state we currently observe. This process is known as cosmic reionization. Understanding how the energy deposition connected with galaxy and star formation shaped the properties of the intergalactic gas is one of the primary goals of present-day cosmology. In addition, reionization back reacts on galaxy evolution, determining many of the properties of the high-redshift galaxy population that represent the current frontier of our discovery of the cosmos. In these two Lectures we provide a pedagogical overview of cosmic reionization and intergalactic medium and of some of the open questions in these fields.    
\end{abstract}

\section{Lecture I -- The Intergalactic Medium} 

In this Lecture we introduce and discuss the physical processes governing the observed properties of the Intergalactic Medium (IGM), i.e. the gas that is not part of collapsed structures. It is important to realize that at any redshift, most of the cosmic baryons reside in this component and therefore its role in cosmic evolution can hardly be overlooked. We start the Lecture with some historical background and then take a phenomenological approach to the observational evidences that will allow us to isolate and discuss in detail the key physical processes required to properly model the IGM properties and evolution.

\subsection{{\bf Historical background}}\label{hist}
The birth of the IGM study canonically goes back to 1965, when J.E. Gunn and B.A. Peterson realized that an expanding universe, filled with gas, would have produced an absorption trough in the spectra of distant quasars (QSO), due to the presence of neutral hydrogen, in the wavelengths blue-wards of the Ly$\alpha$ emission line of the QSO.  In their classic paper, Gunn \atque Peterson  \cite{gp65} showed that the hydrogen in a diffuse uniform IGM must have been highly ionized at $z \approx 2$ in order to avoid complete absorption of the transmitted flux at wavelengths shortwards of the \Lya emission line of the QSO; this is now commonly known as the Gunn-Peterson (GP) effect. Following that suggestion, it was proposed \cite{bs66}  that the GP effect could be used to probe the ionization state of intergalactic hydrogen (and also other elements) at various redshifts. 

\subsubsection{{\bf Early confinement models}}
At the same time, it was also realized that inhomogeneously distributed gas would produce discrete \Lya absorption lines. Initially,
these were identified with either gas clumped into groups of galaxies \cite{bs65}  or low mass protogalaxies \cite{arons72}. However, these proposal were soon found to be unrealistic when different groups \cite{lynds71,sybt80} discovered a very large number of discrete absorption lines in the QSO spectra, which are usually known as the ``\Lya forest". It was shown that these forest lines could not be associated with galaxy clusters; rather they should be of intergalactic origin and arise in discrete intergalactic clouds at various cosmological redshifts along the line of sight \cite{sybt80,wcs81,btn88}. Various arguments (like the apparent lack of rapid evolution in the properties of the forest, the short relaxation time scales for electrons and protons and short mean free path of photons) led to the notion that the clouds were ``self-contained entities in equilibrium" \cite{sybt80}. A two-phase medium was postulated, with the diffuse, very hot, inter-cloud medium (ICM) in pressure equilibrium with the cooler and denser \Lya clouds. In this two-phase scenario, the ICM was identified with the IGM, while the \Lya clouds were treated as separate entities. 

According to the pressure confinement model \cite{sybt80,oi83,io86}, the \Lya clouds are supposed to be in photoionization equilibrium with an ionizing ultraviolet (UV) background. The gas is heated by photoionization and cools via thermal bremsstrahlung, Compton cooling, and the usual recombination and collisional excitation processes. Since the ICM is highly ionized, the photoheating is not efficient and hence the medium cools adiabatically through cosmic expansion. The denser clouds embedded in the hot ICM have a nearly constant temperature fixed by thermal ionization equilibrium ($\approx 3 \times 10^4$ K) \cite{oi83,io86}. The available range of cloud masses is constrained by the requirement that the clouds must be small enough not to be Jeans-unstable but large enough not to be evaporated rapidly when heated by thermal conduction from the ambient ICM \cite{sybt80,oi83}. According to such constraints, clouds formed at high redshifts would survive down to observed redshifts only if
their masses range between $10^{5-6} M_\odot$.

The neutral hydrogen within the confining ICM is expected to cause a residual GP absorption trough between the absorption lines (clouds). However, observations at higher spectral resolution \cite{ss87,gct92,gdf++94} revealed no continuous absorption between the discrete lines, placing strong limits on the GP effect, which in turn, puts a demanding upper bound on the density of the ICM . The ICM temperature has a lower limit from the absorption line width, while the condition that the cloud must be large enough not to evaporate gives an upper limit on the temperature \cite{oi83}. Another independent upper limit on the temperature of the ICM comes from the lack of inverse Compton distortions in the spectrum of the cosmic microwave background \cite{bfr91} through
the Sunyaev-Zeldovich effect \cite{sz80}. In fact, the upper limit of the so-called $y$-parameter \cite{fcg++96} is able to rule out any cosmologically distributed component of temperature $>10^6$ K. When all the limits are combined, only a small region of allowed density-temperature parameter space remains for the ICM. It turns out that, according to the pressure-confinement model, the density of the ICM is too small to be cosmologically significant. Hence, during these early days, the connection between the cosmic reionization and the IGM was not at all obvious as most of the baryons was expected to lie somewhere else. 

The pressure-confinement model ran into severe problems while trying to match the observed column density distribution \cite{pwrcl93}. For example, in
order to reproduce the low column density systems between, say, $13 < \log(\nhi/{\rm cm}^{-2}) <16$ (where $\nhi$ is the column density of neutral hydrogen), the total mass of the system has to vary by 9 orders of magnitude. On the other hand, the
mass is severely constrained in order to ensure cloud survival. Therefore, the only escape route is to invoke pressure inhomogeneities \cite{bchw89}. However, the
\Lya absorbers are found to be weakly clustered over a large range of scales, which thus excludes any significant pressure fluctuations \cite{wb91}.

Similarly, detailed hydrodynamical simulations \cite{wb92} show that the small mass range of
the clouds leads to a failure in producing the column density distribution at
high $\nhi$. In addition, pressure-confinement models predict small cloud sizes
which are incompatible with the observations of multiple lines of sight \cite{dfi++95} .
It was thus concluded that the pure pressure confinement model is unlikely
to explain the \Lya forest as a whole though it is possible that some lines of
sight must go through sites where gas is locally confined by external pressure
(say, the galactic haloes, the likely hosts of the dense Lyman limit absorbing
clouds).

To overcome these difficulties, \emph{self-gravitating baryonic} clouds were suggested by \cite{melott80,black81} as
an alternative to the pressure confinement idea. In this model, the appearance
of the IGM as a forest of lines is because of the variations in the neutral
hydrogen density rather than a sharp transition between separate entities.
In this sense, there is no real difference between an ICM and the clouds in
the gravitational confinement model. This scenario of self-gravitating clouds
predicts larger sizes of the absorbing clouds ($\approx 1$ Mpc) compared to the
pressure-confinement scenario. However, this model, too, runs into problems
while trying to match the observed column density distribution \cite{pbcp93}  as
it predicts a larger number of high column density systems than is observed.
Secondly, the large absorber sizes seemed to contradict observations. Furthermore,
gravitationally confined clouds are difficult to explain theoretically
since the mass of such clouds must lie in a restricted range to maintain the
gas in equilibrium against free expansion or collapse.

As a further alternative, the properties of gas clouds confined by the gravitational
field of dark matter have been investigated \cite{ui85}, more specifically
in terms of the ``minihalo" model \cite{rees86,ikeuchi86}. In this picture, \Lya clouds are a
natural byproduct of the cold dark matter (CDM) structure formation scenario.
Photoionized gas settles in the potential well of an isothermal dark
matter halo. The gas is stably confined if the potential is sufficiently shallow
to avoid gravitational collapse but deep enough to prevent the warm gas from
escaping. CDM minihalos are more compact than the self-gravitating baryonic
clouds \cite{black81} because of the larger dark matter gravity, thus alleviating
the size problem. The detailed structure of the halo depends on the relative
spatial distribution of baryons and CDM. However, the virial radii of the
confining objects ($\approx 10$ kpc) are much lower than the coherence lengths of the \Lya systems as obtained from constraints on absorption line observations
of lensed or paired QSOs \cite{sss++92,srs++95}. It was thus natural to extend the minihalo
model to non-static systems. A non-static minihalo model was proposed by \cite{bss88}, who examined the hydrodynamics of a collapsing spherical top-hat
perturbation and suggested that clouds were in a free expansion phase.

\subsubsection{{\bf IGM as a fluctuating density field}}
Following the non-static models, it was realized that an IGM with the density
fluctuation variance of the order of unity could also produce line-like absorptions in quasar spectra \cite{mcgill90,bbc92}.
 According to such models, the IGM becomes
clumpy and acquires peculiar motions under the influence of gravity, and so
the \Lya (or GP) optical depth should vary even at the lowest column densities \cite{black81,mcgill90,bbc92,mr93,rm95}. In a CDM-dominated structure formation scenario, the accumulation of matter in overdense regions reduces the optical depth for
\Lya absorption considerably below the average in most of the volume of the
universe, leading to what has been called the \emph{fluctuating GP effect}.
Traditional searches for the GP effect that try to measure the amount of
matter between the absorption lines were no longer meaningful, as they were
merely detecting absorption from matter left over in the most underdense
regions. If this is not taken into account, the amount of ionizing radiation
necessary to keep the neutral hydrogen GP absorption below the detection
limits can be overestimated, which would then have severe implications for
reionization studies. In this scenario, the density, temperature and thermal
pressure of the medium were described as continuous fields and could not
be attributed simply to gravitational confinement or pressure confinement.
These studies led to a shift in the paradigm of IGM theories, especially since
they implied that the IGM contains most of the baryons at high redshifts,
thus making it cosmologically significant and hence quite relevant to cosmic
reionization (see Sect. \ref{Mass within the cool phase}).

The actual fluctuation picture was derived from cosmological N-body
and \emph{hydrodynamical simulations} (see Sect. \ref{simulations}).
It was possible to solve hydrodynamical equations from first principles and set up an evolutionary picture of the
IGM in these simulations \cite{cmor94,zan95,hkwm96,mcor96}. 
Although different techniques and cosmological models were used by different groups, all the simulations
indicate a fluctuating IGM instead of discrete clouds.
One of the immediate conclusions was that at redshifts $z \gta 1.5$ 
the volume filling photoionized IGM contains some 90\% of the baryons 
content of the universe, with only 10\% in galaxies and galaxy clusters (see Sect. \ref{Mass within the cool phase}).

Since in this new paradigm, the \Lya forest arises from a 
quasi-linear IGM, it is possible to ignore the high non-linearities. This
made possible to study the IGM through semi-analytical techniques too 
\cite{bbc92,bi93,gh96,bd97,hgz97}. The issue of dealing with quasi-linear densities was dealt
in two ways. In the first method, it was showed that a quasi-linear density
field, described by a lognormal distribution, can reproduce almost all the
observed properties of the \Lya forest \cite{bbc92,bd97}. In fact, this was motivated by
earlier ideas of \cite{cj91} for dark matter distribution. In an alternate method, it
was also possible to obtain the density distribution of baryons from simulations
which could then be used for semi-analytical calculations \cite{mhr00}. Given the
baryonic distribution, the neutral hydrogen fraction was calculated assuming
photoionization equilibrium between the baryons and the ionizing radiation
field. It was also realized that the equilibrium between photoheating and
adiabatic cooling implies a tight relation between the temperature and density
of the gas, described by a power-law equation of state \cite{gh98} (see Sect. \ref{EOS}), which was
used for determining the temperature of the gas. Given such simplifying and
reasonable assumptions, it was possible to make detailed predictions about
the \Lya forest. For example, a relation between column density peaks (``absorption
lines") and the statistics of density peaks was proposed \cite{gh96,hgz97}, and
analytical expressions for the dependence of the shape of the column density
distribution on cosmological parameters were obtained.

Both simulations and semi-analytical calculations have been quite
successful in matching the overall observed properties of the absorption systems.
The shape of the column density distribution and the Doppler parameter 
distribution are reasonably well reproduced by simulations 
\cite{cmor94,hkwm96,mcor96,zan95,mpkr96,zanm97} as well as by semi-analytical calculations 
\cite{hgz97,cps01}  over a wide redshift range. The large transverse sizes of the absorbers seen against background
paired and lensed QSOs are well explained by the coherence length of the
sheets and filaments \cite{mcor96,cs97,cazn97}. In addition, the probability distribution
function and power spectrum of the transmitted flux in the \Lya forest is reproduced
very well by the models \cite{mmr++00,csp01}. The \Lya optical depth fluctuations
were used for recovering the power spectrum of matter density fluctuations
at small scales \cite{cwkh98,cwphk99} and also to obtain various quantities related to the
IGM \cite{csp01,wmhk97}. 

Given the fact that the \Lya can be modeled so accurately, it has become
the most useful tool in studying the thermal and ionization history of the
universe ever since. Subsequently it was realized that this simple description
of the IGM could be coupled to the properties of the ionizing sources and
hence it was possible to compute the reheating and reionization history.
The \Lya forest is also undoubtedly a very powerful test to probe the fundamental properties of the universe. 
Most noticeably it has recently inspired attempts to probe the cosmological structure formation models and, 
more generally to put constraints on the cosmological parameters. Although the unquestionable successes 
achieved by \Lya forest simulations, many aspects require further investigations. These are nicely 
summarized in, e.g. \cite{Finlator(2012)} and \cite{Shapiro et al.(2012)}

\subsection{{\bf IGM phenomenology}}\label{physical}
As previously mentioned, the status and the properties of the IGM can be studied through the analysis 
of the properties of the \Lya forest. Lyman photons are photons with energy sufficient to excite an electron 
in an hydrogen atom from the ground state to a higher energy state. For example, a photon of wavelength 
$\lambda=1216$ \AA\ can induce a electronic transition from the level $n=1$ to $n=2$ (i.e. Ly$\alpha$ transition) of the H atom. The inverse process is possible as well, and implies the emission of a photon with the same energy of the energy difference between the levels. Therefore, the UV light of a distant quasar traversing the IGM towards the observer could be absorbed by intervening bunches of neutral hydrogen atoms in their ground state once the photons are redshifted (due to cosmic expansion) to the proper transition frequency. A re-emission will follow absorption, but as the emission direction is isotropic rather than along the line of sight, this scattering process is equivalent to an effective opacity of the medium. The optical depth, (i.e the amount of light absorbed) is proportional to the cross section (probability that the hydrogen atom will absorb the photon) times the number density of hydrogen atoms along its line of sight, as we will see in more details in  Sect. \ref{lowres}.

\subsubsection{{\bf QSO absorption spectra}}\label{physical}
\begin{figure} \centering
\includegraphics[width=0.9\textwidth]{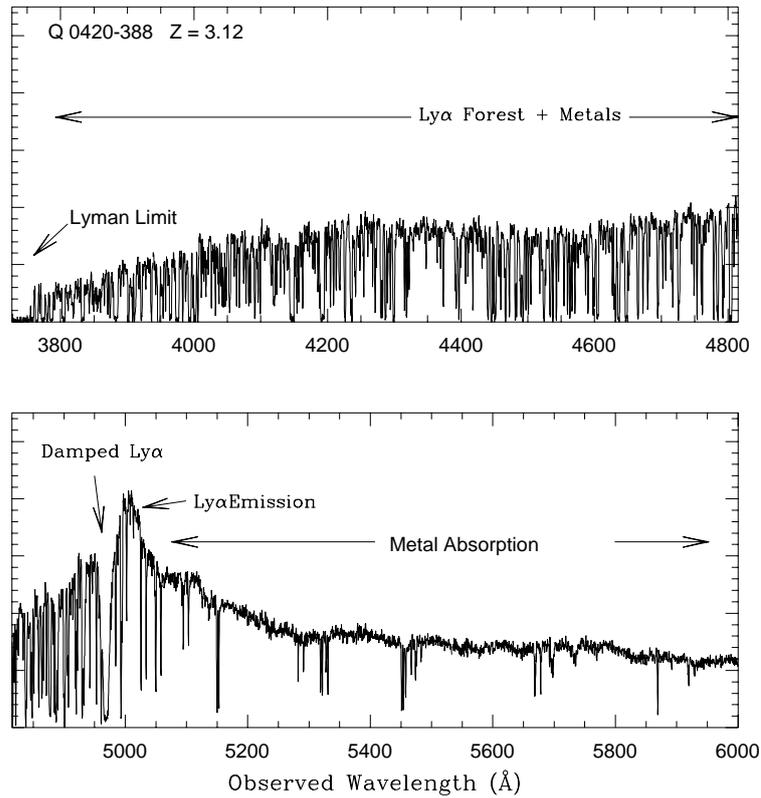}
\vspace{-2.5cm}
\caption{(Taken from \cite{Zaroubi:2012in}) High resolution spectrum of the $z=3.12$ quasar Q0420-388 obtained with the Las Campanas
echelle spectrograph by J. Bechtold and S. A. Shectman. 
The Lyman~$\alpha$  forest is in the upper panel of the figure, bluewards of the quasar rest frame Lyman~$\alpha$ emission features. This figure appeared in~\cite{bechtold03}.
}
\label{fig:Lyaexample}
\end{figure}

All intervening neutral hydrogen over-densities will then produce an absorbing feature in the observed QSO spectrum at a wavelength corresponding to $1216(1+z_a)$ \AA\, where $z_a$ is the redshift of the absorber. Thus, use the \Lya forest to map the positions of the intervening hydrogen regions. In Fig. \ref{fig:Lyaexample} we report a typical spectrum of a distant quasar  
where we can observe Lyman~$\alpha$ forest. The density of weak absorbing lines increases with redshift due to the expansion of the universe and, at redshifts above $z\simeq 4$, the density of the absorption features become so high that it is hard to see them as separate absorption lines (line blending).
The part of the spectrum red-ward of the Ly$\alpha$ line is populated by absorption lines from heavy elements (including carbon, silicon, nitrogen, oxygen, magnesium, iron and others). However, the metal abundances are at most about 10\% of solar at low redshifts, and $\simeq$ 1\% at high redshifts. In addition to raising challenging questions on how these species were carried far away from their production sites, i.e. star forming regions in galaxies,  this shows that the absorption systems are largely made of primordial material. We will come back to the problem of IGM metal enrichment later on.
\begin{figure}
\centering
\includegraphics[width=0.8\textwidth]{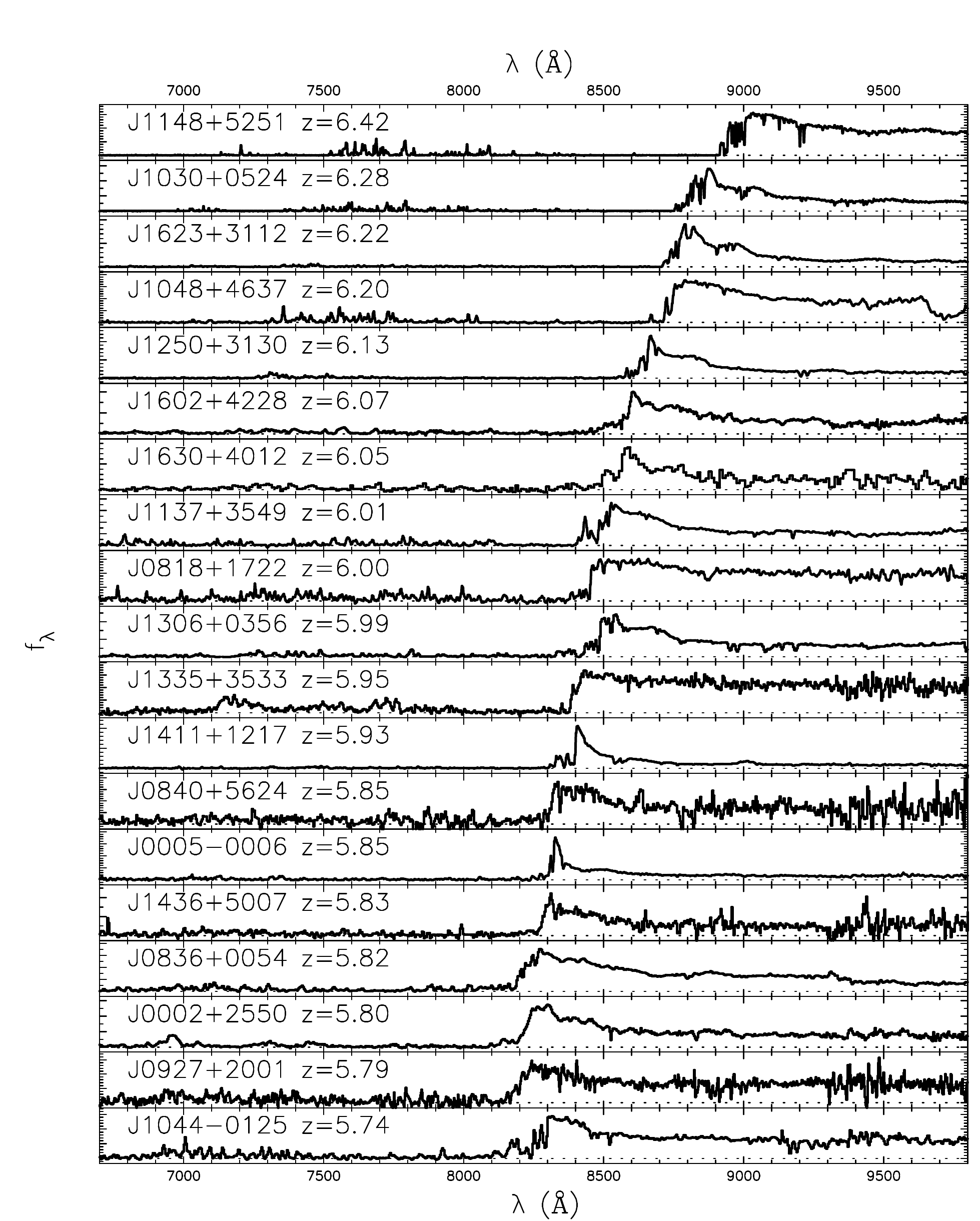}
\caption{(Taken from \cite{Zaroubi:2012in}). Spectra for high redshift SDSS quasars.
The Gunn-Peterson trough bluewards of the QSO Lyman~$\alpha$
emission in the highest redshifts QSO hints to the fact 
that the Universe has become somewhat more neutral at these redshifts.
}
\label{fig:Lya2}
\end{figure}

As the number of absorption systems per path length increases along a line
of sight with redshift, so does the mean flux decrement in
a background QSO spectrum due to \Lya scattering. 
Fig. \ref{fig:Lya2} shows the spectra of  high redshift quasars \cite{fan03, fan06} in the Sloan Digital Sky Survey (SDSS)
\footnote{http://www.sdss.org}.
Notice that the line blending at these high redshifts become so severe that the entire spectrum blue-wards 
of the quasar \Lya restframe emission is absorbed, with only a few sparse transmissivity windows still present.  
This is normally interpreted as a result of the decrease in the ionized fraction of the universe. 
Also note that different quasars at approximately the same redshift show a different number/size of 
transmissivity windows, likely a result of the patchy and inhomogeneous nature of the ionization field. 
This in turn suggests that the epoch of \HI\ reionization may be not far beyond $z\simeq6$. However, 
more detailed reionization models, reviewed in Lecture II, show that this is not necessarily the correct conclusion.

\subsubsection{{\bf Low Resolution Spectroscopy}}\label{lowres}
The most basic observable of the \Lya forest is the flux decrement $D_A$, or the mean fraction of the QSO continuum absorbed, typically measured between the \Lya and $Ly\beta$ emission lines. This is defined as
\be
D_A = \langle 1- { f_{\rm obs}\over f_{\rm cont}}\rangle = \langle 1- e^{-\tau}\rangle = 1- {e^{-\tau_{eff}}}
\ee
where $f_{\rm obs}$ is the observed (=residual) flux, $f_{\rm cont}$ the estimated flux of the unabsorbed continuum, and $\tau$ is the resonance line optical depth as a function of wavelength or redshift \cite{Rauch:1998xn}. The absorption is measured against a continuum level usually taken to be a power law in wavelength extrapolated from the region red-ward of the \Lya emission line. 

With $D_A$ measurements available over a range of redshifts the redshift evolution of the \Lya forest can be investigated. If we characterize a \Lya forest as a random distribution of absorption systems in column density $N$, Doppler parameter $b$ (see Sect. \ref{highres}), and redshift $z$ space, such that the number of lines per interval $dN$, $db$ and $dz$ is given by $F(N, b, z)dN db dz$ (see Sect. \ref{Column}), then
\be
\tau_{eff}=\int_{z_1}^{z_2}\int_{b_1}^{b_2}\int_{N_1}^{N_2}dN db dz 
(1- e^{-\tau(N,b)}) F(N, b, z)
\ee

Assuming that the $N$ and $b$ distribution functions are independent of redshift (see again Sect. \ref{Column}), and that the redshift evolution of the number density of lines can be approximated by a power law, we can write $F(N, b, z) = (1 + z)F(N, b)$, and
\be
\tau_{eff}(z) = (1 + z)^{\gamma+1}\lambda_0^{-1} \int_{b_1}^{b_2}\int_{N_1}^{N_2}dN db ~~e^{-\tau(N,b)} F(N, b)W(N,b)
\ee
where $W$ is the rest frame equivalent width.
At high redshifts, the blending of the absorption feature makes it difficult to assign the absorption to separate systems, 
and indeed at redshifts $z>5.5$ the lines are all merged into the so-called Gunn-Peterson trough. 
Absorption line at intermediate redshifts can be described through the rest frame equivalent width $W$.
This quantity is defined as the width, expressed in units of wavelength, a square-well
absorption feature with zero flux at its bottom must have to match the
integrated area of the detected feature under the continuum. 
For a Voigt profile the equivalent width is related to the column density through a relation 
that is known as ``the curve of growth''.  
This relation enables us to measure the redshift evolution of the number density forest clouds, $dN/dz \propto (1 + z)^\gamma$, from the redshift dependence of the effective optical depth ($\tau_{eff} \propto (1 + z)^{\gamma+1}$) even if individual absorption 
lines cannot be resolved.

\subsubsection{{\bf High Resolution Spectroscopy}}\label{highres}
In order to proper resolve the lines high resolution spectroscopy is required. 
In general, scatterers will not be at rest in reference frame of the scattering photons, and at least they will 
have a thermal component described by a Maxwellian velocity distribution corresponding
to their temperature $T$. Moreover, there could be a so-called micro-turbulence component 
in the medium, in collapsed or shocked regions. 
For this reasons, the line profile is found convolving the resonant line scattering cross-section
with a Maxwellian. The result is called the Voigt profile. 
Indeed, the Voigt function 
accounts for the frequency shift from the line center in units of the Doppler width, 
through the Doppler parameter
\begin{equation}
b_{th}=\sqrt\frac{2k_{\rm B}T}{m}
\label{eq:bparam}
\end{equation}
where $m$ is the mass of the scatterer and $k_{\rm B}$ is the Boltzmann's constant.
Moreover, an additional term is added in quadrature for accounting for a 
kinematic component such as the micro-turbulence
\begin{equation}
b= \sqrt{b_{\rm th}^2 + b_{\rm kin}^2}.
\label{eq:bbparam}
\end{equation}

If the absorber is a gas cloud with a purely Gaussian velocity dispersion (a thermal Maxwell-Boltzmann distribution, plus
any Gaussian contributions from turbulence) a Voigt profile provides an exact description of the absorption line shape.

The standard approach to Voigt profile fitting relies on $\chi^2$ minimization in order to achieve a complete decomposition of the spectrum into as many independent Voigt profile components as necessary to make the $\chi^2$ probability consistent with random fluctuations.
For stronger \Lya lines the higher order Lyman lines can provide additional constraints when fitted simultaneously. Given sufficient spectral resolution, and assuming that \Lya clouds are discrete entities (in the sense of some of the models discussed in Section \ref{hist}) the profile fitting approach is the most physically meaningful way of extracting information from the \Lya forest. 

Unfortunately, in more realistic models of the absorbing gas finite velocity and density gradients invalidate the assumptions underlying Voigt profile fitting, and the line parameters may have less immediate physical meaning. Departures of the absorption line shape from a Voigt profile  may contain valuable information about the underlying nature of the absorption systems, and
different scenarios may have quite different observational signatures. 

\subsubsection{{\bf Absorption line properties}}

The absorption features of the \Lya forest are broadly
classified into three main types: \Lya forest systems, Damped \Lya
Absorbers (DLAs), and Lyman Limit Systems (LLSs).

The classification is based primarily on the physical origin of the features, but is not strictly
exclusive. The \Lya
forest systems are generally well-fit by
Doppler line profiles, while the rarer DLAs show the radiation damping wings
of the Lorentz profile for having high
hydrogen column densities: they require the Voigt line profile for
accurate fitting. 

The LLSs instead have a sufficiently large
column density to absorb photons with energies above the photoelectric
edge (Lyman limit).  


\subsubsection{{\bf Number density evolution of the \Lya forest}}
\label{subsubsec:evolution}
\begin{figure}[!h]
\centering
\includegraphics[height=5.5cm]{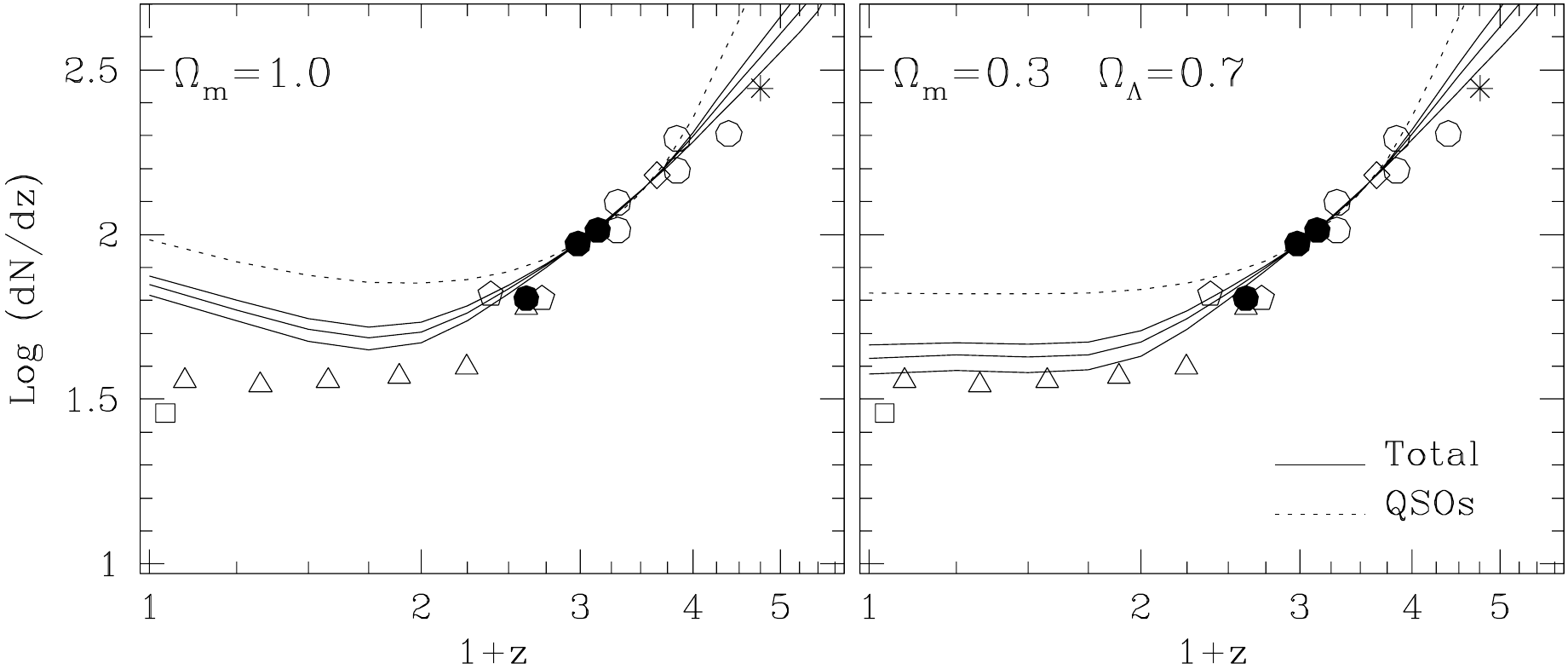}
\caption{Number density evolution of the \Lya forest with $\nhi = 10^{13.64-16} {\rm cm}^{-2}$ . Dotted lines refer to the evolution compatible with an ionizing UV background due only to QSOs. Solid lines show the expected evolution when both
QSOs and galaxies contribute to the background, for models with fesc=0.05 (upper line), 0.1 and 0.4 (lower line). Data points come from several observations in the literature (see \cite{Cristiani:2001xa}).}
\label{cristiani_fig1.eps}    
\end{figure}

The most impressive feature of the \Lya forest is the rapid increase of the number of lines with redshift, shown in Fig. \ref{cristiani_fig1.eps}. 
The expected number of absorbers per unit proper length is
$dN/dl_p=n_a(z)\sigma_a(z)$ with $n_a(z)$ as the number density of absorption
systems at redshift $z$. As $dl_p/dz = c/[H(z)(1+z)]$, where $H(z)=H_0E(z)$
is the Hubble parameter, the evolution in the number density is given by
\begin{equation}
\frac{dN}{dz} \simeq
(2100\,{\rm Mpc})n_{a, {\rm c}}(z)\sigma_a(z)(1+z)^{1/2}
\biggl[1+\frac{2.3}{(1+z)^3}\biggr]^{-1/2},
\label{eq:dNdz}
\end{equation}
if we assume a flat Universe ($\Omega_K=0$) and standard cosmological parameters,
$E(z)\simeq0.55(1+z)^{3/2}[1+2.3/(1+z)^3]^{1/2}$. For absorber with no intrinsic evolution, only moderate evolution is expected,
$dN/dz\propto(1+z)^{1/2}$. The maximum-likelihood fit to the data at $z > 1.5$ with the power-law parametrization discussed above gives $N(z) = N_0(1 + z)^\gamma = (6.5\pm 3.8) (1+z)^{2.4\pm 0.2}$. The UVES observations imply that the turn-off in the evolution does occur at $z \approx 1$. While the opacity is varying so fast, the column density distribution stays almost unchanged. 

\subsubsection{{\bf Column density distribution}}\label{Column}

The differential density distribution function measured by UVES, that is, the number of lines per unit redshift path and per unit $\nhi$ as a function of $\nhi$, follows a power-law: $f(\nhi) \propto \nhi^{-1.5}$. It extends over 10 orders of magnitude with little, but significant deviations: the slope of the power-law in the range $14 < \log\nhi  < 16$ goes from about $-1.5$ at $\langle z \rangle = 3.75$ to $-1.7$ at $z < 2.4$. This trend continues at lower redshift, with a slope of $-2.0$ at $z < 0.3$. Physically, the evolution of the $N(z)$ is governed by two main factors: the Hubble expansion and the metagalactic UV background (UVB). At high $z$ both the expansion, which decreases the density and tends to increase the ionization, and the UVB, which is increasing or non-decreasing with decreasing redshift, work in the same direction and cause a steep evolution of the number of lines. At low z, the UVB starts to decrease with decreasing redshift, due to the reduced number and intensity of the ionizing sources, counteracting the Hubble expansion. As a result, the evolution of the number of lines slows down. Up to date, numerical simulations have been remarkably successful in qualitatively reproducing the observed evolution, although more precise predictions are jeopardized by the yet poor knowledge of the evolution of the UV background (UVB), and in particular of the relative contribution of different sources (i.e. quasars vs. galaxies).  
One of the most widely used techniques used so far to determine the amplitude of the UVB is the so-called proximity effect (Bajtlik et al. 1988). The almost perfect power law distribution over a great dynamic range of values strongly suggests a single formation mechanism. 

\subsection{{\bf Basic physical processes}}
In this section we discuss the basic processes required to model the IGM and to interpret quasar absorption line experiments. A more detailed treatment can be found in \cite{Meiksin:2007rz}, \cite{Choudhury:2009kk}, \cite{Ciardi:2004ru}, \cite{Barkana:2000fd}.

\subsection{{\bf Photoionization}}
In the presence of an ionizing radiation, the evolution of the mean neutral 
hydrogen density $n_{\rm HI}$ is given by 
\be
\dot{n}_{\rm HI} = -3 H(t) n_{\rm HI} 
- \Gamma_{\rm HI} n_{\rm HI} + {\cal C} \alpha(T) n_{\rm HII} n_e
\label{eq:dnhidt_global}
\ee
where overdots denote the total time derivative $\de/\de t$, 
$H \equiv \dot{a}/a$ is the Hubble parameter, 
$\Gamma_{\rm HI}$ is the photoionization rate per hydrogen atom, 
$\alpha(T)$ is the recombination rate coefficient,  
$n_e$ represents the mean electron density, and $\cal C$ is the clumping factor.
This latter term accounts for the fact that the recombination rate in a inhomogeneous, clumpy IGM
is higher than a medium of uniform density. 
The first term on the r.h.s of Eq. \ref{eq:dnhidt_global} accounts instead for the dilution
in the density due to the expansion of the universe; the second term is the photoionization 
by the ionizing flux; the third term accounts for the recombination processes.

The photoionization rate factor $\Gamma_{\rm HI}$ represent 
the rate at which an hydrogen atom gets ionized per second, and it is
given by the number of  photons with energies exceeding the ionization potential of a 
bound electron in a hydrogen atom per neutral atom:
\begin{equation}
\Gamma_{\rm HI} = c\int_{\nu_{\rm T}}^\infty\,d\nu
\frac{u_\nu}{h_{\rm P}\nu}a_\nu^{\rm HI},
\label{eq:GHI}
\end{equation}
where $a_\nu^{\rm HI}$ is the hydrogen photoelectric cross section,
$\nu_{\rm T}$ is the threshold frequency required to ionize hydrogen
and $u_\nu$ is the specific energy density of the ultraviolet (UV) 
background radiation field (see Sect. \ref{UV}). 
The third term in Eq. \ref{eq:dnhidt_global} is the total recombination rate coefficient. 
Free electrons can be radiatively captured by protons 
at the rate per proton $n_e\alpha_A(T)$ where $\alpha(T)$ 
is the total rate coefficient for radiative capture
summed over recombinations to all energy levels.

If we indicate the ionization fractions with $x_{\rm HI}=n_{\rm HI}/ n_{\rm H}$ and
$x_{\rm HII}=n_{\rm HII}/ n_{\rm H}$, where $n_{\rm H}= n_{\rm HI} +
n_{\rm HII}$ is the total hydrogen number density, we can write the ionization rate equations as

\begin{eqnarray}
\frac{dx_{\rm HI}}{dt} &=& -x_{\rm HI}\Gamma_{\rm HI}
+x_{\rm HII}n_e\alpha_A(T), \nonumber\\
\frac{dx_{\rm HII}}{dt} &=& -\frac{dx_{\rm HI}}{dt}.
\label{eq:phionizeH}
\end{eqnarray}

In equilibrium, the neutral fraction will be
\begin{equation}
x_{\rm HI}^{\rm eq}=\frac{n_e\alpha_A(T)}{\Gamma_{\rm HI} + n_e\alpha_A(T)}.
\label{eq:phionizeHI-eq}
\end{equation}

Deriving the photoionization equations for helium is straightforward and the 
derived equations are analogous to Eqs.~(\ref{eq:phionizeH}).
If the fraction of \HeI, \HeII, and
\HeII \, are given by $x_{\rm HeI}$, $x_{\rm HeII}$ and $x_{\rm HeIII}$,
the equations look like:
\begin{eqnarray}
\frac{dx_\HeIs}{dt}  &=& -x_\HeIs \Gamma_\HeIs + x_\HeIIs n_e\alpha_\HeIIs, \nonumber \\
\frac{dx_\HeIIs}{dt}&=& -\frac{dx_\HeIs}{dt} - \frac{dx_\HeIIIs}{dt}, \nonumber \\
 \frac{dx_\HeIIIs}{dt}&=&  x_\HeIIs \Gamma_\HeIIs - x_\HeIIIs n_e\alpha_\HeIIIs, \label{eq:phionizeHe}
\end{eqnarray}
with obvious meaning of the terms. The total electron density is $n_e = n_{\rm HII} + n_{\rm HeII}
+ 2n_{\rm HeIII}$.

As the recombination rate coefficient is a function of the temperature, 
one needs to know the temperature of IGM. Photoionization will raise the temperature of the gas, by providing heat through the
excess energy absorbed by an electron above the ionization threshold.
The photoheating rate is
\begin{equation}
G_i = n_i c\int_{\nu^i_T}^{\infty}\frac{d\nu}{\nu}u_\nu \sigma_i(\nu)
(\nu-\nu^i_T),
\label{eq:PhotoionizationHeating}
\end{equation}
where $h_{\rm P}\nu^i_T$ is the ionization potential of species $i$.
The total heating rate from all species is
\begin{equation}
G = G_\HIs + G_\HeIs + G_\HeIIs.
\label{eq:netHeating}
\end{equation}

If we now indicate $L$ as the thermal loss function per unit volume, and with $n$ the particle density, the net heat transfer per particle is of $(G-L)/n$.  The rate of change of entropy per particle $s$ is given by the 
second law of thermodynamics:
\begin{equation}
\frac{ds}{dt}=\frac{1}{nT}(G-L),
\label{eq:dsdt}
\end{equation}
where $T$ is the temperature of the system.  For an ideal gas, we have $s=(\gamma-1)^{-1}k_{\rm B}\ln(p/\rho^\gamma)
+ s_0$, where $p$ is the gas pressure, $\rho$ is the mass density,
$\gamma$ is the heat capacity ratio (equal to 5/3 for a monatomic gas), 
$k_{\rm B}$ is Boltzmann's constant, and $s_0$ is an arbitrary additive constant. 

It is often useful to introduce the the entropy parameter 
\begin{equation}
S_E \equiv \frac{p}{\rho^\gamma}.
\label{eq:S}
\end{equation}
and to rewrite Eq.~(\ref{eq:dsdt}) as 
\begin{equation}
\frac{dS_E}{dt} = (\gamma-1)\rho^{-\gamma}(G-L);
\end{equation}
the gas temperature is related to $S_E$ by
\begin{equation}
T = \frac{\bar m}{k_{\rm B}} S_E \rho^{\gamma-1} 
\label{eq:TS}
\end{equation}
where $\bar m$ is the mean mass per particle.

Several radiative processes contribute to the loss function $L$,
including recombinations, collisional excitation of the
excited levels in neutral hydrogen, and inverse Compton scattering off
CMB photons. At high temperatures, free-free losses and collisionally
excited line radiation from atoms and ions other than neutral hydrogen
may become important contributions as well. 

\subsubsection{{\bf The UV background}}\label{UV}
As we have seen before, the term $u_\nu$ in Eq. \ref{eq:GHI} represents the specific energy density of the UV metagalactic background ionizing radiation field (UVB) . As the UVB intensity produced by astrophysical sources is absorbed and spectrally filtered by the IGM itself, one must account for these effects by solving the  full radiative transfer problem. These effects affect only mildly the UVB after reionization is complete as the mean free path of photons becomes very large (typically $\approx 30-100$ Mpc, set by the neutral islands leftover from reionization, often referred to as Lyman Limit Systems). However, during the various stages of reionization, the patchy nature of the ionization field induces very large spatial fluctuations in the intensity and spectral shape of the UVB which are crucial for many problems, as for example the detectability of the \HI 21 cm line form the Dark Ages. Hence in the following we will concentrate on the simples post-reionization uniform UVB case that is relevant to interpret the \Lya forest data.

In a homogeneous and isotropic expanding universe, we can write the integrated specific energy
density produced by sources with a total proper emissivity of 
$\epsilon_\nu(z)=4\pi j_\nu(z)$
as:
\begin{equation}
u_\nu(z) = \frac{1}{c}\int_z^\infty\, dz^\prime \frac{dl_p}{dz^\prime}
\frac{(1+z)^3}{(1+z^\prime)^3}\epsilon_{\nu^\prime}(z^\prime)
\exp[-\tau_{\rm eff}(\nu,z,z^\prime)],
\label{eq:unu}
\end{equation}
where $\tau_{\rm eff}(\nu,z,z^\prime)$ is an effective optical depth
due to absorption by the IGM and $\nu^\prime=\nu(1+z^\prime)/(1+z)$. The emissivity $\epsilon_\nu$ is the sum of two contributions: (a) isolated radiation sources, and (b) diffuse emission from the IGM itself. 
The source of the diffuse emission are \cite{Haardt:1995bw}: recombinations to the ground state of \HI\ and \HeII\ , 
\Lya recombination radiation of \HeII\ , \HeII\ two-photon continuum emission, and \HeII\  Balmer continuum emission. 
For what concern the discrete sources, these are mainly of two types: QSO-like and stellar-like sources. 
At $z< 2.5$, the UV ionizing background is dominated by quasar and Active Galactic Nuclei (AGN)
\cite{Haardt:1995bw}, while at redshift $z>3$, the density of the luminous quasar decreases faster than that 
of star-forming galaxies, and therefore at high redshift the ionizing background is dominated by stars. 

The most direct way to estimate the ionizing flux of both QSOs and galaxies is to integrate their luminosity function and
convert the total luminosity into the ionizing rate assuming the knowledge of the intrinsic spectrum. Although straightforward
in principle, in practice considerable uncertainties are present. These descend from the fact that a large fraction of the UV photons come from low luminosity QSOs that are below the current detection threshold. Based on these arguments and on the observed luminosity function at $z=6$, \cite{Fan:2001ff} showed that quasars could not have maintained the IGM ionized at that redshift. 

\begin{figure}
\centering
\includegraphics[width=8cm, height=8cm]{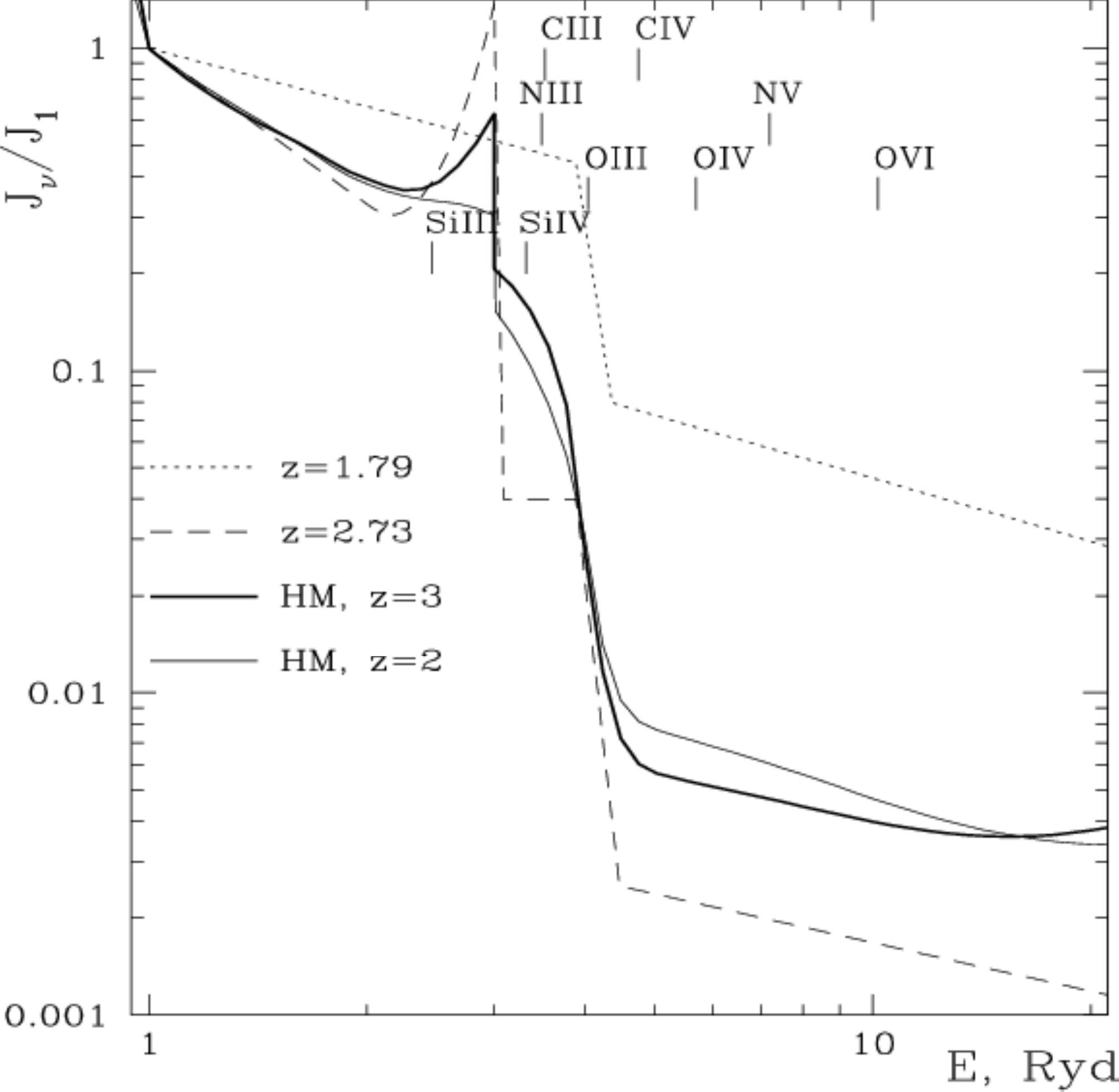}
\caption{(Taken from \cite{Meiksin:2007rz}  The UV metagalactic ionizing background intensity $J_\nu$,
normalized by the intensity $J_1$ at 1~Ry, from \cite{Haardt:1995bw} (HM lines) at $z
= 3$ (thick line) and $z=2$ (thin line).}
\label{fig:UVBG}
\end{figure}

In Fig. \ref{fig:UVBG} we show the angle-averaged intensity $J_\nu=cu_\nu/4\pi$ of the UV background
computed by \cite{Haardt:1995bw} at redshifts $z=2$
and $z=3$ (HM lines). The steps in the spectrum are produced by photoelectric absorption 
of the QSO radiation by intervening absorber, while the ``horns'' are due
to the re-processing of the ionizing radiation by the IGM into redshifted 
\HI\ and \HeII\ \Lya photons. At higher redshifts the UV background is resembles more closely  
the spectral shape of the sources as the IGM opacity is so large that it is dominated by local sources. 

Due to the rapid demise of the QSO population at high redshifts the most likely source of hydrogen-ionizing photons 
are the hot-young stars in star-forming galaxies, even if the spectra are too soft to produce appreciable \HeII-ionizing photons.
However there is still uncertainty in estimating the total UV 
photon emissivity from these star-forming galaxies, as such an estimate 
requires a bunch of theoretical assumptions regarding the Initial Mass Function, 
populations, ages and spectra of the massive stars, which dominate the radiation
at early times. 
A  large uncertainty is the star-formation rate at high redshifts: observational estimates 
are plagued by sample variance at $z\simeq 6$, due to the small volume of deep surveys, 
as well as extrapolations toward faint luminosities. 
Moreover, another major uncertainty is the escape fraction $f_{\rm esc}(\nu)$ of
ionizing radiation, as this must multiply the emissivity based on
galaxy counts and finally governs the contribution of the star-forming
galaxies to the UV background (see Sect. \ref{escape}).

The contribution of the galaxies to the UV background is still uncertain due to
several factors. On top of the ionizing radiation escape fraction
and UV-to-Lyman limit conversion factor, the effects of internal
galactic extinction, uncertainty in the amount of intergalactic
extinction by the IGM  and the minimum source luminosity on the UV luminosity density and its evolution
are source of uncertainties in the net contribution of the galaxies.
All errors combined together still leave
a range of uncertainties of about a factor of 4 for the contribution of the
galaxies to $\Gamma_{\rm HI}$, excluding the uncertainty of the escape
fraction.

\subsubsection{{\bf Proximity Effect}}\label{UV2}
The UV radiation from QSOs has been considered as the most natural origin for the ionization of the intergalactic gas. The finite number density of
QSOs suggests that there may be inhomogeneities in the ionization state of the \Lya clouds near each QSO. The term ``Proximity Effect'' refers to a relative lack of \Lya absorption in the vicinity of the background QSO. The effect was first discussed in the early '80s, when it was also suggested the currently accepted explanation of increased ionization of the clouds by the nearby QSO. It was then realized that the general increase of the absorption line density $dN/dz$ with redshift was accompanied by a simultaneous decrease of $dN/dz$ in each individual QSO spectrum when approaching the QSO's emission redshift. 

If the proximity effect is indeed caused by enhanced ionization measuring the intensity of the ionizing UV background becomes possible from observations of the density of lines, $dN/dz$, as a function of the distance from the QSO. Let us assume that in the vicinity of a QSO $dN/dz$ is reduced, presumably owing to the excess ionization of the gas. With increasing distance from the
emission redshift the QSO's ionizing flux decreases until the UV background intensity begins to dominate the ionization of the intergalactic gas. For example, at the point where the background intensity equals the QSO flux, $L_Q/(4\pi r_L)^2$ (known from photometry), the neutral column density
of a cloud should be lower by a factor of one half, with a corresponding decrease in $dN/dz$ for lines above a given detection threshold. In this way  the first crude measurements of the UV background radiation field were carried on, obtaining $J_{21}=3$, where $J=J_{21} \times 10^{-21}$ erg/cm$^2$/s/Hz/sr is the intensity at the Lyman limit, 912 \AA. This result was later confirmed using larger low resolution samples, obtaining $J_{21}=1^{+3.2}_{-0.7}$. 
Their measurement procedure (adopted by most later studies) consists of fitting the number density of lines per unit redshift distance $X = \int(1 + z)^\gamma dz$
\be
{dN\over dX} = \left({dN\over dX}\right)_0 \left(1+{L_Q\over 16\pi^2 r_L^2 J }  \right)^{1-\beta}
\ee 
as a function of the luminosity distance $r_L$, where the background intensity $J$ is the quantity desired. The quantity $\beta$ is again the exponent of the power law distribution of column densities. The largest compilations of high resolution data gave $J_{21}=0.5\pm 0.1$. None of the studies has found
evidence for a significant change with redshift (for $1.6 < z < 4.1$). However, more recent data (see review by Fan, Carilli \& Keating 2006) indicate a sharp drop of the UVB intensity for redshift above $z =5$, perhaps signaling the approach to the end of the reionization epoch. 

A more modern version of the proximity effect has been pioneered in the last few years by Adelberger and his group (Adelberger et al. 2003). In brief such technique uses a survey of the relative spatial distributions of galaxies and intergalactic neutral hydrogen at high redshift. By obtaining high-resolution spectra of bright quasars at intermediate redshifts ($z=3-4$) and spectroscopic redshifts for a large number of Lyman break galaxies (LBGs) at slightly lower redshifts, and by  comparing the locations of galaxies to the absorption lines in the QSO spectra shows that the intergalactic medium (at least for a good fraction of the galaxies belonging to the sample) contains less neutral hydrogen than the global average within about 1 comoving Mpc of LBGs and more than average at slightly larger distances. 

Although the interpretation of the lack of \HI absorption at small distances from LBGs as a result of a galaxy (as opposed to the previously discussed QSO) proximity effect is quite tempting. However, such explanation might be affected by the presence the galaxies' supernova-driven, which could produce a similar effect by heating the gas via their powerful shock waves. In any case this new technique is likely to open new avenues for the future better understanding of the very dynamic interplay between galaxies and the intergalactic medium.

\subsection{{\bf Gunn Peterson Effect}}\label{GP}
We have seen in the previous Sections that the primary evidence
for the ionized IGM at $z < 6$ comes from the 
measurements of GP optical depth in the spectra of QSOs.
Under the assumptions of photoionization equilibrium and 
a power-law relation between temperature and density, 
the Ly$\alpha$ optical depth $\tau_{\rm GP}$ arising from a region 
of overdensity $\Delta$ at a redshift $z$ can be written as

\begin{equation}
\tau_{\rm GP} = \frac{\pi e^2}{m_e c} f_{\alpha} \lambda_{\alpha} H^{-1}(z) n_{\rm HI },
\end{equation}
where $f_{\alpha}$ is the oscillator strength of the \Lya transition,
$\lambda_\alpha$ = 1216\AA, $H(z)$ is the Hubble constant at redshift
$z$, and $n_{\rm HI }$ is the density of neutral hydrogen in the IGM.
At high redshifts:
\begin{equation}
\tau_{\rm GP} (z) = 4.9 \times 10^5 \left( \frac{\Omega_m h^2}{0.13} \right)^{-1/2}
\left( \frac{\Omega_b h^2}{0.02} \right)
\left ( \frac{1+z}{7} \right )^{3/2}
\left( \frac{n_{\rm HI}}{n_{\rm H}} \right ),
\end{equation}
for a uniform IGM. From the above expression is clear that, even a tiny amount 
of neutral fraction $x_{HI} \sim 10^{-4}$, would produce an optical depth 
of the order of unity and hence would show
clear GP absorption trough in the spectra. 
Therefore, even if the source is located beyond the reionization epoch, 
a GP trough could be caused by some neutral residual fraction in
otherwise ionized regions, or maybe from individual damped Ly$\alpha$ absorbers.

Therefore, this test is mainly sensitive to the end of the reionization 
when the IGM is almost completely ionized, and it saturates for 
higher neutral fractions in earlier stage. 
As such absorption is not observed for QSOs at $z < 6$, 
this is a clear indication of the fact that the universe is highly ionized at $z < 6$.
The GP optical depth depends on the local IGM density, 
and by studying the evolution of the transmitted flux or the effective 
optical depth, one can trace the evolution of the neutral fraction 
of the IGM. 
A strong evolution of the \Lya absorption at redshifts $z> 5$ is evident from the study of \cite{Songaila:2004gk}, 
the transmitted flux approaches zero at $z> 5.5$, and for redshifts $z> 6$ complete absorption troughs 
begin to appear. 

However, the low sensitivity of the \Lya forest to small changes into the neutral hydrogen fraction, 
leaves opened a question: is the evolution a smooth transition or it is the result 
of a phase change in the status of the IGM?

For answering this question higher order information are needed. Indeed, 
for a source that is beyond the reionization epoch, but close to it, the GP trough
split into disjoint \Lya and \Lyb (and possibly higher Lyman series lines),  
with some transmitted flux in between them, that can be detected for
bright sources. 

The GP optical depth is proportional to $f\lambda$, where $f$ and $\lambda$
are the oscillator strength and rest-frame wavelength of the transition.
For the same neutral density, the GP optical depth of \Lyb and \lyc are
factors of 6.2 and 17.9 smaller than that of Ly$\alpha$.



\begin{figure}[thb]
\centering
\vspace{-0.8cm}
\includegraphics[width=0.8\textwidth]{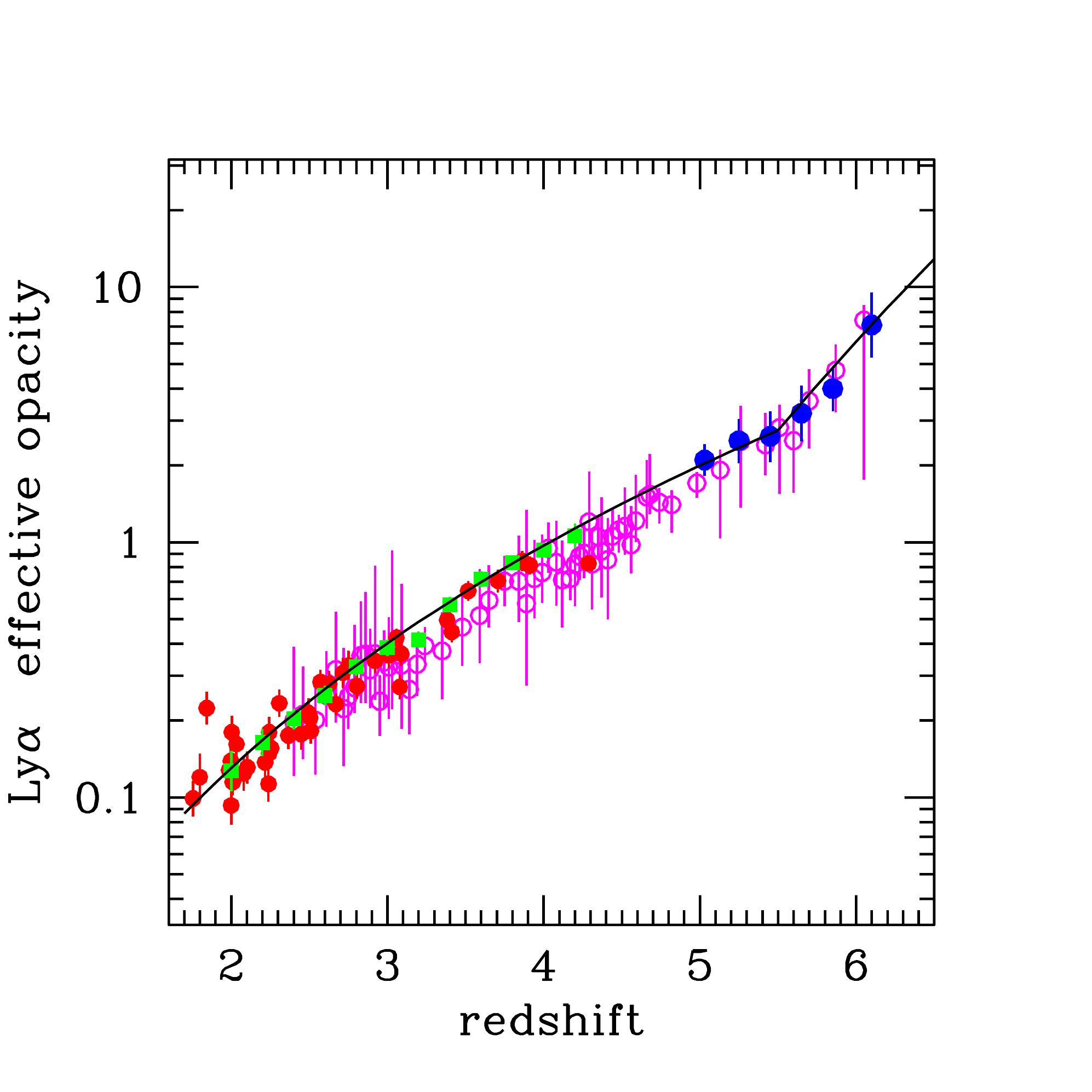}
\caption{(Taken from \cite{Haardt:2011xv}.) Evolution of the observed effective \Lya\ optical depth,
Data points are from \cite{Schaye2003}; {\it red filled circles}, \cite{Songaila2004}; {\it magenta empty squares}, 
\cite{Faucher2008}; {\it green filled squares}, and \cite{Fan:2005eq}; {\it blue filled circles}}. 
\label{fig2}
\end{figure}

Fan and collaborators found, in 2006 \cite{Fan:2005eq}, from the SDSS
survey of nineteen $z>5.7$ quasars, that for $z<5.5$ 
the optical depth can be fitted as $\tau \propto
(1+z)^{4.3}$, while for $z >5.5$ the evolution increases as  
the best fit is given by $\tau \propto (1+z)^{>10}$ (See Fig.~\ref{fig2}, taken from \cite{Haardt:2011xv}, for a collection of recent data on the evolution of the optical depth).

The increase in the optical depths as a function of redshift is larger than expected 
from a passive evolution of the density of the universe (dashed lines).
This could be easily interpreted as a feature of an increase of the neutral fraction
at high redshifts, a sign of the end of the reionization process, but the subject is 
controversial. The debated point is how much the neutral fraction increases
(see Sect. \ref{HIEvol}).

\subsubsection{{\bf \HI\ {\bf Evolution}}}\label{HIEvol}
From measurement of the GP optical depth one can infer also 
the (volume averaged) neutral hydrogen fraction, and therefore 
the ionization state of the IGM, through estimates of parameters like the mean UV
background and the mean free-path of UV photons.

Several studies (\cite{Fan:2001vx}, \cite{Songaila:2002kt}, \cite{Cen:2001us})
pointed out that the mean free path of UV photons at $z>6$ is of the same order
of magnitude of the clustering scale of the star-forming galaxies at the same 
redshifts, and therefore that the UV background is no more homogeneous
while approaching the beginning of the reionization.
Controversial results on the size of the ionized bubbles around quasars
may bring to the conclusion that the size of the observed
ionized regions cannot put stringent limits on IGM ionized fraction \cite{Ciardi:2004ru}.

\begin{figure}
\centering
\includegraphics[width=0.7\textwidth]{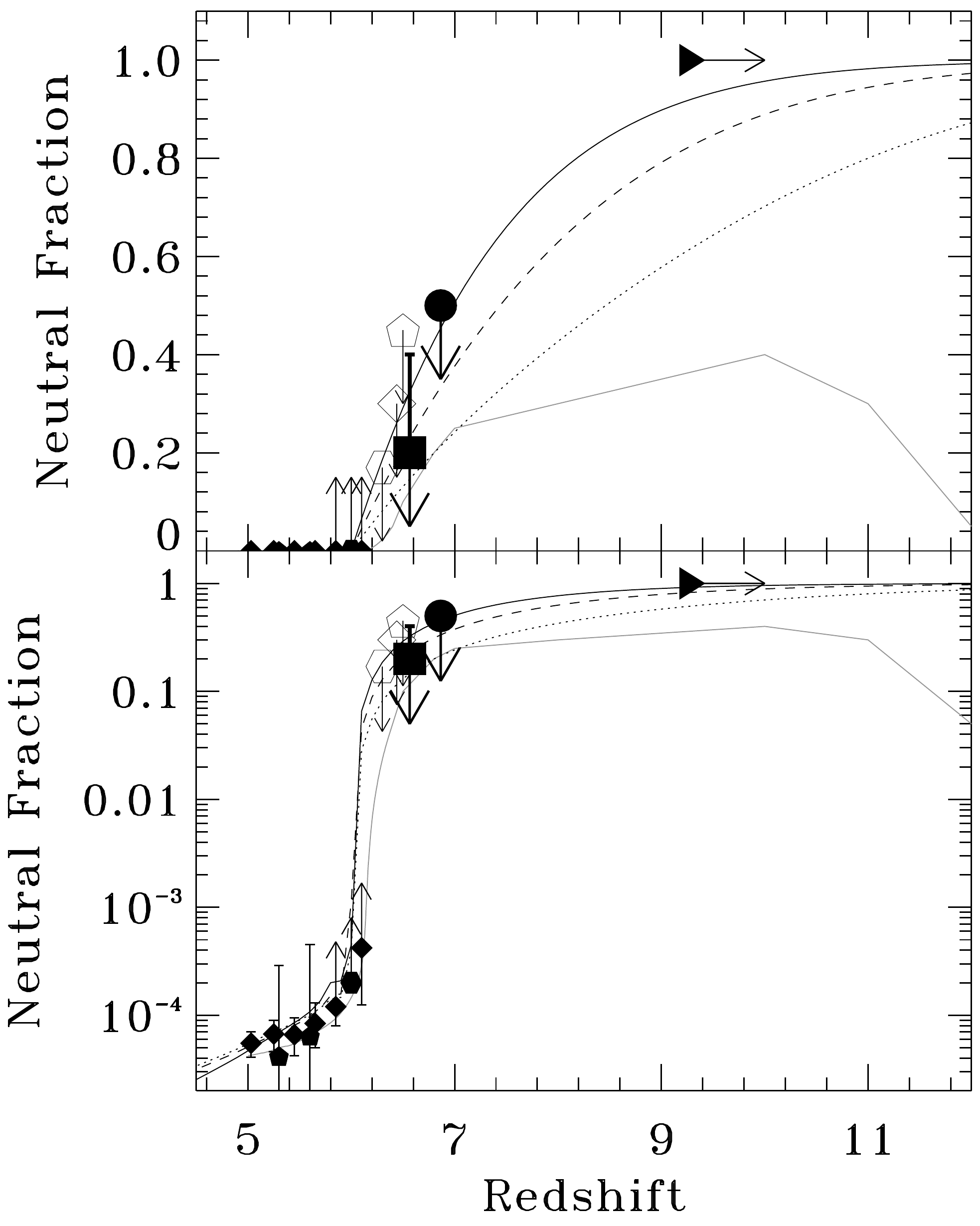}
\caption{(Taken from \cite{Ouchi:2010wd}) Neutral hydrogen fraction, $x_{\rm HI}$,
of IGM as a function of redshift. Top and bottom panels are the same, but with a vertical axis of linear and log scales, respectively. Filled square and circle are the upper limits of $x_{\rm HI}$ obtained in \cite{Ouchi:2010wd} from the evolution of Ly$\alpha$ LF and clustering, respectively. Open diamond and pentagon denote the upper limits from Ly$\alpha$ LF at $z=6.5$ given by \cite{malhotra2004} and \cite{kashikawa2006}. Open hexagon is the upper limit estimated from the constraints of Ly$\alpha$ damping wing of GRB at $z=6.3$ \cite{totani2006}. Filled hexagon and pentagons indicate constraints given by GRB spectra \cite{gallerani2008b} and QSO dark gap statistics \cite{gallerani2008a}, respectively. Filled diamonds represent the measurements from GP optical depth of SDSS QSOs \cite{Fan:2005eq}. Triangle plots the $1\sigma$ lower-limit of redshift of a neutral universe  given by WMAP7 \cite{larson2010} in the case of instantaneous reionization. Dotted, dashed, and solid lines show the evolution of $x_{\rm HI}$ for minihalo, small, and large halo cases, respectively, predicted by \cite{choudhury2008}. Gray solid line presents the prediction in the double reionization scenario suggested by \cite{cen2003}.
\label{fig:z_xHI}}
\end{figure}

Finally, some authors (see e.g. \cite{Fan:2001vx}, \cite{Cen:2001us}, \cite{Fan:2005es})
find that at $z>6$ the volume-averaged neutral
fraction of the IGM has increased to $> 10^{-3.5}$ (Fig. \ref{fig:z_xHI}, taken from \cite{Ouchi:2010wd}), 
although the poor sensitivity of the \Lya line could lead us to take it as only
an upper limit constraint.


\subsubsection{{\bf Primordial Cooling Function}}
Solar composition gas contains a mass fraction in heavy element that is about 2\%.
However, heavy element such as C$^+$, O, CO and dust grains are very effective at radiating thermal energy and therefore this small fraction of ``metals'' (as heavy elements are usually called in the astrophysical jargon) can control the gas thermal process.
On the other hand, a gas of primordial composition essentially does not contain heavy elements in the primordial gas. Because H and He atoms are very poor radiators for temperatures below $\approx 10^4$~K, if the gas were to remain purely atomic, the cloud would follow an almost adiabatic evolution. 

The main cooling processes in primordial gas clouds are   
\begin{itemize}
\item {\it Radiative recombinations} The thermal energy loss associated with the recombination of proton with electron is caused by the photon emitted during the process. The recombined atom which is in an excited state in general, eventually decays to the lowest energy level by emitting photons. Accordingly, the energy loss per one recombination process is the difference between the energy of the electrons at a bound state of hydrogen atom and the kinetic energy of the free electron.
\item {\it Collisional ionizations} The collisional ionization of the hydrogen atom is also a cooling process. The energy loss in this process is related to the ionization potential energy: the thermal energy of electrons is converted into the ionization energy by this process.
\item {\it Bound-bound transitions} This is the most important cooling process around 10,000 K. Collisionally excited atoms 
emit radiation of energy equal to the energy difference between two levels when electrons decay. The level population must be determined by solving the detailed excitation/de-excitation balance equation rates for each level.   
\item {\it Bremsstrahlung emission} Radiation due to the acceleration of a charge in Coulomb field of another charge is called bremsstrahlung or free-free emission.
\end{itemize}

The cooling function is dominated by bremsstrahlung for $T > 10^{5.5}$ K and by collisional excitation
of H and He in the range $10^4 < (T/{\rm K}) < 10^{5.5}$. However, below $10^4$ K
the cooling rate drops off very sharply. 
Therefore, for primordial gas clouds with temperature below $10^4$ K, the drop of the cooling function causes the cooling time to increase dramatically, thus preventing their collapse. The collapse condition required to form the first stars can then only be achieved thanks to the only other available coolant in the early universe: molecular hydrogen.
The process of \HH radiative cooling is the key ingredient to allow primordial gas to cool below 10,000 K. The hydrogen molecules have the energy levels corresponding to the vibrational transitions and the rotational transitions. 
Vibrational transitions are more important at high temperatures $10^3 < (T/{\rm K}) < 10^4$; rotational ones are more significant for $T < 10^3$ K. As \HH is present only in negligible abundances in the IGM we will not enter into details here, and refer the reader for an extended discussion on cooling by molecular hydrogen to \cite{Ciardi:2004ru} or \cite{Barkana:2000fd}.

Knowing all (i.e. for both H and He) rates, and further assuming thermal ionization-recombination equilibrium (although this assumption is not adequate for some applications;  full time-dependent equations must be solved to determine the ionization level in that case), we can estimate the total cooling rate, $L(T)/n_H^2$, where $T$ and $n_H$ are the gas temperature and hydrogen density, respectively:
\begin{equation}
L = L_\HIIs + L_\HeIIs + L_\HeIIIs + L_{eH} + L_C  + L_{ff}.
\end{equation}
where the first three terms are the radiative recombination rates,
the term $L_{eH}$ is the cooling rate due to the collisional excitation of \HI\ by
electrons. Cooling is also produced by the Compton scattering of CMB photons off the electrons, which is especially important at high redshifts. Finally, cooling by thermal bremsstrahlung radiation, or free-free emission, due to the acceleration of a charge in the Coulomb field of another charge may be significant.

\subsubsection{{\bf Equation of State}}\label{EOS}
\begin{figure}
\centering
  \includegraphics[width=12cm]{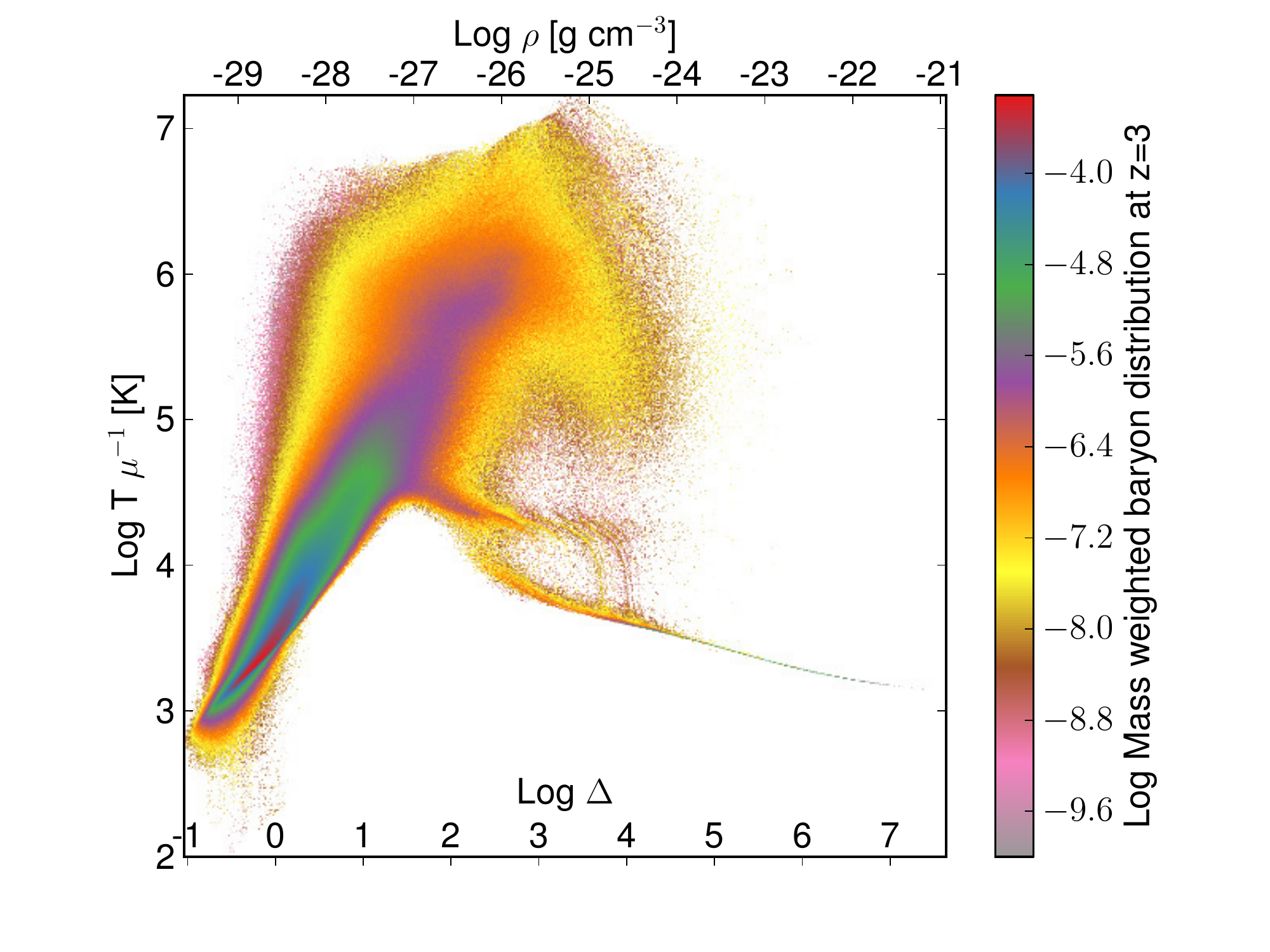}
\caption{IGM equation of state at $z=3$ (left). The colorbar represents the differential mass-weighted probability function; temperature is expressed in molecular weight units}.\label{stato_fig_03}
\end{figure}

Fig. \ref{stato_fig_03} shows the so-called phase diagram of the IGM (in this case, at redshift $z=3$) taken for a recent cosmological hydrodynamical simulation of \cite{Pallottini:2013rja}. One can immediately appreciate that the photo-ionized IGM, i.e. the \Lya forest is described by a well-defined temperature-density relation (Equation of State, EOS); a hotter and rarefied component, the ``warm-hot" phase is also present. Such gas has been shock-heated by either supernova energy injections and/or virialization shock in collapsing large-scale structures. The full IGM lies in a confined region in the $(\rho, T)$ plane, and its boundaries can be analytical well-defined by physical considerations \cite{Valageas:2001wh}.

Studying the thermal history of the IGM, through observations or cosmological simulations, can help
putting important constraints on reionization. Due to its low density, the IGM cooling time is long
and can retain some signatures of its past ionization and/or heating history.  Measuring the IGM temperature at high ($z \gsim 3.5$) redshifts can help in reconstructing, under certain assumptions, its thermal history up to the initial  
reionization stages. For example, the authors of \cite{theuns02, hui03} used the measured temperature at redshift $z\simeq3$ to set $z\approx 9$ as an upper limit for the reionization epoch (somewhat simplistically assuming an instantaneous reionization)  and \cite{bolton10} confirmed these findings with higher redshift quasars. The measured temperatures of the IGM at redshift $z\approx 3$ and $z\approx 6$ 
are too high for the bulk of reionization to have occurred at redshift $ z \gsim 10$. 

In the  linear and quasi-linear density regime the temperature-density relation for the photo-ionized forest gas follows a power-law, 
\begin{equation} 
T=\bar T \left(\frac{\rho}{\bar\rho}\right)^{\gamma-1},
\label{eq:IGMtemp}
\end{equation} 
where $\bar T$ is the temperature of the IGM at the mean density of the universe and $\gamma$ is the adiabatic power law index.
This relation follows from theoretical considerations \cite{hui97} and numerical simulations \cite{theuns98}, 
and describes a photo-ionized gas heated up to $T \sim 10^4$ K by the background UV flux. It corresponds
to the moderate density fluctuations which form the \Lya forest. This phase can be found in the left lower corner of the $(\rho, T)$ diagram, in the relatively low densities part of the diagram ($\Delta\lesssim 10^{2}$), and for temperatures $T\mu^{-1}\lesssim 10^{5}$ K, where $\mu$ is the mean molecular weight.

A second  ``warm'' phase is made of higher density regions, which are linked to the gravitational structures, and therefore are more scattered in the phase diagram. This ``warm'' IGM no longer follows the power-law equation of state, its temperature is governed by shock-heating due to gravitational dynamics, and depends on the properties of its neighboring dark matter density field. Therefore, it gains a stochastic component which 
which enlarges and spreads the density-temperature relation. This phase of the IGM is located at high temperatures $T >10^{5}$ K 

Finally, a third phase often denoted as CircumGalactic Medium (CGM) is located in the right lower corner of the $(\rho, T)$ diagram ($T\mu^{-1}\lesssim 10^{4}$ K and $\Delta\gtrsim 10^{4}$), and it refers to the gas which is in the process of cooling and falling to the center of the gravitational wells of dark matter haloes to form a disk and stars. The CGM works as an interface between the IGM and the galaxies and therefore cannot be strictly considered as parto fo the IGM; it is made of  high-density, relatively cold gas, which form pressure-supported clumps that can host star formation. This material is actually at the primary material out of which galaxies form. 

\subsubsection{{\bf Temperature Evolution}}
Associated with IGM reionization there is also 
a raising of its temperature (see e.g. \cite{MiraldaRees94}, \cite{Theuns02a}, \cite{RicottiOstriker04a}).
Because its cooling time is long, the low-density IGM
retains some memory of when and how it was reionized. The postionization
temperature is generally higher for harder spectra of the
ionizing radiation, but it depends also on the type of source.

\begin{figure}
\centerline{\includegraphics[width=15cm]{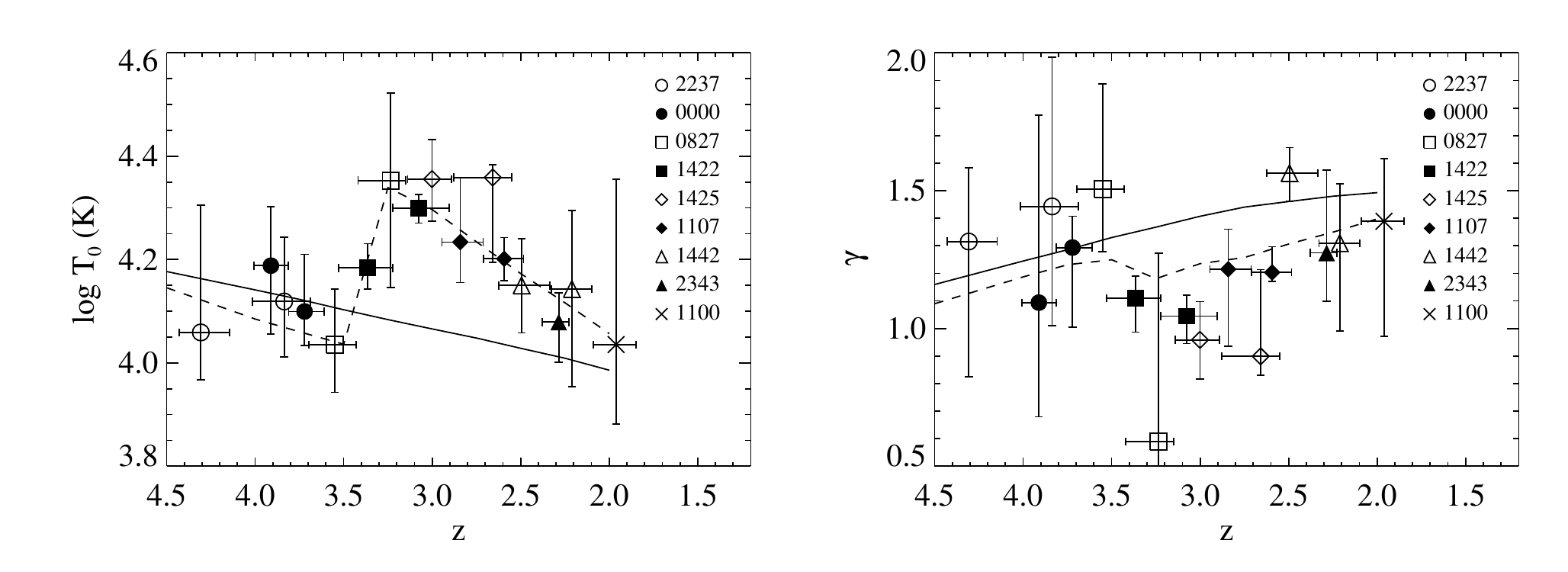}}
\caption{(Taken from \cite{Ciardi:2004ru})  Temperature at the mean density (left panel) and slope of the effective
equation of state (right panel) as a function of redshift. The dashed lines are
from a simulation designed to fit the data.}
\label{temp}
\end{figure}

There have also been suggestions (Ricotti, Gnedin \& Shull 2000;
Schaye et al. 2000b; Theuns, Schaye \& Haehnelt 2000) for a relatively sudden
increase in the line widths between $z=3.5$ and 3.0, which could be associated
with entropy injection resulting from the reionization of \HeII. While the
IGM can be heated up to T $\approx$ 10, 000 K during hydrogen reionization,
it can reach T$\approx$20, 000 K during helium reionization (see also Fig. \ref{temp}).

However, the existence of  the peak in the IGM temperature evolution is still
quite uncertain, as the data are also consistent with no feature at $z=3$.
From Fig.~\ref{temp} it is clear that the temperature peaks at $z \approx 3$
and that the gas becomes close to isothermal (adiabatic index $\gamma \approx 1$), but the present
constraints are not sufficient to distinguish between a sharp rise (as indicated
by the dashed line) and a more gradual increase.

Before completing our discussion on the QSO absorption lines, it is
worth mentioning a set of indirect constraints on reionization based on the
temperature of the IGM at $z \approx 2-4$. Using various techniques,
like, for example, the lower envelope of the neutral hydrogen column density and 
velocity width scatter plot \cite{stres00,rgs00}, or wavelet transforms
\cite{tszktc02}, one can infer the 
temperature of the IGM from absorption lines. These analyses
suggest that $T_0 \sim 1-2 \times 10^4$ K at $z \approx 3$, which 
in turn implies that hydrogen reionization must occur at $z < 9$ or, otherwise,
the temperature would be too low to match the observations. However, 
one should keep in mind that the analyses has large uncertainties, 
like, for example, the dust photoheating of the IGM
could give rise to high temperatures at $z \approx 3$ \cite{fnss99,nss99,ik03,ik04}. 
Furthermore,  a complex ionization history of helium could relax considerably
the constraints obtained from $T_0$
on the reionization epoch.

\subsubsection{{\bf IGM mass content}} \label{Mass within the cool phase} 
In~\cite{Fukugita:1997bi}, the authors presented an estimate of the baryon content of the universe at redshift 
$z=3$ and at redshift $z=0$. Stars and their remnants provide a very small contribution 
to the total budget, while the dominant baryonic mass component at redshift $z=3$ is in the IGM, in the intra-cluster medium in the form of \Lya forest gas, and in the high column density  damped-\Lya absorbers. The remaining part is in the form of cool plasma between the forest clouds.
In 2003 the authors of \cite{Fukugita:1997bi} revisited the cosmic baryon budget using the modern observation that in 
the meantime have become available \cite{Fukugita:2003gk}, focusing on the
present day ($z=0$) baryon content of the universe. 
Their conclusion was that the mean value of the data do not greatly 
differ from their previous estimate, but the increased accuracy reveals
a missing baryon component in the local universe, of about the $\simeq$ 35 \% of the total.

The ``missing baryon" problem remains an open issue (see e.g. \cite{Shull:2011aa}). In fact, the baryon census suffers of many uncertainties. The estimate of the state of the gas is made difficult by many factors, but mainly by the formation of galaxies and large-scale structure, and feedback from star formation. 
As the authors of \cite{Shull:2011aa} summarize, $\simeq$ 60\% of the baryons are in the 
IGM; 7\% is in galaxies and 4\% in clusters; an additional 5\% may be in the so-called circumgalactic 
gas, but this inventory leaves a fraction of $\simeq$ 30\% which is unaccounted for (see Fig. 10 of  \cite{Shull:2011aa}).
However, as the authors of  \cite{Shull:2011aa} also notice, most numerical simulation 
suggest that a substantial fraction of baryons exist in the so-called warm-hot IGM (see Sect. \ref{EOS})

The next question we can ask is about the redshift evolution of the fraction of the matter enclosed in the various IGM components. 
\begin{figure}[!htb]
\begin{center}
\label{frac}   
\includegraphics[width=8cm]{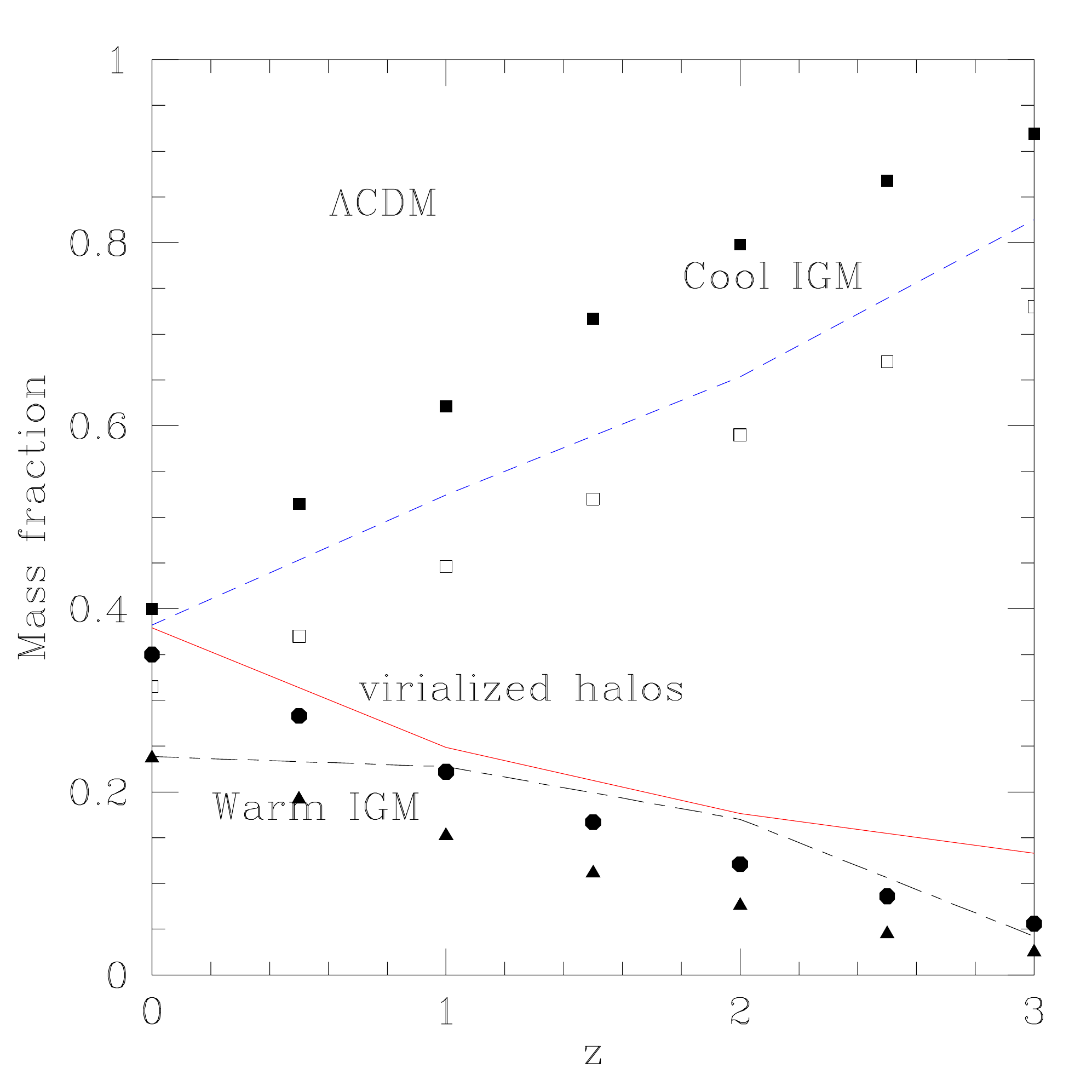}
\caption{ The distribution of baryonic matter. The dashed line which increases with redshift is the fraction of matter within the ``cool'' IGM phase, as computed in an approximate, analytical way. The dot-dashed line gives the mass fraction associated with the ``warm'' IGM component. The solid line is the fraction of matter within collapsed objects. The filled symbols show the results of the numerical simulations from \cite{Dave2} and  \cite{Dave1}. Adapted from \cite{Valageas:2001wh}.} 
\end{center}
\end{figure}
This problem has been addressed, e.g. by \cite{Valageas:2001wh}; their results are summarized in Fig. \ref{frac}. As we can see, at high redshifts most of the baryon content is in the ``cool'' phase of the IGM, i.e. the \Lya-forest, while a small fraction of the mass is located in collapsed objects. The mass fraction within collapsed halos and the ``warm'' IGM are of the same order of magnitude, as both component are related to the formation of non-linear structures. At lower redshifts, in the present universe, the three components contain approximately equal amounts of mass. Explaining these evolutionary trends is a necessary requirement for any viable IGM model.

\subsection{{\bf Galaxy-IGM interplay}}
Interactions between the IGM gas, galaxies and QSOs require a unified and complete treatment
of the IGM as a dynamic medium: the separation of these systems into distinct and isolated entities is  clearly an artificial
construct, as galaxies and QSOs originate from the IGM, and their radiation and outflows dramatical affect the intergalactic gas. 
Moreover, interpreting the increasingly refined observations requires detailed modeling using numerical simulations implementing
gas hydrodynamics, feedback, metal enrichment, radiative processes, and gravity (see Sect. \ref{simulations}).

\subsubsection{{\bf Ab Initio Models}}\label{abinitio}
The word ``feedback'' is by far one of the most used ones in modern cosmology where
it is applied to a vast range of situations and astrophysical objects. However, for
the same reason, its meaning in the context is often unclear or fuzzy. 
Let's start from setting the definition of feedback on a solid basis.  
We have found quite useful to this aim to go back to the Oxford Dictionary from where
we take the following definition:

{\it {\bf Feedback} n. 1. (Electr.) Return of fraction of output signal from one 
stage of circuit, amplifier, etc. to input of same or preceding stage 
({\bf positive, negative}, tending to increase, decrease the amplification, 
etc). 2. (Biol., Psych., etc) Modification or control of a process or system 
by its results or effects, esp. by difference between the desired and actual 
results.}

In spite of the broad description, we find this definition quite appropriate 
in many ways. First, the definition outlines the fact that the 
concept of feedback invokes a back reaction of a process on itself or on the causes that have produced it.
and the character of feedback can be either negative or positive. Moreover, and most
importantly, the idea of feedback is intimately linked to the possibility that a system
can become self-regulated. Although some types of feedback processes are disruptive, 
the most important ones in astrophysics are probably those that are able to drive the systems towards 
a steady state of some sort. To exemplify, think of a galaxy which is witnessing a burst of star formation. 
The occurrence of the first supernovae will evacuate/heat the gas thus suppressing the star formation
activity. Such feedback is then acting back on the energy source (star formation); it is of a 
negative type, and it could regulate the star formation activity in such a way that only 
a sustainable amount of stars is formed (regulation). However, feedback can fail to produce
such regulation either in small galaxies where the gas can be ejected by the first 
SNe or in cases when the star formation timescale is too short compared to the feedback one.   
As we will see there are at least three types of feedback, and even the mechanical feedback 
described in the example above is part of a larger class of feedback phenomena related to the 
energy deposition of massive stars. We then start by briefly describing the key physical ingredients of 
feedback processes in cosmology.
	
One of the main aims of physical cosmology is to understand in detail
galaxy formation starting from the initial density fluctuation field
with a given spectrum (typically a CDM one). Such {\it ab initio} computations
require a tremendous amount of physical processes to be included before it
becomes possible to compare their predictions with experimental data. In particular,
it is crucial to model the interstellar medium of galaxies, which is know to be
turbulent, have a multi-phase thermal structure, can undergo gravitational instability 
and form stars. To account for all this complexity, in addition one should treat 
correctly all relevant cooling processes, radiative and \index{shock} shock heating (let alone 
magnetic fields!). This has proven to be essentially impossible even with present day
best supercomputers. Hence one has to resort to heuristic models where simplistic 
prescriptions for some of these processes must be adopted. Of course such an approach
suffers from the fact that a large number of free parameters remains which cannot be
fixed from first principles. These are essentially contained in each of the ingredients
used to model galaxy formation, that is, the evolution of dark halos, cooling and star
formation, chemical enrichment, stellar populations. Fortunately, there is a large
variety of data against which the models can be tested: these data range from the
fraction of cooled baryons to cosmic star formation histories, the luminous content
of halos, luminosity functions and faint galaxy counts. The feedback processes enter the 
game as part of such iterative try-and-learn process to which we are bound by our
ignorance in dealing with complex systems as the galaxies. Still, we are far from 
a full understanding of galaxies in the framework of structure formation models.      
The hope is that feedback can help us to solve some ``chronic'' problems found in
cosmological simulations adopting the CDM paradigm.  Their (partial) list includes:

\begin{enumerate}
\item Overcooling: the predicted cosmic fraction of cooled baryons is larger
than observed. Moreover models predict too many faint, low mass galaxies;
\item Disk Angular Momentum: the angular momentum loss is too high and 
galactic disk scale lengths are too small;
\item Halo Density Profiles: profiles are centrally too concentrated;
\item  Dark Satellites: too many satellites predicted around our Galaxy.
\end{enumerate}

\subsubsection{{\bf Overcooling problem}}
Among the various CDM problems, historically the overcooling has been the most
prominent and yet unsolved one. In its original formulation it has been first spelled out by White \& Frenk (1991).
Let us assume that, as a halo forms, the gas initially relaxes to an isothermal
distribution which exactly parallels that of the dark matter. The virial
theorem then relates the gas temperature $T_{vir}$ to the 
circular velocity of the halo $V_c$,

\be
kT_{vir}=\frac{1}{2}\mu m_p V_c^2 \mbox{ or } T_{vir}=36 V^2_{c,km/s} \mbox{K},
\ee

\noindent
where $\mu m_p$ is the mean molecular weight of the gas. At each radius in 
this distribution we can then define a cooling time as the ratio of the 
specific energy content to the cooling rate,

\be
t_{cool}(r)=\frac{3\rho_g(r)/2\mu m_p}{n_e^2(r)\Lambda(T)},
\ee

\noindent 
where $\rho_g(r)$ is the gas density profile and $n_e(r)$ is the electron 
density. $\Lambda(T)$ is the \index{cooling!function} cooling function. The \index{cooling!radius} cooling radius is defined
as the point where the cooling time is equal to the age of the universe, i.e.
$t_{cool}(r_{cool})=t_{Hubble}=H(z)^{-1}$.

Considering the virialized part of the halo to be the region encompassing a
mean overdensity is 200, its radius and mass are defined by

\be
r_{vir}=0.1H_0^{-1}(1+z)^{3/2}V_c,
\ee
\be
\label{eq:mvir}
M_{vir}=0.1(GH_0)^{-1}(1+z)^{3/2}V_c^3.
\ee

Let us distinguish two limiting cases. When $r_{cool}\gg r_{vir}$ (accretion
limited case), cooling is so rapid that the infalling gas never comes to 
hydrostatic equilibrium. The supply of cold gas for star formation is then
limited by the infall rate rather than by cooling. The accretion rate is obtained
by differentiating equation~(\ref{eq:mvir}) with respect to time and multiplying
by the fraction of the mass of the universe that remains in gaseous form:

\be
\label{eq:macc}
\dot{M}_{acc}=f_g\Omega_g \frac{d}{dt} 0.1 (GH_0)^{-1}(1+z)^{3/2}V_c^3= 
0.15 f_g \Omega_g G^{-1}V_c^3.
\ee

\noindent 
Note that, except for a weak time dependence of the fraction
of the initial baryon density which remains in gaseous form, $f_g$, this
infall rate does not depend on redshift. 

In the opposite limit, $r_{cool}\ll r_{vir}$ (quasi-static case), the 
accretion \index{shock} shock radiates only weakly, a quasi-static atmosphere forms, and
the supply of cold gas for star formation is regulated by radiative losses
near $r_{cool}$. A simple expression for the inflow rate is 
\be\label{eq:mqst}
\dot{M}_{qst}=4\pi\rho_g(r_{cool})r_{cool}^2 \frac{d}{dt}r_{cool}.
\ee
In any particular halo, the rate at which cold gas becomes available for star
formation is  the minimum between $\dot{M}_{acc}$ and $\dot{M}_{qst}$.
Integrating the gas supply rates $\dot{M}_{cool}$ over redshift and 
halo mass distribution, we find that for $\Omega_b=0.04$ {\it most of the gas
is used before the present time}. This is unacceptable, since the density contributed
by the observed stars in galaxies is less than 1\% of the critical density. 
So, the star formation results to be too rapid without a regulating process,
i.e. feedback (``cooling catastrophe''). 

To solve this puzzle, Larson (1974) proposed that the energy input from young stars
and supernovae could quench star formation in small protogalaxies before more
than a small gas fraction has been converted into stars. 
Stellar energy input would counteract radiative losses in the cooling gas and
tend to reduce the supply of gas for further star formation. 

We can imagine that the star formation process is self-regulating in the sense
that the star formation rate $\dot{M}_\star$ takes the value required for heating to balance 
dissipation in the material which does not form stars. This produces the 
following prescription for the star formation rate:

\be
\dot{M}_\star(V_c,z)& =& \epsilon(V_c) \mbox{min}(\dot{M}_{acc},\dot{M}_{qst}),\\
\epsilon(V_c)& =& [1+\epsilon_0(V_0/V_c)^2]^{-1}.
\ee

\noindent
For large $V_c$ the available gas turns into stars with high efficiency because
the energy input is not  sufficient to prevent cooling and fragmentation; for
smaller objects the star formation efficiency $\epsilon$ is proportional to 
$V_c^2$. The assumption of self regulation at small $V_c$ seems plausible 
because the time interval between star formation and energy injection is much
shorter than either the sound crossing time or the cooling time in the 
gaseous halos. However, other possibilities can be envisaged. For example, 
Dekel \& Silk (1986) suggested that supernovae not only would suppress cooling
in the halo gas but would actually expel it altogether. The conditions leading 
to such an event are discussed next. 

\subsection{{\bf Heating and Metal Enrichment}}
One of the most obvious signatures of  (mechanical) feedbacks is the metal enrichment
of the intergalactic medium.  The same stars that ionize the IGM are responsible for its metal enrichment.
A number of studies (for a review see Section 5.4 of  \cite{Ciardi:2004ru}) reached 
the conclusion that metal pollution from stellar wind is negligible for low metallicity
stars, as therefore Pop III stars contribute to the metal pollution
mainly from the later stages of their evolution.  
\begin{figure}
\begin{center}
\includegraphics[width=11cm]{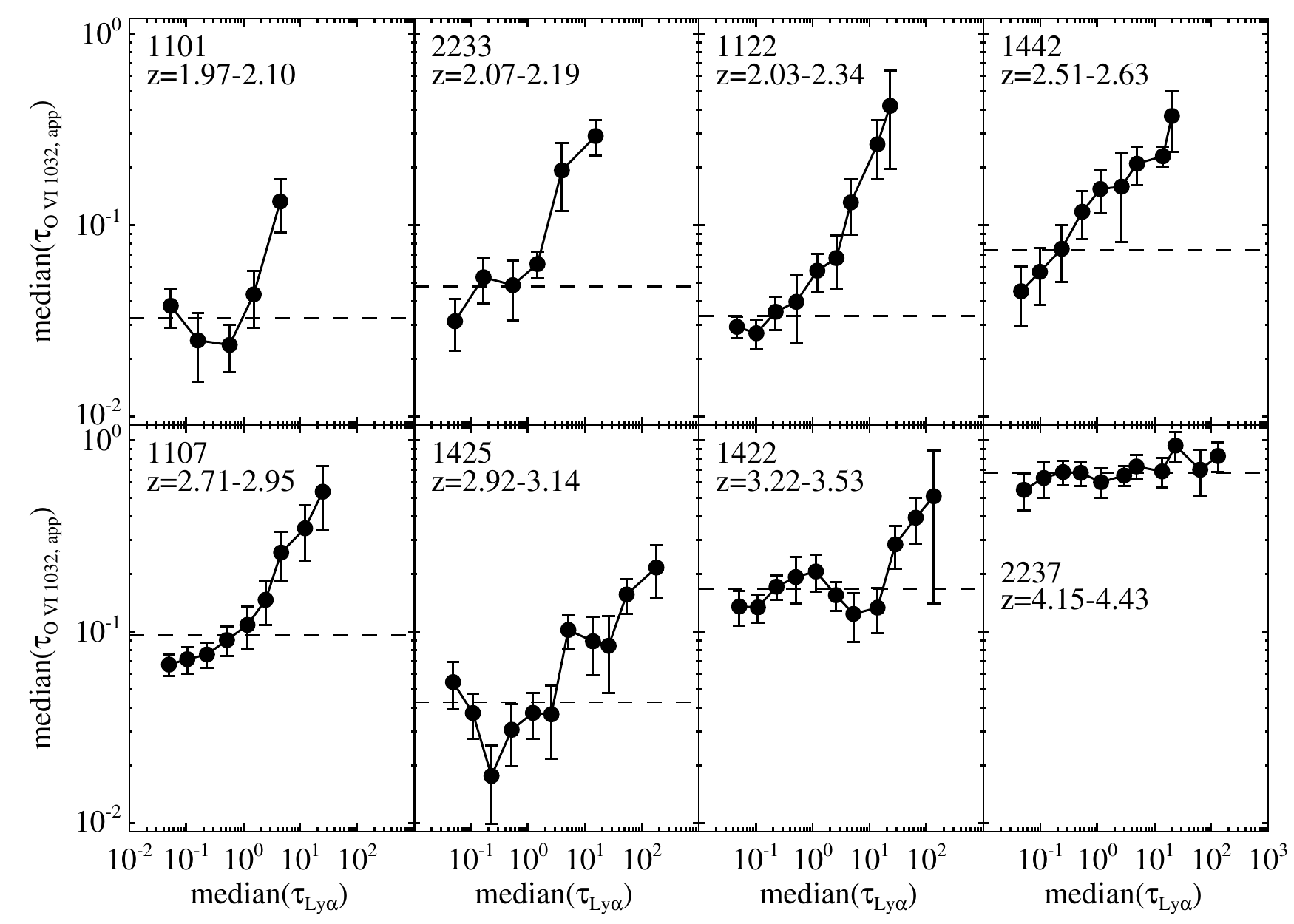}
\end{center}
\caption{(Taken from \cite{schaye00}) IGM Ly$\alpha$-\OVI optical depth correlation from the pixel analysis
of the spectra of 8 QSOs. Vertical error bars are $1\sigma$ errors,
horizontal error bars are smaller than the symbols and are not
shown. For $z\simlt 3$ $\tau_{\rm \OVI, app}$ and $\tau_{\rm HI}$ are
clearly correlated, down to optical depths as low as $\tau_{\rm HI}
\sim 10^{-1}$. A correlation between $\tau_{\rm \OVI, app}$ and
$\tau_{\rm HI}$ implies that \OVI absorption has been detected in the
\index{Ly$\alpha$!forest} Ly$\alpha$ forest.}
\label{fig:ovi}
\end{figure}
One would expect a positive correlation between density and metallicity, 
as stars, as the production locus for metals, are almost 
always associated with high density regions.
Although such positive correlation is seen, 
it is not as strong as expected from the above simple argument. 
Several different mechanisms have been suggested to remove metals 
from the galaxies in which they are produced, among which we can list:
dynamical encounters between galaxies, ram-pressure stripping, 
supernova-driven winds.

If powerful winds are driven by supernova explosions,
one would expect to see widespread traces of heavy elements away from their
production sites, i.e. galaxies. 
Schaye et al. (2002) have reported the detection of \OVI in the low-density
\index{IGM} IGM at high redshift. They perform a pixel-by-pixel search for \OVI 
\index{absorption} absorption
in eight high quality quasar spectra spanning the redshift range $z=2.0-4.5$.
At $2\simlt z\simlt 3$ they detect \OVI in the form of a positive correlation
between the HI Ly$\alpha$ \index{optical depth} optical depth and the optical depth in the
corresponding \OVI pixel, down to $\tau_{\rm HI}\sim 0.1$ (underdense regions),
that means that metals are far from galaxies (Figure~\ref{fig:ovi}). 
Moreover, the observed narrow widths of metal \index{absorption} absorption lines (CIV, SiIV)
lines lines imply low temperatures $T_{\rm \index{IGM} IGM}\sim {\rm few}\times 10^4$ K.

A natural hypothesis would be that the  Ly$\alpha$ forest has been enriched by metals ejected
by Lyman Break Galaxies at moderate redshift. The density of these objects is
$n_{\rm LBG}=0.013$ $h^3$ Mpc$^3$. A (metal) filling factor (that is, the fraction of cosmic volume
occupied by ionized (metal) bubbles produced by sources)
of $\sim 1$\% is obtained
for a \index{shock} shock radius $R_s=140$ $h^{-1}$ kpc, that corresponds at $z=3$ ($h=0.5$) 
to a \index{shock} shock velocity $v_s=600$ km s$^{-1}$. In this case, we expect a 
post\index{shock} shock gas temperature larger than $2\times 10^6$ K, that is around hundred times what
we observed.
So the metal pollution must have occurred earlier than redshift 3,
resulting in a more uniform distribution and thus enriching vast regions of 
the intergalactic space. 
\begin{figure}
\centering
\includegraphics[width=8cm]{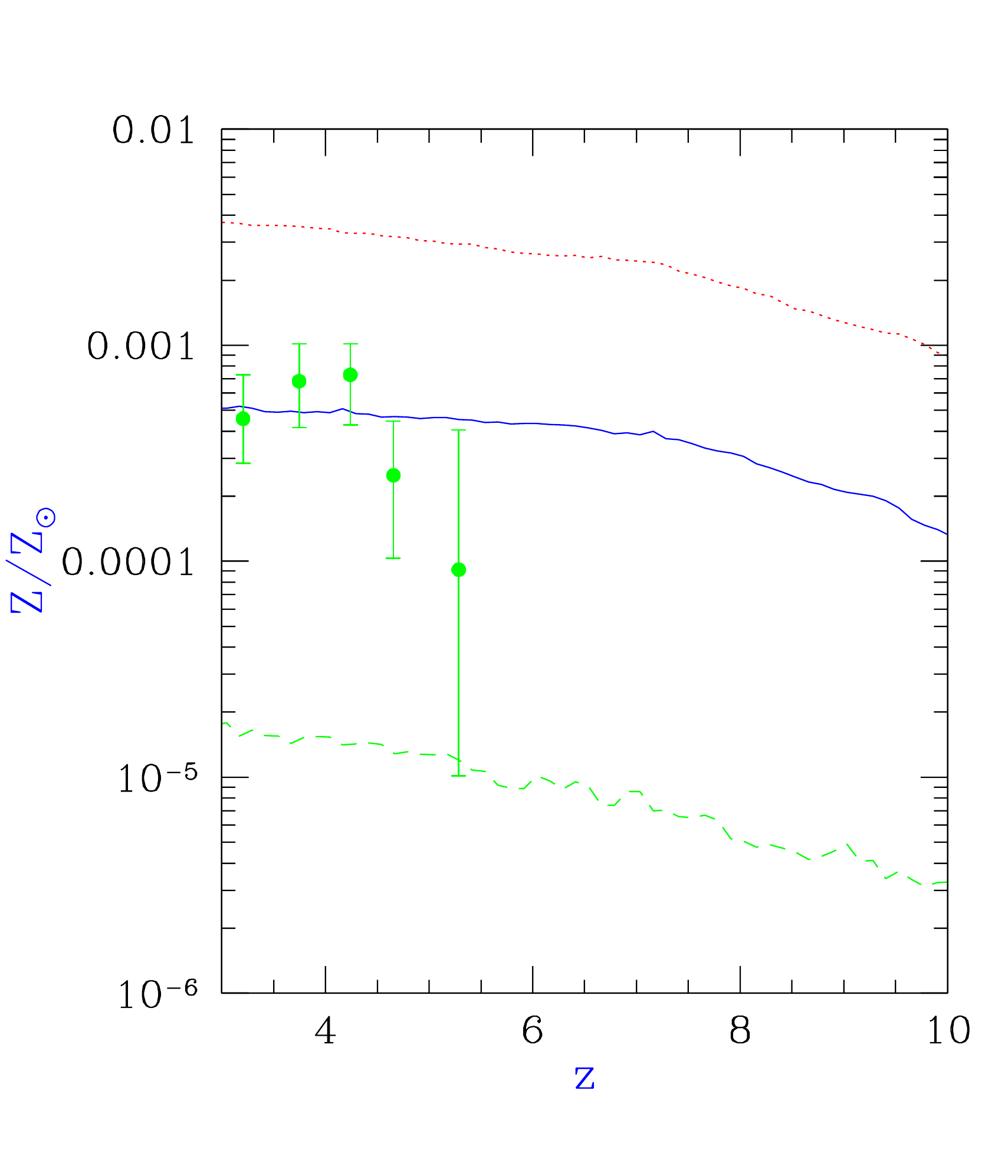}
\caption{ Metallicity evolution for low-density ($12  < \log N_{HI} < 15$ ) IGM. 
Points are the data of Songaila (2001); dotted, solid, and dashed lines are the 
predicted values for star formation
efficiencies 0.5, 0.1, and 0.01, respectively (see \cite{Scannapieco:2002ad})}
\label{smf3}
\end{figure}

Metal ejection by galactic winds is one of the most popular mechanism
for the IGM pollution. 
Madau et al. in 2001 (\cite{Madau:2000pa}) found that pre-galactic outflows from the same primordial halos that reionize the IGM,
could also pollute it to a mean metallicity of $Z \gsim 10^{-3}~Z_\odot$.
Metals produced in these low-mass halos can more easily escape from 
their shallow gravitational potential wells, they have to travel much shorter 
distances between halos and therefore it might be easier to create
non-negligible metal filling factors. On top of that, the (relatively) small velocities
associated with these outflows may not affect the thermal status of the IGM. 
The volume filling factor of the ejecta can be up to 30\% or higher, 
according to the star formation rate, and the majority of the metal 
enrichment occurs relatively early, between redshifts $z\simeq 14$ and $z\simeq 5$.
This allows the \index{Ly$\alpha $!forest} Ly$\alpha$ forest to be hydrodynamically
`cold' at low redshift, as intergalactic baryons have enough time to relax again
under the influence of dark matter gravity only \cite{Scannapieco:2002ad}.

Fig \ref{smf3} shows the metallicity evolution as a function of redshift and as a function of star
formation efficiency (taken from \cite{Scannapieco:2002ad}).
This early enrichment model can explain quite well the observed redshift evolution of 
elemental abundances derived from QSO absorption lines experiments.

%
%
\section{Lecture II -- Cosmic Reionization}
\subsection{\bf{Introduction}}
Given the observational constraints discussed in the previous Section (but see also Sect. \ref{other}), is important to develop models which can explain every data, 
and which can be constrained by data. The main difficulty lies into the complexity of the process:  the reionization process is tightly
connected to the properties and evolution of star-forming galaxies and QSOs (see Sect. \ref{global}). 
Moreover, very little is known about the first generation of stars and galaxies: there are strong theoretical and numerical studies that 
the first generation stars were massive and metal-free, with a very different kind of spectrum than the stars 
we observe today, known as the PopIII stars (see Sect. \ref{sources}).
Similarly, there is still a lively debate on the existence of  undetected low-luminosity
QSOs powered by intermediate mass black holes at high redshift
\cite{ro04b} (see again Sect. \ref{sources}).

Other difficulties arise in the issue of a self-consistent formulation of the reionization process, among 
which we can list: (i) Escape fraction: we still do not have a clear idea on how the photons ``escape'' from 
the host galaxy to the surrounding IGM (see Sect. \ref{escape}) and (ii) Feedbacks effects: mechanical and radiative feedback processes can alter the (hierarchical) structure
formation sequence of structure (see Sect. \ref{abinitio}).

Analytical models of reionization parametrize the different contribution of the sources
through various free parameters (like efficiency of star formation, 
fraction of photons which can escape to the IGM, etc. see Sect. \ref{constraining}).
Although these models can obtain a good general picture of the process 
and can span a larger range of the physical plausible values for the parameters compared to numerical 
simulations (see Sect. \ref{simulations}), is difficult through them to picture the details of the reionization, especially those of the 
pre-overlap phase (see Sect. \ref{prop}), such as the shape of ionized region around sources and their overlapping.
The alternative way, i.e. numerical solutions, is a demanding task because of the 
large range of temporal and spatial scales that are needed in order to follow the
ionization history (again, see Sect. \ref{prop}), from the IGM inhomogeneities, 
to the sizes of the ionized regions, and this is still far beyond our computational capabilities.

However, one must realize that in spite of these 
difficulties in modeling reionization
there have been great progresses in recent years, 
both analytically and 
through numerical simulations, in different aspects of the process.

\subsection{{\bf Basic Theory}}
In this section we will examine all the ingredients which are needed to be considered in order to build a 
global model for the reionization process, and we will highlight some of the successes of our understanding
in this field. 

\subsubsection{{\bf Propagation of ionization fronts in the IGM}}\label{prop}
\begin{figure}
\begin{center}
\includegraphics[width=5cm]{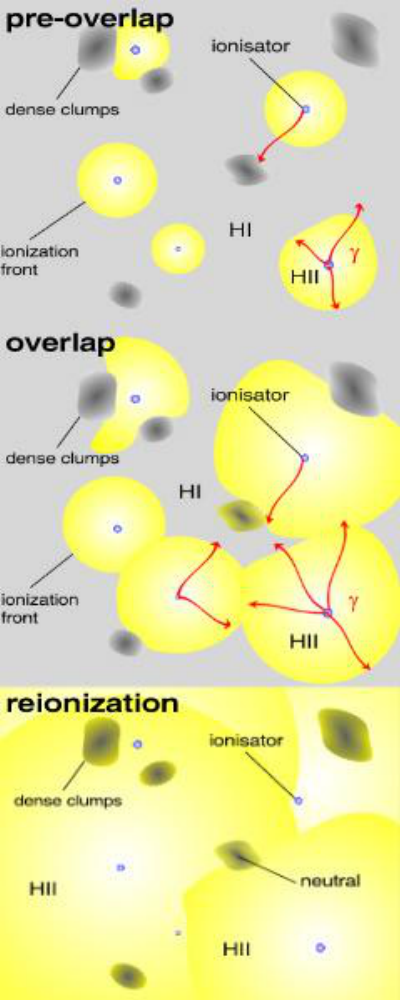}
\end{center}
\caption{Schematic picture of the reionization phases according to the terminology introduced by \cite{Gnedin:1999fa}. Picture taken from http://www.wissenschaft-online.de/astrowissen}
\label{figI4new}
\end{figure}

According to the terminology introduced by \cite{Gnedin:1999fa} the reionization 
process can be divided in three phases (see Fig. \ref{figI4new}): ``pre-overlap'', ``overlap'', and ``post-overlap'' phase.
In the initial, pre-overlap phase each individual source creates a ionizing shell in its surrounding, 
as it turns on. As we saw in the previous Sect. and in Sect. \ref{UV}, the first galaxies formed in the most 
massive halos, in the region with the highest density. At this stage the IGM is a two-phase medium, 
with ionized regions separated from the neutral ones by the ionization fronts. The degree of ionization
is very inhomogeneous and its intensity depends on the distance to the source and its efficiency. 
At high redshifts, the mean free path of ionizing photons is so small that we can safely treat the radiation locally: it means that 
we can assume that all the photons are absorbed shortly after being emitted and therefore that the background 
intensity depends only on the instantaneous value of the emissivity, and not on his history. 

In the so-called ``overlap'' phase,  the ionized regions starts to overlap and eventually ionize the whole IGM.
The mean free-path of the UV photons increases dramatically, and the universe eventually become transparent to UV radiation.
This stage is predicted to occur very rapidly (or, in other terms, to be a phase transition): the ionizing intensity
in overlapping $\HII\ $ regions increases rapidly for the contribution of two (or more) sources, and so it 
is able to expand faster in the neutral regions which was previously characterized by an high recombination 
rate as the ionizing intensity has been low.
At the end of this stage the IGM is highly ionized everywhere except in the self-shielded, high density clouds.

Finally, there is a ever-continuing ``post-overlap phase''  in which the ionization fronts propagate into the neutral 
high-density regions, and this phase continues indefinitely as collapsed objects retain neutral hydrogen even in the 
present universe. 
%
%

In this Section we deal with the pre-overlap phase, and so our aim would be
to describe how each ionization front propagates, and how distinctive ionized regions
evolve and expand according to the different sources, eventually overlapping with the neighboring ones, 
marking the end of the pre-overlap phase. 
Assuming a spherical ionized (comoving) volume $V$, an ionized proper volume $V_p$
is given by the following equation
\beq 
\nb_H V_p=\Ng\ 
\eeq 
where $\nb_H$ is the mean number density of hydrogen and $\Ng$ is the total number
of ionizing photons produced by the source.
To make a step further, if we instead consider also recombinations  (and neglecting the expansion of the universe), 
in case of a steady ionizing source, the resulting steady ionized volume in which recombinations balance ionization is given
by the Str\"{o}mgren sphere:
\beq 
\alpha_B \nb_H^2 V_p=\frac{d\, \Ng}{dt}\ ,
\eeq
where the recombination rate $\alpha_B$ depends on the square of the density.
Finally, including in the equation the expansion of the universe, 
the exact evolution of an expanding $\HII\ $region is given by
\beq 
\nb_H\left( \frac{dV_p}{dt}-3 H
V_p\right)= \frac{d\, \Ng}{dt} - \alpha_B \left<n_H^2\right> V_p\ , 
\label{front} 
\eeq
in which the the mean density $\nb_H$ varies
with time as $1/a^3(t)$.

As the recombination rate is dependent on the square of the density,
we should introduce a factor able to take into account the 
non-uniformity of the early universe IGM, where the gas is mostly distributed in 
high-density clumps. This issue is loaded on  the so-called clumping factor, 
that is volume averaged, and in general time-dependent, defined as:
 \beq C=\left<n_H^2\right>/\nb_H^2 \label{clump}\
. \eeq

If we insert Eq. \ref{clump} into Eq. \ref{front}, and write it in comoving coordinates 
we find:
\beq \frac{dV}{dt}=
\frac{1}{\nb_H^0} \frac{d\, \Ng}{dt}- \alpha_B \frac{C}{a^3} \nb_H^0
V\ , \label{HIIreg} 
\eeq 
where the present number density of hydrogen
is \beq \nb_H^0=1.88\times 10^{-7} \left(\frac{\Omega_b
h^2}{0.022}\right)\ {\rm cm}^{-3}\ . \eeq 

The solution of this equation for the volume $V(t)$ at the time $t$ around a source which turns on at the time
$t=t_i$ is:
\beq
V(t)=\int_{t_i}^t \frac{1}{\nb_H^0} \frac{d\, \Ng}{dt'} \,e^{F(t',t)}
dt'\ ,\label{HIIsoln} 
\eeq 
where 
\beq F(t',t)=-\alpha_B \nb_H^0
\int_{t'}^t \frac{C(t'')} {a^3(t'')}\, dt''.\
\label{Fgen} 
\eeq 
The volume of the ionized  $\HII\ $region depends on the type and luminosity  of the source
which produces it. 

We can define the overall number of ionizations per baryon as 
\beq \label{Nion}
\Ni \equiv N_{\gamma} \, f_{\rm star}\, f_{\rm esc}\ . 
\eeq 
where $ f_{\rm star}$ is the efficiency with which baryons are incorporated into sources (star formation efficiency), and $f_{\rm esc}$ is the escape fraction for the resulting radiation, that is, how many baryons can escape into the IGM.
For QSOs, the production efficiency of ionizing photons is lower, but the escape fraction is likely to be an order of magnitude higher; 
thus, we obtain similar results for the total number of ionizing photons per baryon produced by different sources.
Therefore, we get similar results for the evolution of the hydrogen ionization front if we consider quasars rather than stars, 
while for what concerns the second reionization of helium the propagation of ionization fronts for the two 
types of source dramatically changes: normal stars are hardly able to produce any photons above the 54.4 eV threshold needed to 
tear apart the second electron from a He atom.

\subsubsection{{\bf Statistical Approach}}
Now that we have discussed the evolution of each individual ionization 
bubble, the next step is to describe statistically the global distribution 
of the ionized volumes in the pre-overlap phase. 
The statistical quantity normally used is the volume filling factor
of ionized regions, $Q_I$, i.e. the fraction of cosmic volume filled 
by $\HII \ $ regions. 

Starting from Eq.  \ref{HIIreg}, assuming that the sources of radiation
are distributed uniformly over the volume and a common clumping factor $C$
for all the $\HII \ $ regions, we can sum the terms of the equation 
over all bubbles in a given large volume, and divide by this volume. 
Therefore, we replace $V$ with the filling factor and $N_{\gamma}$
with the total number of ionizing photons produced up to some 
time $t$, per unit volume.

We can also re-write Eq. \ref{Nion} as 
 \beq
\frac {\nb_\gamma} {\nb_b}= \Ni F_{\rm col}\ , \label{ngnb} 
\eeq
where ${\nb_b}$ is the mean density of baryons, 
and this quantity is the number of ionizing photons
per baryons per unit volume, and $F_{\rm col}$ is the fraction of all the baryons
in the universe which are in galaxies. 

Under these assumption we can therefore re-write Eq.  \ref{HIIreg}
as 
\beq \frac{dQ_{\rm H\ II}}{dt}=\frac{\Ni}{0.76} \frac{dF_{\rm
col}}{dt}- \alpha_B \frac{C}{a^3} \nb_H^0 Q_{\rm H\ II}
\label{QIIeqn}\ , \eeq
where 0.76 is the assumed primordial mass fraction og hydrogen. 

The solution  is \beq Q_{\rm H\ II}(t) =\int_{0}^t
\frac{\Ni} {0.76} \frac{dF_{\rm col}}{dt'}\,e^{F(t',t)} dt'\ , \eeq
where $F(t',t)$ is determined by
equation \ref{Fgen}.

\subsubsection{{\bf ``Hello world'' Reionization Model}}
Rough estimates for the quantities into play can lead to 
the building of a first, simple, reionization picture.  

An estimate for $F_{col}$ at high redshifts is the mass fraction in halos above the cooling
threshold, that is above the minimum mass of an halo in which the baryonic gas can 
efficiently cool.  Assuming a Press-Schechter model, this mass corresponds
approximatively to a mass into a halo of virial temperature 
$T_{\rm vir}=10^4$ K.  The number of photons per baryon $N_{\gamma}$
is dependent from the stellar Initial Mass Function (IMF): assuming an Salpeter-like
IMF similar to the one measured locally, we have $N_{\gamma}\approx 4000$.
Assuming a 10\% efficiency for  both   $f_{\rm star}$ and  $f_{\rm esc}$, $N_{ion}$ 
results to be $N_{ion}\approx 40$. 

The value of $C$ is relatively difficult to calculate and is usually assumed to be
somewhere between a few and 100. In this case, let's assume that the clumping 
factor is constant and between 1 and 30. The actual clumping factor depends instead on the
density and clustering of sources, and on the distribution and
topology of density fluctuations in the IGM.  

\begin{figure}[htbp]
\centering
\includegraphics[width=8cm]{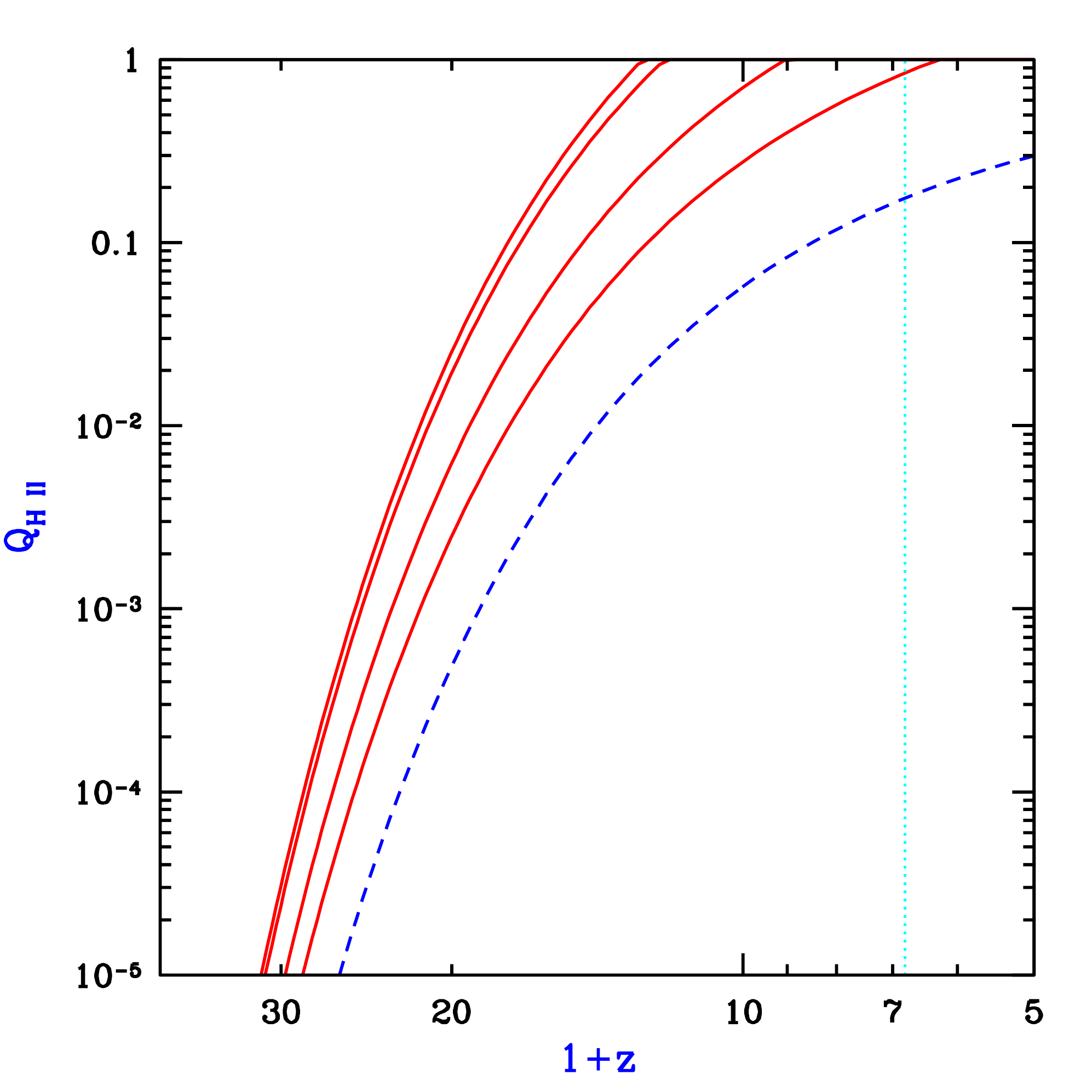}
\caption{Semi-analytic calculation of the reionization of the IGM (for
$\Ni=40$), showing the redshift evolution of the filling factor
$Q_{\rm H\ II}$. Solid curves show $Q_{\rm H\ II}$ for a clumping
factor $C=0$ (no recombinations), $C=1$, $C=10$, and $C=30$, in order
from left to right. The dashed curve shows the collapse fraction
$F_{\rm col}$, and the vertical dotted line shows the $z=5.8$
observational lower limit (Fan et al.\ 2000) on the reionization
redshift.}
\label{fig6c}
\end{figure}

In this simple picture, by definition, reionization is said to be complete when $Q_{\rm H\ II}$ 
reaches unity. Fig. \ref{fig6c} show the simplest reionization histories we can build following
the above explained arguments, as a function of $C$: $C=0$ (no recombinations), $C=1$, $C=10$, 
and $C=30$, in order from left to right.  If $C \sim 1$, recombinations are not important,
but if $C \geq 10$ then recombinations significantly delay the
reionization redshift (for a fixed star-formation history). 
The dashed curve is the collapsed fraction $F_{\rm col}$ .

Even though in the majority of the models reionization occurs at redshift around 7--12, the
exact value depends on a number of uncertain parameters, such as the formation
efficiency of stars and quasars and the escape fraction of ionizing
photons produced by these sources (see Sect. \ref{escape} and following), as they may also 
be reduced by feedback effects (see Sect. \ref{abinitio}), and the clumping factor.


\subsubsection{{\bf Cosmological Radiative Transfer}}

The propagation of the radiation through a medium is governed by the equation of radiative transfer
through the IGM for a specific intensity of a background radiation source $I_{\nu} \equiv I(t,{\bf x},\nu,{\bf \hat{n}})$. 
It is defined as the rate at which energy crosses a unit area per unit solid angle per unit time, per unit frequency range.
It is a function of time and space $(t,{\bf x})$ and the direction of propagation ${\bf \hat{n}}$. 
Its evolution can be written as 

\begin{eqnarray}
\f{\del I_{\nu}}{\del t} + \f{c}{a(t)} {\bf \hat{n} \cdot \nabla_x} I_{\nu}
-H(t) \nu \f{\del I_{\nu}}{\del \nu} 
+ 3 H(t) I_{\nu}
\nline
= -c \kappa_{\nu} I_{\nu} + \f{c}{4 \pi} \epsilon_{\nu},
\label{eq:radtrans_local}
\end{eqnarray}
where $\kappa_{\nu}$ is the absorption coefficient and $\epsilon_{\nu}$ is the
emissivity. This latter term describes the local specific luminosity per solid angle per unit volume emitted by sources, 
while the absorption coefficient accounts for absorption and may also account for the scattering out of the beam of the photons (which is a rarer event, though).
The third term on the left hand side of Eq. (\ref{eq:radtrans_local}) is the dilution of the intensity and the fourth term accounts for the shift of frequency of photons due to the expansion of the universe. 
If the medium has a number density of
$n_{\rm abs}$ with a cross-section of $\sigma_{\nu}$, the absorption
coefficient is given by $\kappa_{\nu} = n_{\rm abs} \sigma_{\nu}$. The mean specific intensity is an average of $I_{\nu}$ over a large volume and over all possible directions:
\begin{equation}
J_{\nu}(t)
\equiv \int_V \f{\de^3 x}{V} \int \f{\de \Omega}{4 \pi}  
I_{\nu}(t,{\bf x},{\bf \hat{n}}).
\end{equation}
The quantity $J_{\nu}$ is then the energy per unit time per unit area
per frequency interval per solid angle. Eq (\ref{eq:radtrans_local}) becomes:
\begin{equation}
\dot{J}_{\nu} \equiv
\f{\del J_{\nu} }{\del t} 
-H(t) \nu \f{\del J_{\nu} }{\del \nu} 
= - 3 H(t) J_{\nu} 
-c \kappa_{\nu} J_{\nu}  + \f{c}{4 \pi} \epsilon_{\nu}
\end{equation}
where the coefficients $\kappa_{\nu}$  and $\epsilon_{\nu}$ 
are now assumed to be averaged over the large volume. 

The formal solution along a line of sight is:
\begin{equation}
J_{\nu}(t) = \f{c}{4 \pi} \int_0^t \de t' 
\epsilon_{\nu'}(t')
\left[\f{a^3(t')}{a^3(t)}\right] 
{\rm e}^{-\tau(t,t';\nu)}
\label{eq:jnu_intsol_tau}
\end{equation}
where $\nu' = \nu a(t)/a(t')$,  $\nu'' = \nu a(t)/a(t'')$, and 
\begin{equation}
\tau(t,t';\nu) \equiv c \int_{t'}^t \de t''
\kappa_{\nu''}(t'')
= c \int_{t'}^t \f{\de t''}{\lambda_{\nu''}(t'')}
\end{equation}
is the optical depth along the line of sight from $t'$ to $t > t'$.

This solution indicates that the intensity at a give epoch is proportional to the integrated
emissivity with an exponential attenuation due to absorption in the medium, and it attenuated 
by a factor of $1/\exp$ when the radiation has travelled one mean free-path distance., $\lambda_{\nu}(t) \equiv 
\kappa^{-1}_{\nu}(t)$. When the mean free path is much smaller than the horizon size of the universe, $\kappa^{-1}_{\nu} \gg H(t)/c$,
the absorption is said to be ``local''. If we also assume that the emissivity is constant over a small time interval $\lambda/c$, 
we can write 
\begin{equation}
J_{\nu}(t) \approx \f{\epsilon_{\nu}(t) \lambda_{\nu}(t)}{4 \pi}
= \f{\epsilon_{\nu}(t)}{4 \pi \kappa_{\nu}(t)}
\label{eq:jnu_mfp}
\end{equation}
In this approximation, all the photons are absorbed shortly after being emitted and therefore the background intensity
depends only on the instantaneous value of the emissivity. This is a good approximation for the IGM at redshifts $z\gtrsim 3$.

\subsubsection{{\bf Numerical Solution Techniques}}\label{simulations}
The radiative transfer equation lives in a seven-dimensional space: 3 position parameters, 2 angular coordinates, time and frequency. Moreover, the problem does not have any obvious symmetry and is highly inhomogeneous. A radiative transfer modeling should follow the evolution of the radiation field, taking into account emission, absorption and scattering. 

Finding a solution of the complete radiative transfer equation is a task that can be addressed only with numerical 
simulations.  Even so, the radiative transfer is computationally extremely demanding for the above reasons, and also because in order to model reionization one has also to take into account on the top of radiative transfer, physics of gravitation and gas dynamics.  

The problems arise, for example, in dealing with many orders of a magnitude difference between the lengths scales involved in the process: on one side, the ionizing photons during early stages of reionization originate mostly form smaller haloes which are more numerous than the larger galaxies at high redshift, and on the other side, the ionized regions they create are expected to overlap and grow to very large sizes. Note, however that if the contribution of small objects to the production of ionizing photons might be negligible due to feedback effects, they are nevertheless sinks of ionizing radiation (\cite{BarkanaLoeb02}).

Therefore many different approximations and numerical schemes have been proposed for simulate the complex interaction between
dark matter, gas, stars, and radiation. Radiative transfer algorithms should scale close to linearly with the number of resolved elements to be implemented in N-body simulation (which take into account the collisionless  Dark Matter) and/or hydrodynamic algorithms (which model the collisional dynamics of the baryonic cosmic gas). Therefore, some level of physical approximation and computational optimization is needed.  The existing algorithms can be classified as follows (we only give reference to the original papers presenting the code; some of these codes are constantly updated and we refer the reader to a literature search for the latest versions):
\begin{itemize}

\item Ray Tracing/Long characteristics \emph{Abel, Norman \& Madau 1999; Razoumov \&
 Scott 1999; Sokasian, Abel \& Hernquist 2001; Razoumov et al 2002}
\item Ray Tracing/Short characteristics \emph{Umemura et al. 1999; Rijkhorst et al. 2005}
\item Flux-Eddington tensor \emph{Gnedin \& Abel 2001}
\item Flux-limited diffusion \emph{Turner \& Stone 2001; Whitehouse \& Bate 2004}
\item Fourier transforms \emph{Cen 2002}
\item Unstructured grids \emph{Ritzerveld et al. 2004}
\item Statistical (MC) methods \emph{Ciardi et a.l 2001; Maselli, Ferrara \& Ciardi 2003}
\end{itemize}

For reviews on the subject we refer the reader to Refs. \cite{Shapiro:2012xha} - \cite{Finlator:2012hm}.
Given the importance of radiative transfer simulations for modeling the reionization process, the availability on the market of such a large number of different codes makes the selection for a given problem of the right code a difficult task. 
Therefore, in 2006,  a ``Cosmological Radiative Transfer Codes Comparison Project" started, with the aim of determining the ranges of ``validity, accuracy and performances" of a large number of radiative transfer codes on five different tests (see Refs. \cite{Iliev:2006vh} and \cite{Iliev:2009du}).

\begin{figure}
\hspace{4.4mm}
  \includegraphics[width=3.4cm]{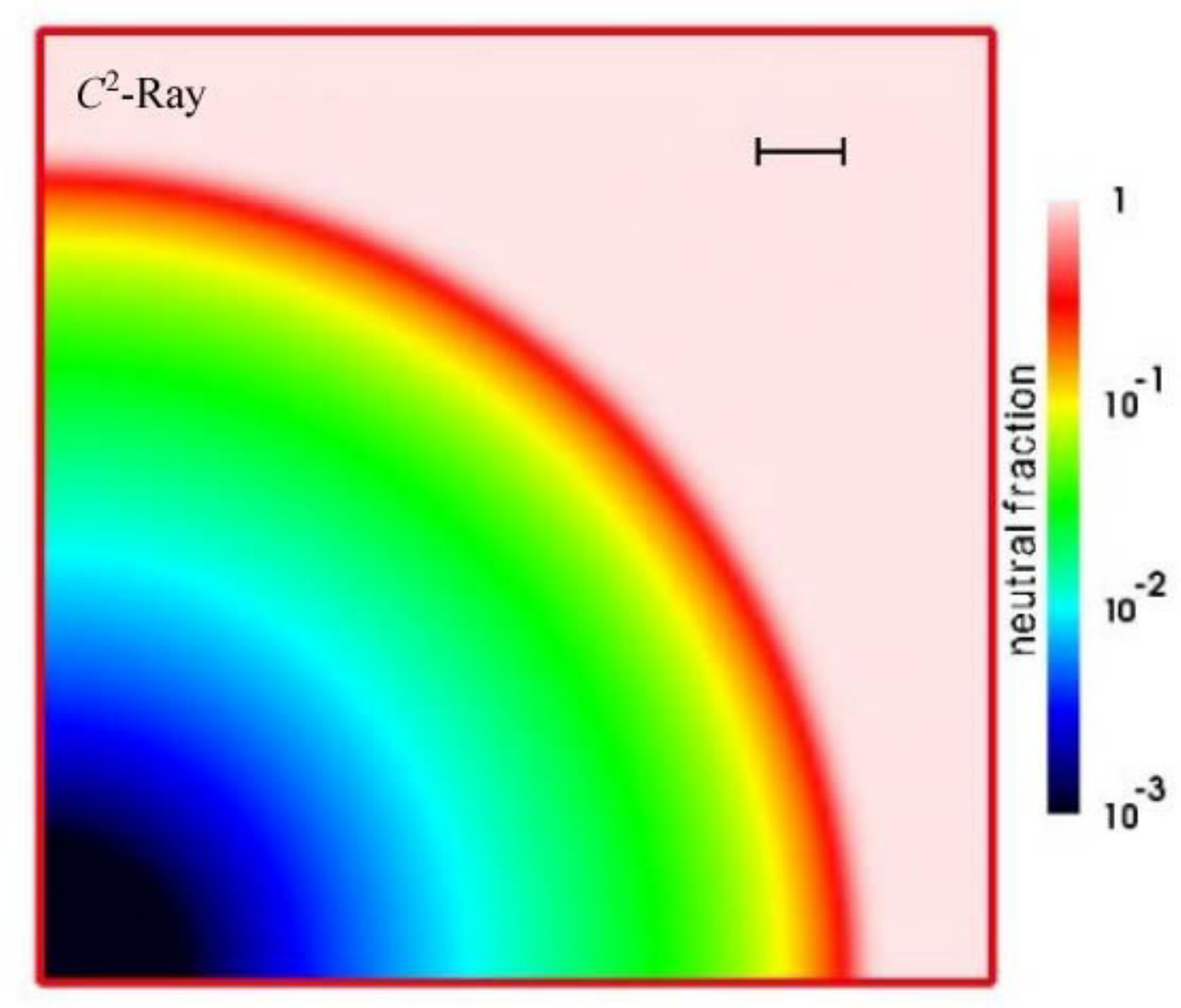}
  \includegraphics[width=4cm]{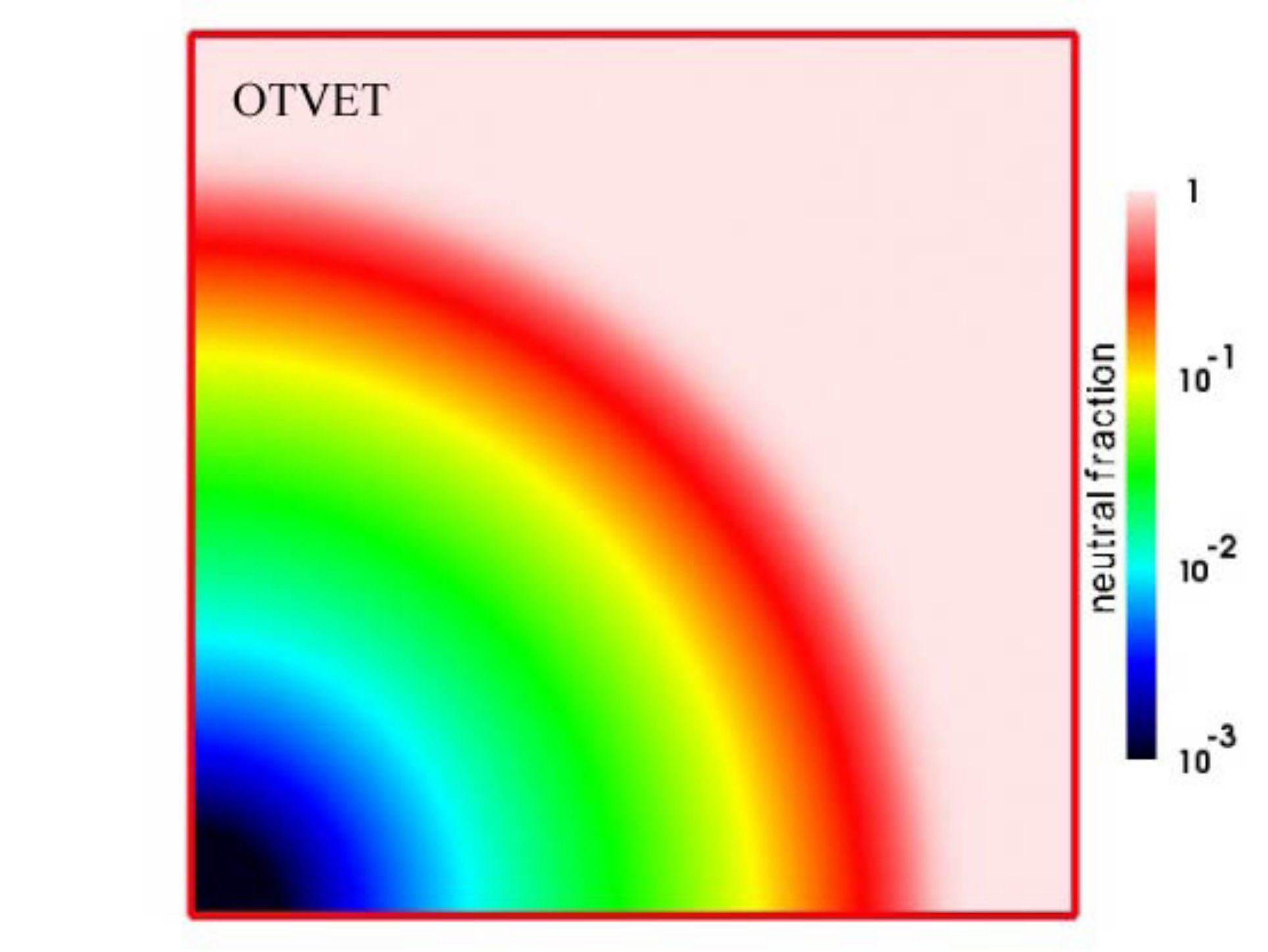}
  \includegraphics[width=4cm]{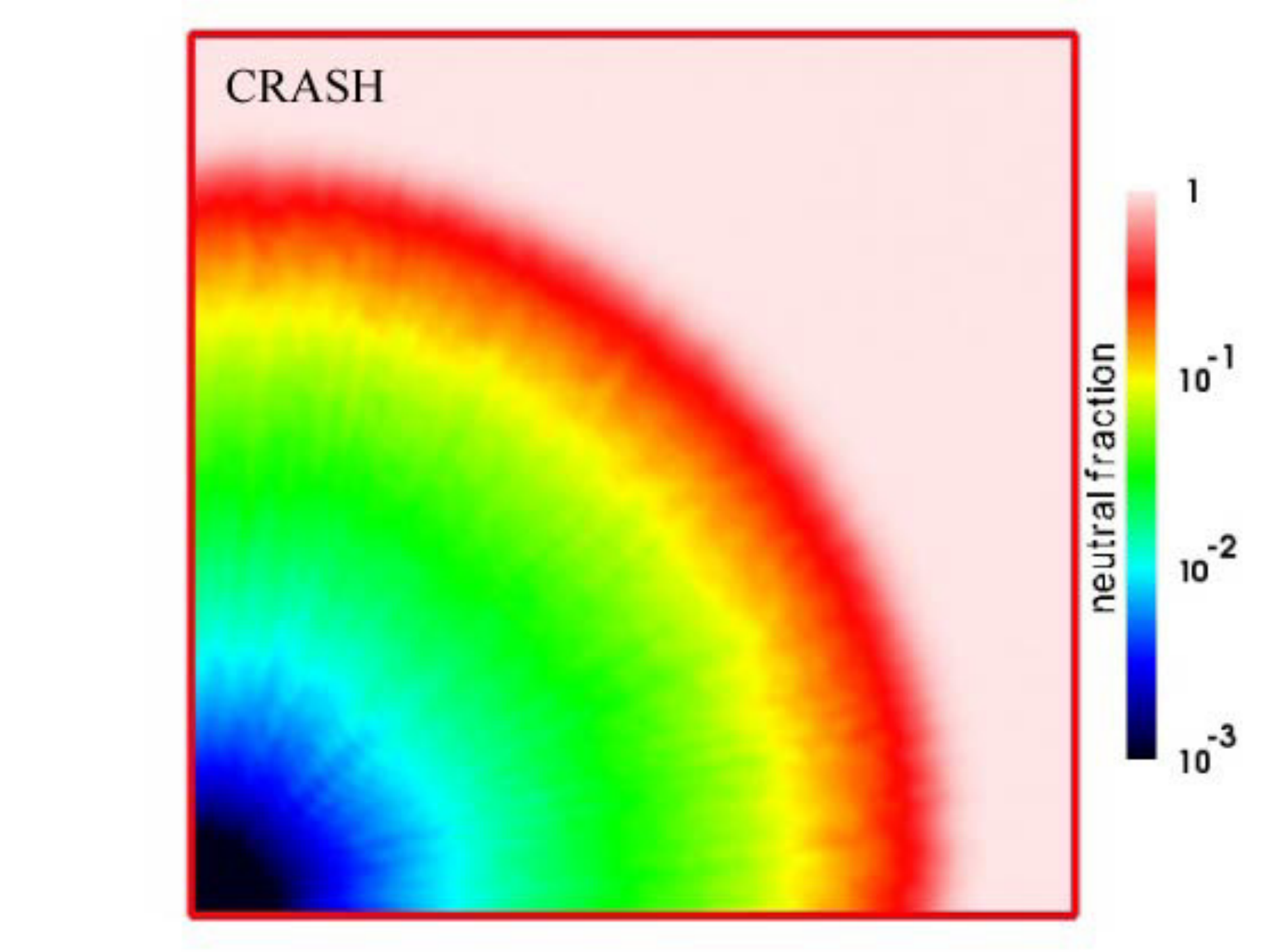}\\
  \includegraphics[width=4cm]{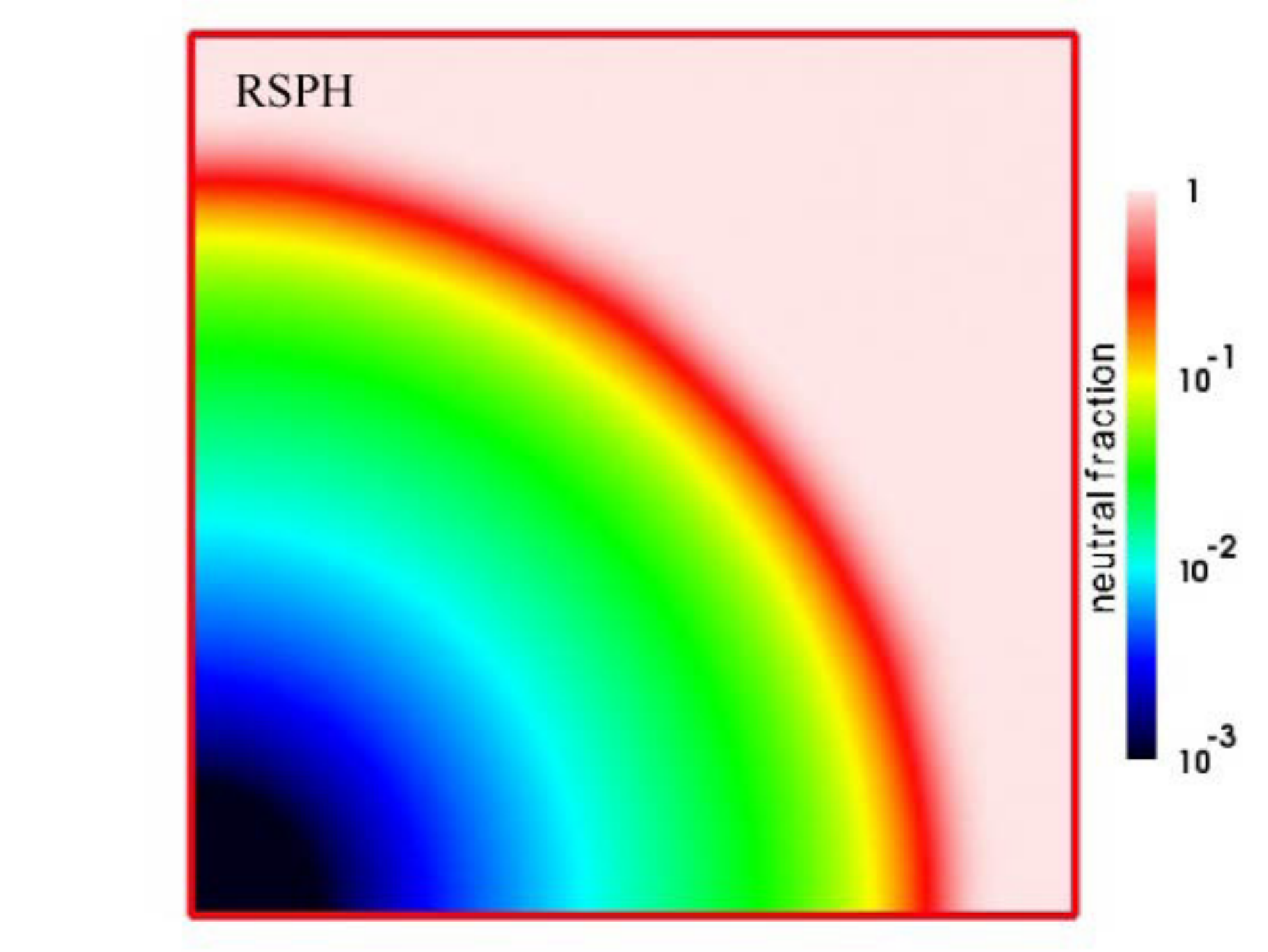}
  \includegraphics[width=4cm]{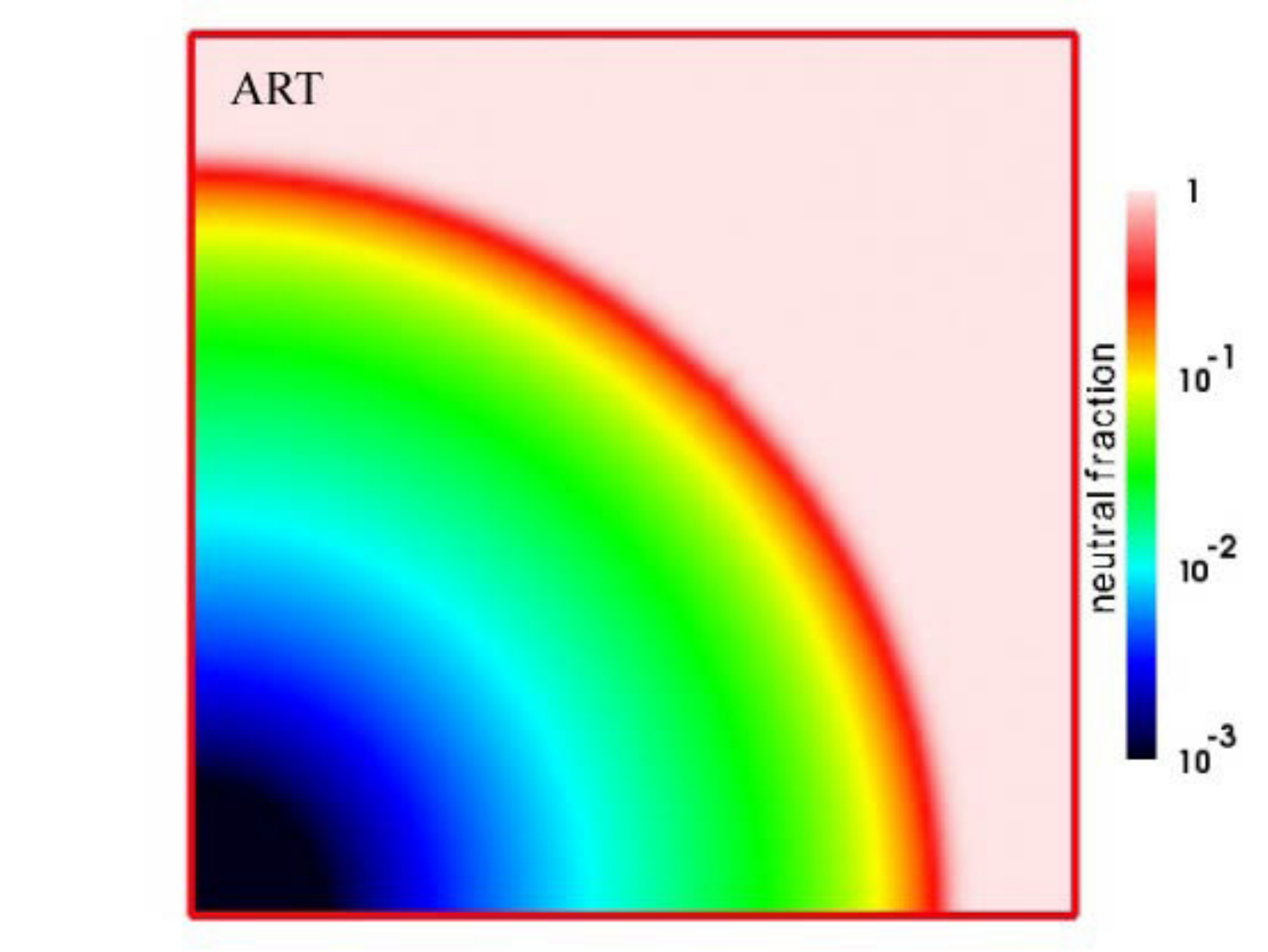}
  \includegraphics[width=4cm]{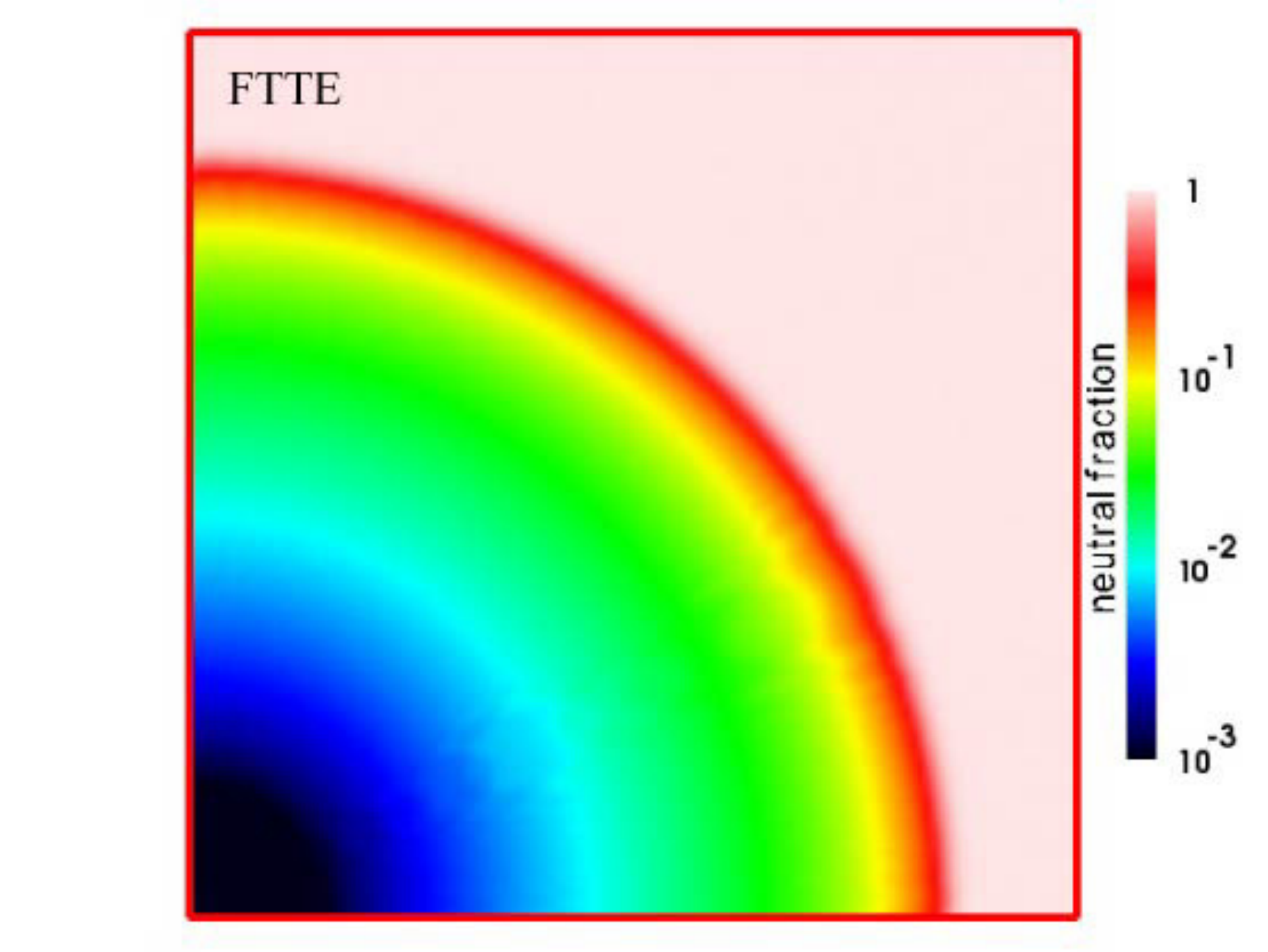}\\
  \includegraphics[width=4cm]{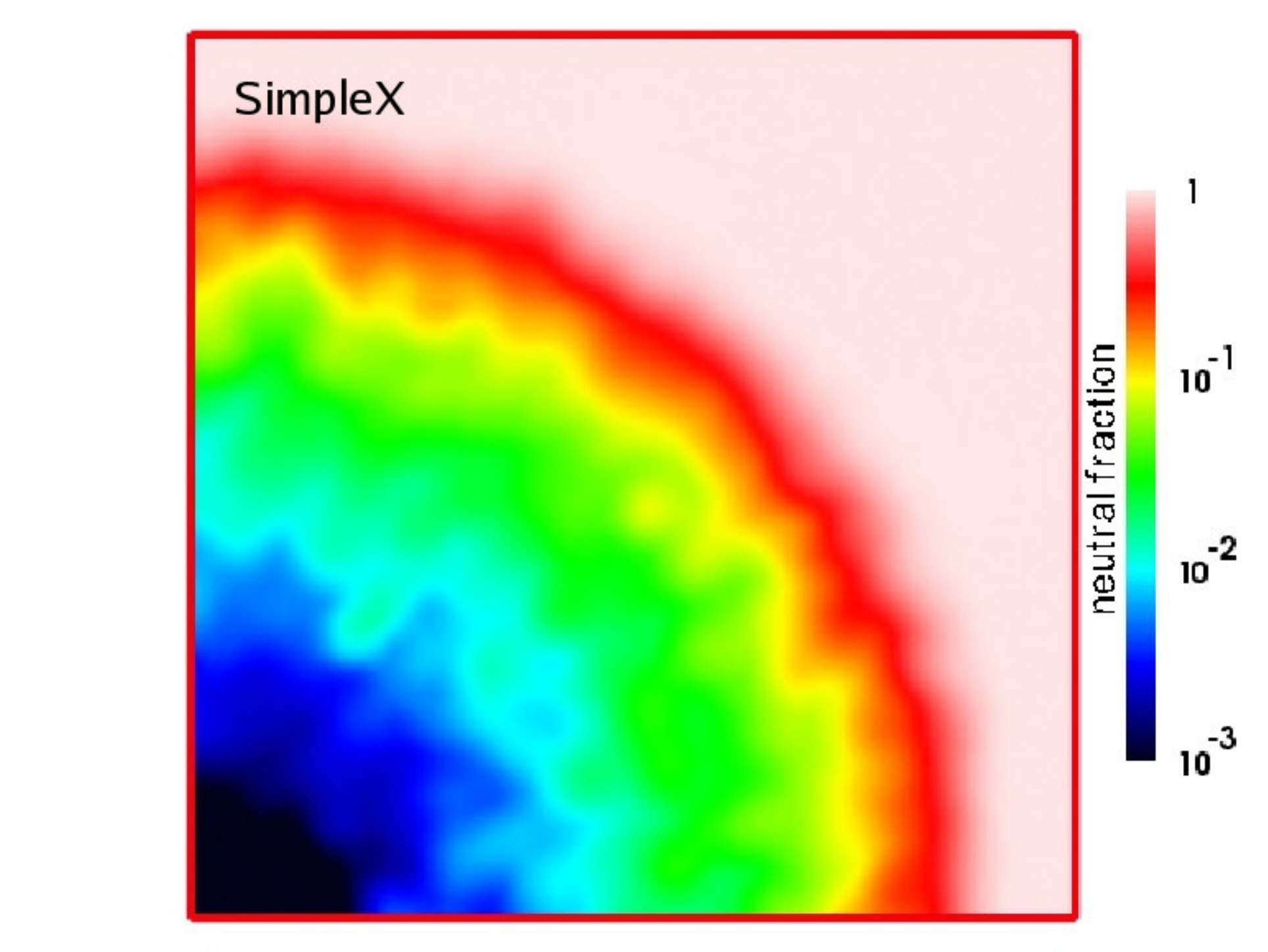}
  \includegraphics[width=4cm]{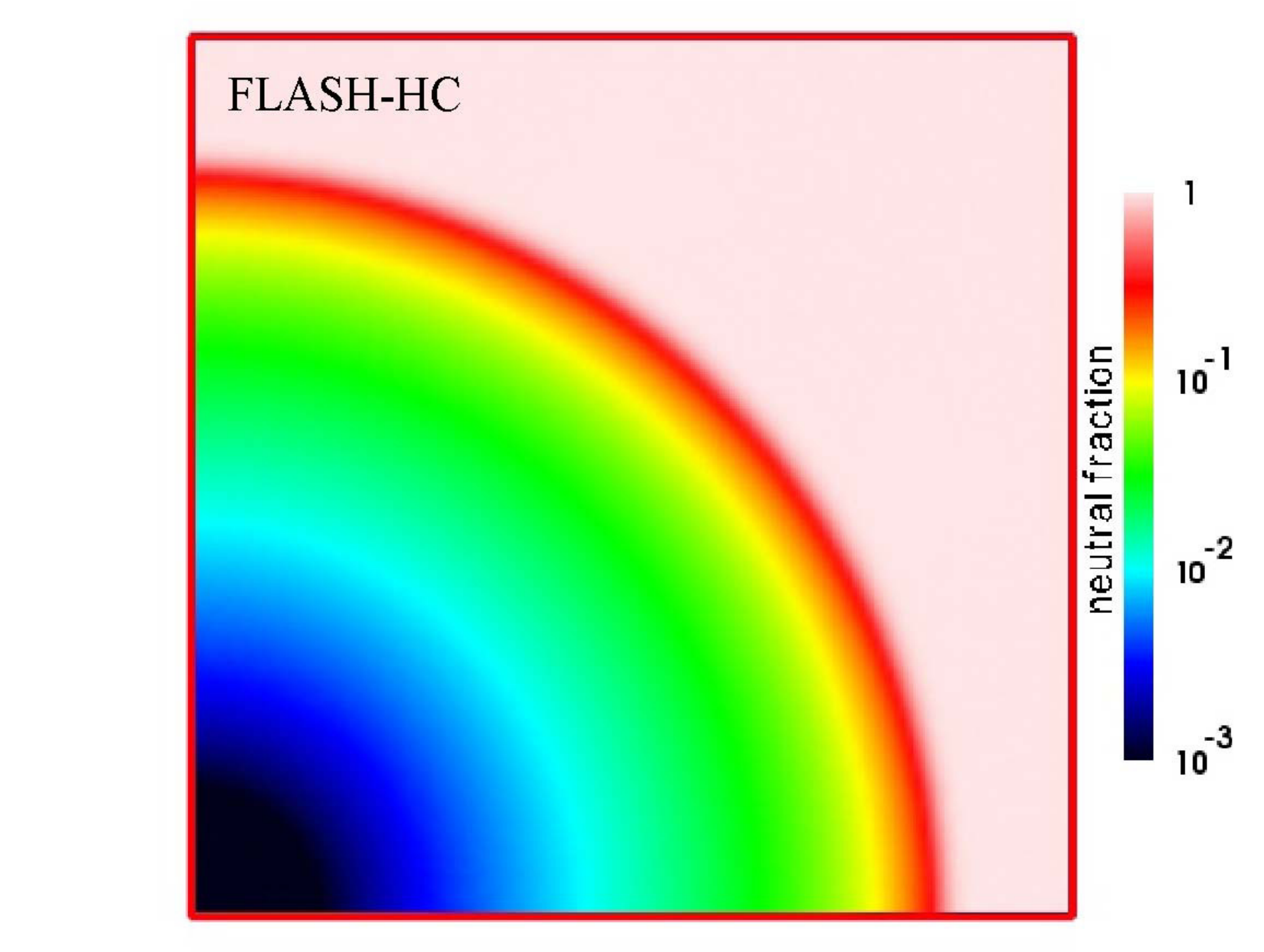}
  \includegraphics[width=4cm]{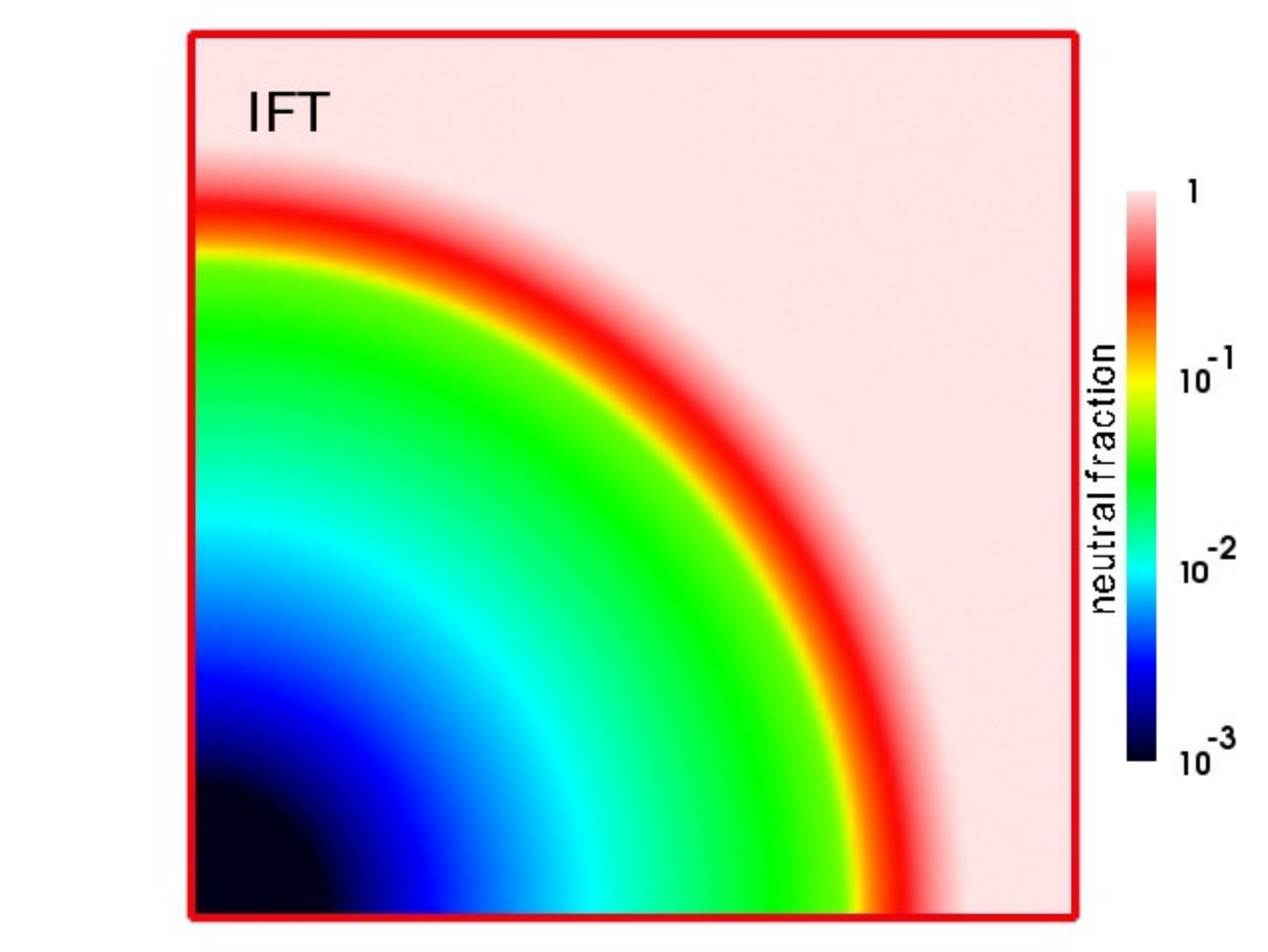}
\caption{Figure taken from \cite{Iliev:2006vh}. Comparison between 9 radiative transfer codes on the  final Str\"omgren sphere.
\label{T1_images_fig}}
\end{figure}

\subsection{{\bf Cosmic Reionization}}
 
How does reionization proceed? We already briefly mentioned the different phases of the reionization 
history of the IGM in Sect. \ref{prop}, but here we will now
described the evolution of the reionization process as it appears in cosmological 
hydrodynamical simulations.

In Fig.~\ref{Gnedina} we report 8 of $15 h^{-1}$ comoving kiloparsecs deep slices of the simulations performed in \cite{Gnedin:1999fa}
They are taken at different epochsand the epoch is measured by the scale factor $a\equiv1/(1+z)$, at the center of each 
4-plots block. 
In the four panels for each redshift, from the top-left, in counter-clockwise order, are shown:
the neutral hydrogen fraction, (upper left panel), the gas density (lower left), 
the gas temperature (lower right), while in the upper right panel there is the 
UV ionizing background intensity averaged over the entire volume $J_{21}$, 
defined as the radiation intensity at the Lyman limit, expressed in unit of 
10$^{-21}$ erg cm$^{-2}$ s$^{-1}$ sr$^{-1}$ Hz$^{-1}$, logarithmically scaled. 
When the reionization starts, the ionization fronts propagate from the first galaxies 
located in the high-density regions. The ionization front moves on leaving the 
high-density outskirts of an object still neutral as at high densities the recombination rate is very high 
and there are not enough photons to reionize them (see Sect. \ref{prop} and  Figs.~\ref{a} -\ref{d}).

In this ``pre-overlap'' phase, the high-density regions just around the source
slowly become ionized, but the the high-density regions far from the source remain neutral. 
The ionizing intensity, as an integrated quantity, is slightly mis-defined as the ionizing intensity
in this phase is highly inhomogeneous. Anyway, at this time it remains low and it is increasing with time. 
By redshift $z \approx 7 $ the $\HII\ $ regions starts to overlap (Fig.\ref{d})
and the ionizing intensity starts to rise rapidly (Fig.\ref{e}) as at each point the number of
ionizing sources increases. This phase is quite rapid, the last neutral low-density regions are quickly ionized, 
and the mean free path of UV photons rapidly rises. 

After this phase, the high-density regions far from the sources are still neutral, as the 
number of ionizing photons are not yet sufficient.  This is the so-called ``post-overlap" 
phase (Figs.~\ref{e} - \ref{g}). As time goes on, more and more ionizing photons are emitted 
and the high-density regions become gradually ionized. The spatially averaged UV intensity 
continues to rise until it saturates at $z \sim 5$. 

As the authors of \cite{Gnedin:1999fa} noticed this saturation may just be an artifact due to the finiteness of the simulation box, as the mean free path exceed by this time the box size by a factor of 10, and therefore the simulations stop to correctly reproduce reality.

\begin{figure}[!h]
\centering
\subfloat[]{\label{a}\includegraphics[width=0.5\textwidth]{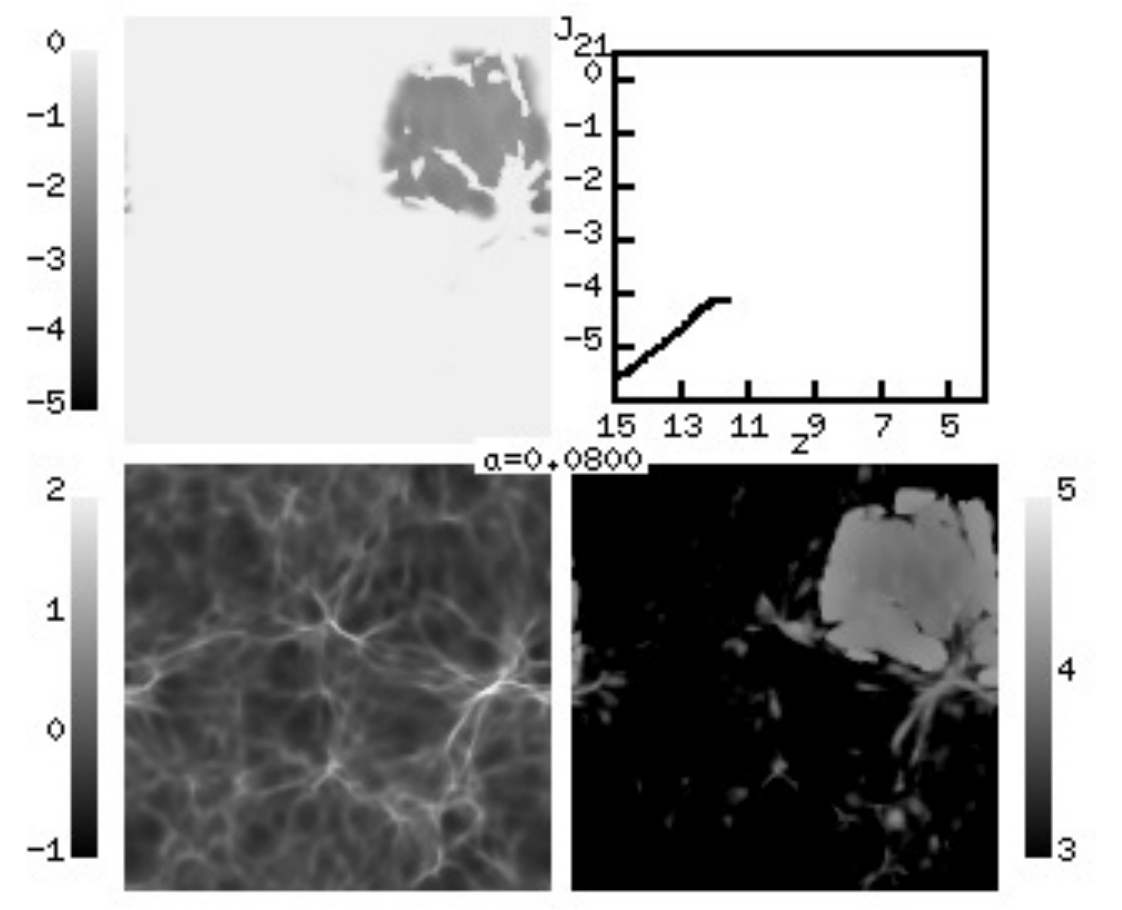}}
\subfloat[]{\label{b}\includegraphics[width=0.5\textwidth]{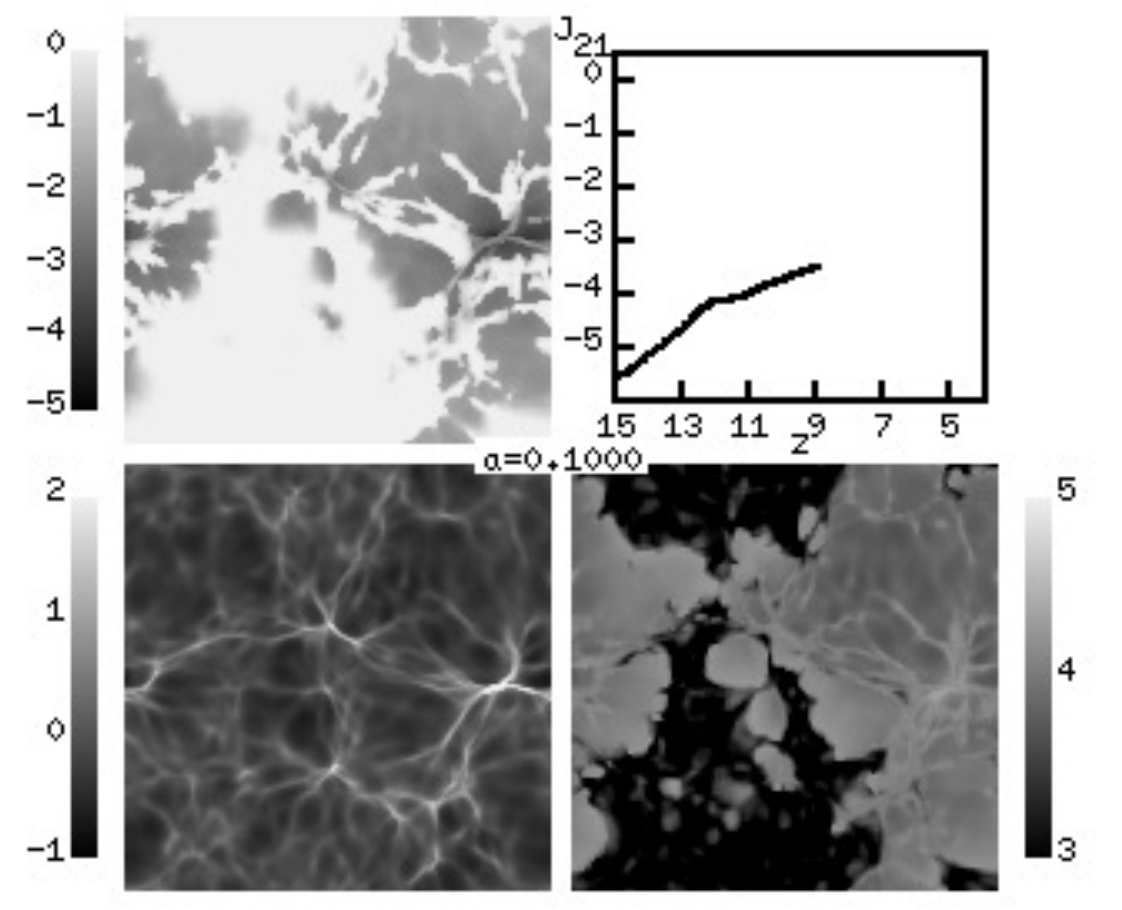}}

\subfloat[]{\label{c}\includegraphics[width=0.5\textwidth]{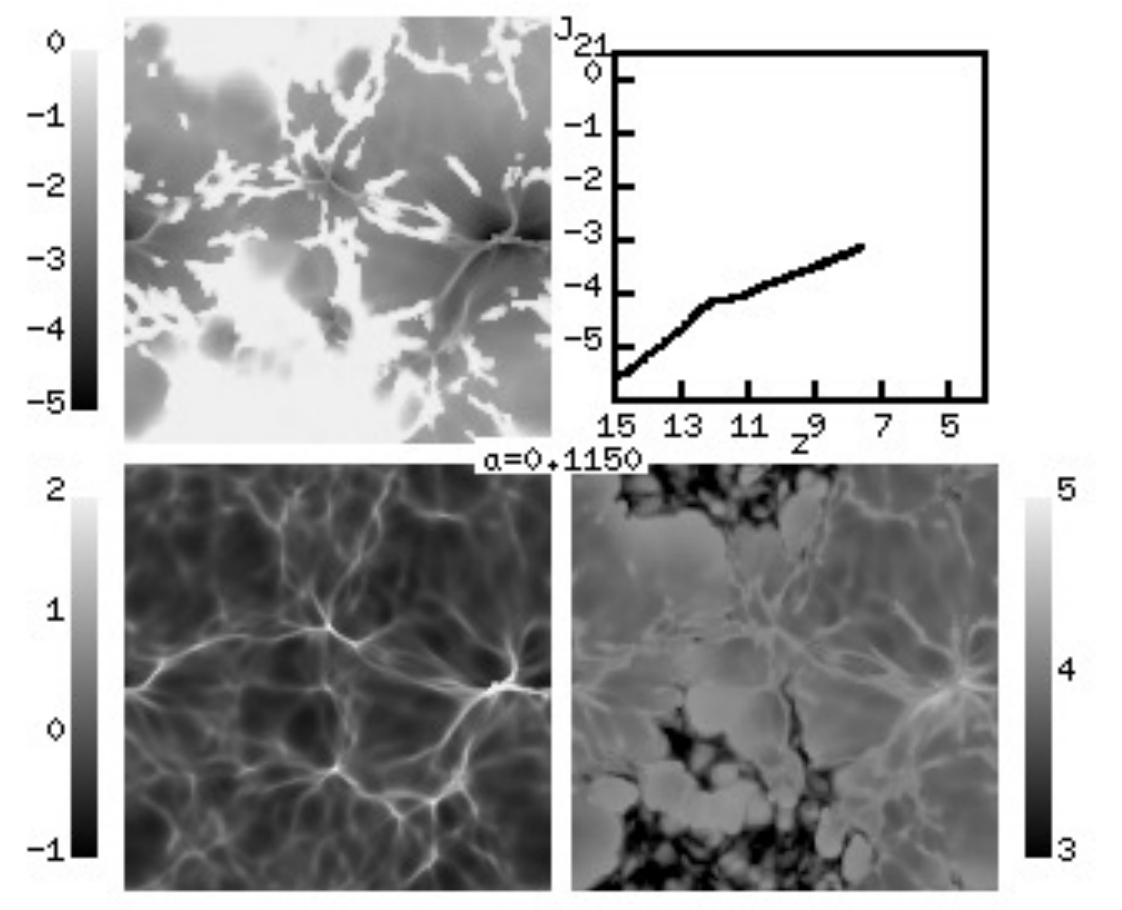}}
\subfloat[]{\label{d}\includegraphics[width=0.5\textwidth]{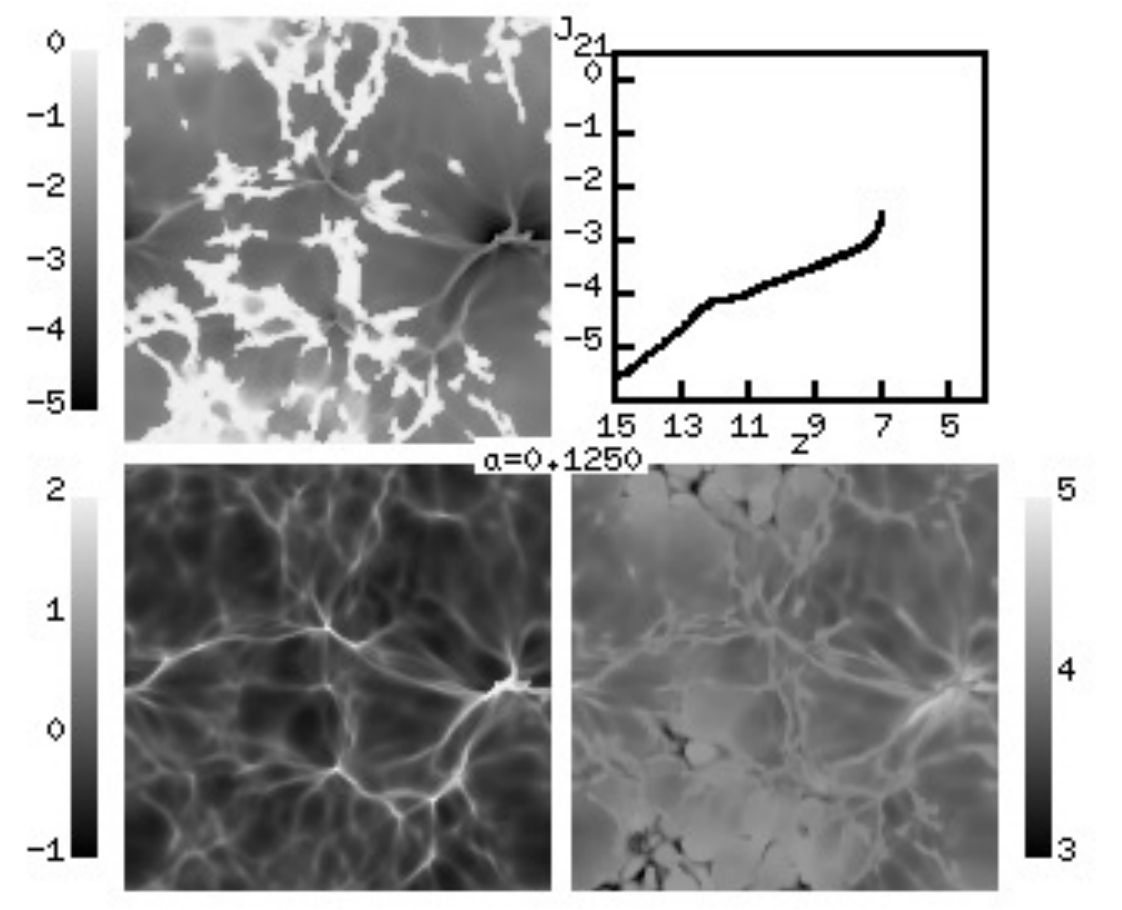}}
\caption{Slices of 4$h^{-1}$ Mpc taken from \cite{Gnedin:1999fa} from his numerical simulation of reionization,  at 4 earlier different epochs, 
({\it a\/}) $z=11.5$, 
({\it b\/}) $z=9$, 
({\it c\/}) $z=7.7$, 
({\it d\/}) $z=7$.
Every group of four panels shows: the neutral hydrogen fraction, (upper left panel), the gas density (lower left), 
the gas temperature (lower right),  
UV ionizing background intensity averaged over the entire volume $J_{21}$ (upper right panel).}
\label{Gnedina}
\end{figure}  

\begin{figure}[!h]
\centering
\ContinuedFloat
\subfloat[]{\label{e}\includegraphics[width=0.5\textwidth]{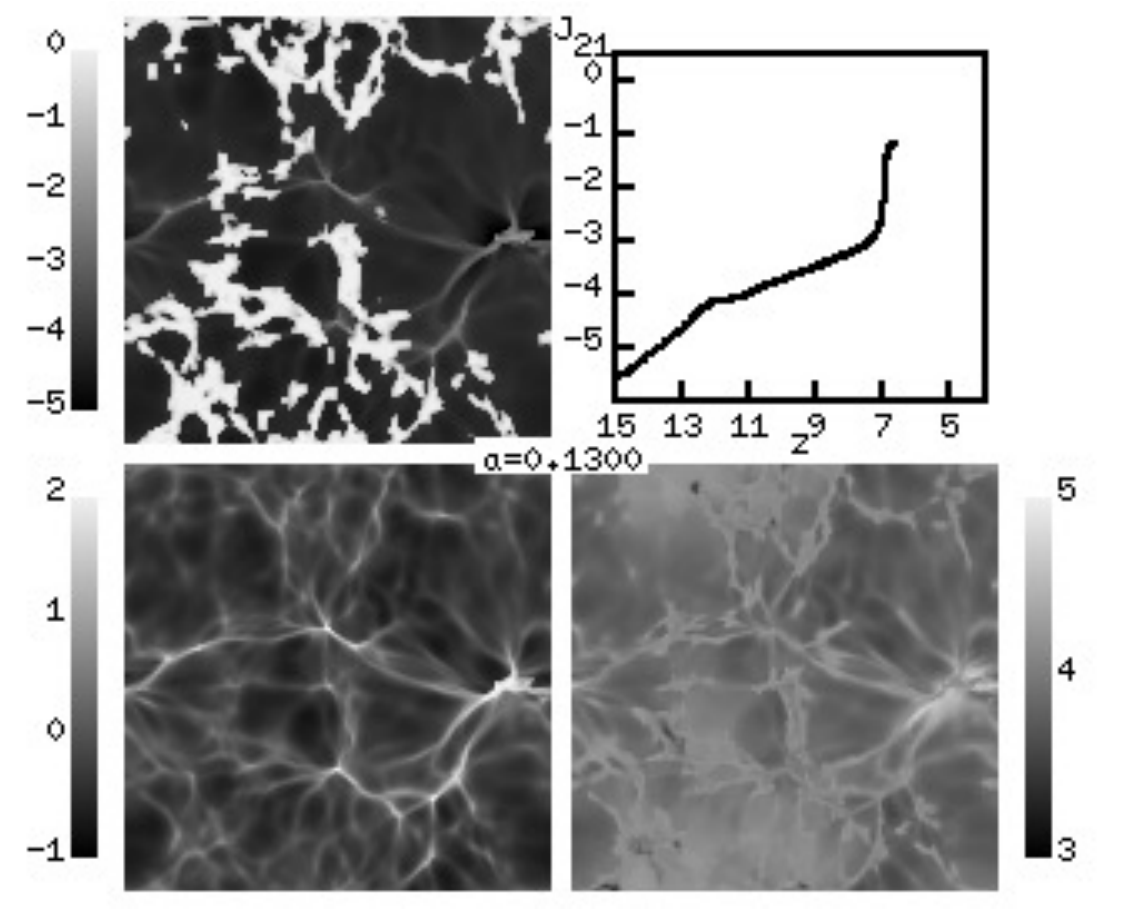}} 
\subfloat[]{\label{f}\includegraphics[width=0.5\textwidth]{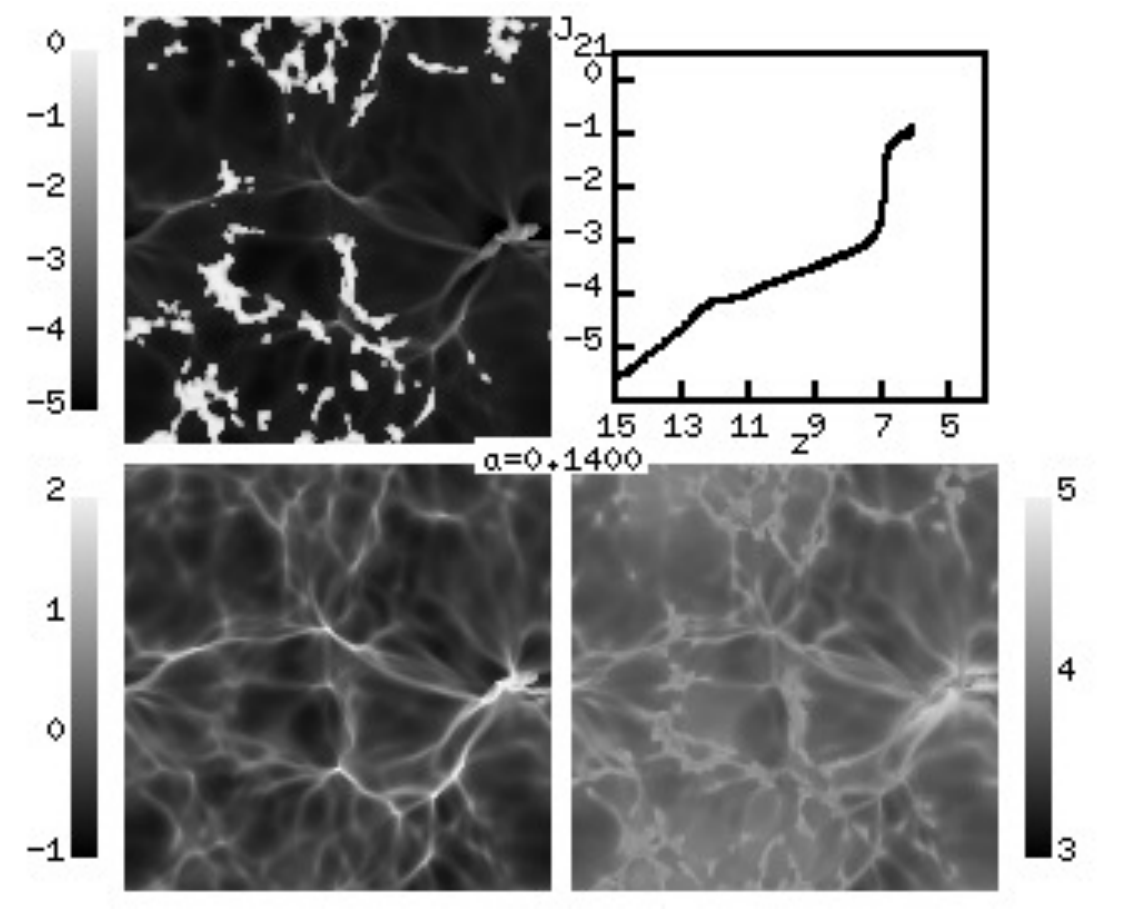}} 

\subfloat[]{\label{g}\includegraphics[width=0.5\textwidth]{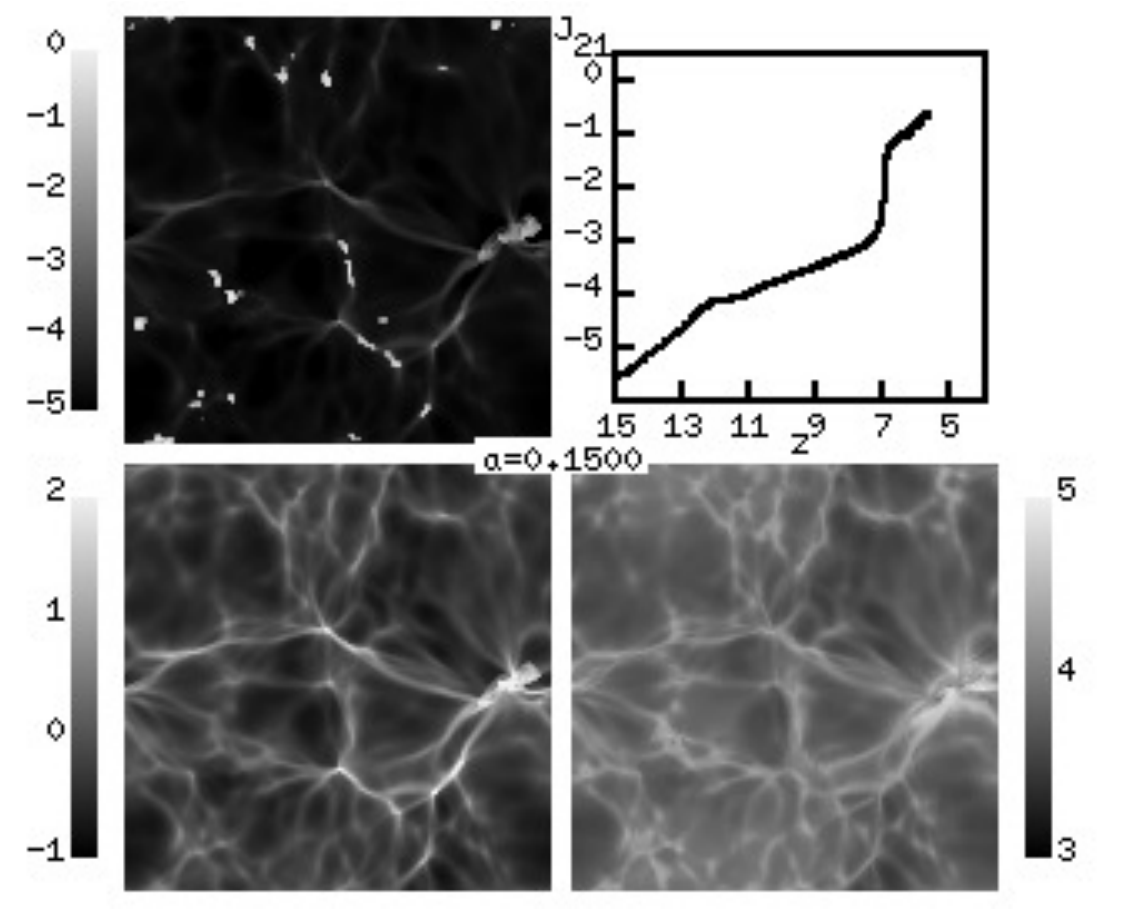}} 
\subfloat[]{\label{h}\includegraphics[width=0.5\textwidth]{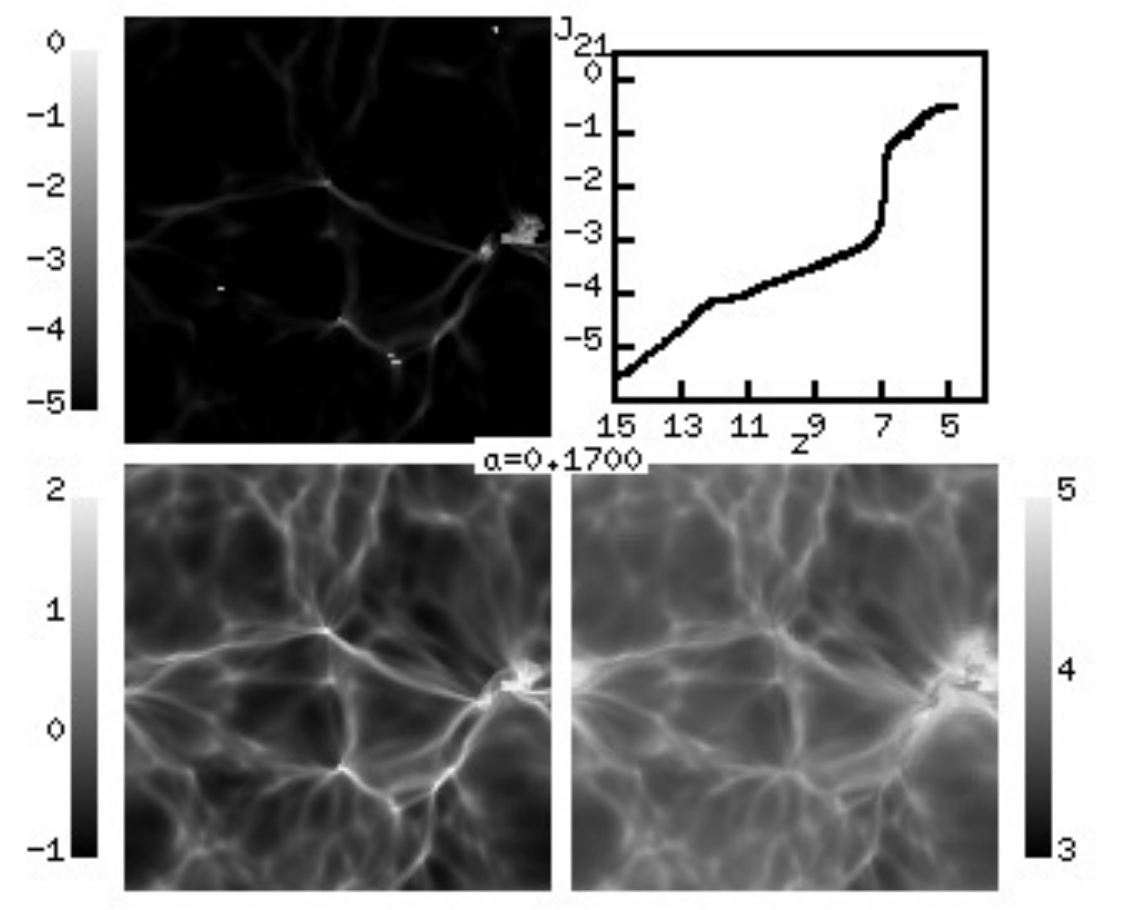}} 

\caption{Slices of 4$h^{-1}$ Mpc taken from \cite{Gnedin:1999fa} from their numerical simulation of reionization, for 4 different further epochs, 
({\it e\/}) $z=6.7$, 
({\it f\/}) $z=6.1$, 
({\it g\/}) $z=5.7$, 
({\it h\/}) $z=4.9$. 
Every group of four panels shows: the neutral hydrogen fraction, (upper left panel), the gas density (lower left), 
the gas temperature (lower right),  
UV ionizing background intensity averaged over the entire volume $J_{21}$ (upper right panel).}
\end{figure}


\subsection{{\bf Reionization Sources}}\label{sources}
 
It was soon realized that the observed population of QSOs could not produce enough UV flux to reionize the IGM at redshifts $z\approx 3$ \cite{ShapiroGiroux8796}. Therefore the majority of theoretical models adopt stellar type sources with different spectra and initial mass function (IMF), which are constrained through: (i) the value of the IGM temperature from the \Lya forest at $z\approx 2-4$ (see e.g. \cite{Zaldarriaga:2000mz}), (ii) the abundance of neutral hydrogen from the high-redshifts QSOs spectra, (iii) the optical depth from CMB measurements.

Quasars and Active Galactic Nuclei (AGN) are effective emitters of UV photons, as the escape fraction of photons is generally assumed to be of the order of unity (see Sect. \ref{prop}), but their density declines exponentially towards high redshifts. Another crucial point about QSOs is their short lifetime, which might affect the growth of their ionized regions. Moreover, unless they are extremely X-ray quiet (to avoid overproducing the X-ray background (see Sect. \ref{xray})), their contribution to the reionization history should be $< 30 \%$ with respect to star-forming galaxies \cite{Salvaterra:2005as}. They are instead they key sources for Helium reionization, occurring at $z <3$, as we mentioned in Sect. \ref{prop}.
However, there is still uncertainty on the estimate of the total UV photon emissivity from star-forming galaxies at high redshifts as there are still unknown factor like the stars formation rate, the clumping factor and the UV escape fraction (see Sect. \ref{prop}). 

The contribution of very massive metal-free stars is very uncertain, as producing a number of photons able to ionize the IGM would lead to an overproduction of metals. This, in turn, implies that the transition to metal-enriched Pop II stars would occur earlier (see e.g. \cite{Ricotti:2003mx}). The proposed picture includes a differentiation of very-massive metal free stars and metal-free stars with masses $M<100\ M_{sun}$, as in this way the formation of lower mass Pop III stars would stop at lower redshifts due to metal enrichment and feedback, and therefore their contribution would be more prominent \cite{Schneider06a}.

Other UV photons sources could have been: (i) mini-quasars powered by intermediate mass black holes \cite{Madau04}; (ii) decays/annihilations from Dark Matter particles like decaying sterile neutrinos (\cite{Hansen:2003yj}), even if the authors of \cite{MapelliFerrara05} concluded that they must have played a minor role; (iii) X-ray photons (see Sect. \ref{xray}); (iv) enhanced structure formation from; (v) a non-scale-free isocurvature power spectrum in addiction to the scale-free adiabatic one \cite{Sugiyama:2003tc}; (vi) non-Gaussian density fluctuations \cite{Chen03}.

\subsection{{\bf (More) Reionization Tests}}\label{other}
There are a variety of observational tests which can put constraints on the various phases of the 
reionization process of the universe. While the presence of a Gunn-Peterson trough in the spectrum of high-redshift QSO
carries an evidence of the beginning of the reionization process, patchy reionization absorption is a probe for the overlap phase, and the \Lya forest absorption observations is instead a test for the never-ending post-overlap phase. 
In what follows we will more extensively describe several different tests for the reionization history, other than the \Lya forest absorption. 

\subsubsection{{\bf Cosmic Microwave Background}}\label{CMB}
The Cosmic Microwave Background radiation (CMB) is the second major probe of the reionization 
epoch after the Gunn-Peterson probe. 
Reionization has a three-fold effect on the CMB: 
(i) it damps primary anisotropies at all scales, (ii) it creates an additional feature
in the large-scale modes of the polarization power spectra, and  (iii) it affects the small-scale modes of the 
temperature power spectra via the production of secondary anisotropies. 

As for the first effect, Thompson scattering of the CMB photons on top of the free electrons released by the 
reionization process damps the fluctuations on all the scales smaller than the horizon at reionization. 
The amount of this damping is ${\rm e}^{-\tau_{\rm el}}$ where 
\begin{equation}
\tau_{\rm el} = \sigma_T  \int  ~ n_e ~ 
(1+z)^{-1}[c/H(z)]~{\rm d} z
\end{equation}
is the optical depth, measured at present epoch, and where 
$n_e$ is the  average value of the comoving electron density and $\sigma_T$ is 
the Thomson scattering cross-section. 
While the Gunn-Peterson optical depth is proportional to the neutral gas fraction the
CMB optical depth $\tau_{\rm el} $ is proportional to the density of free electrons, 
and therefore it is proportional to the ionized gas fraction. 

This effect on the temperature power spectrum is completely degenerate with
the primordial power spectrum amplitude, $A_s$ for scales smaller than the 
horizon at last scattering, and therefore the optical depth is only mildly constrained 
by the temperature fluctuations because of this strong degeneracy. 
If a model of sudden reionization is assumed there is one-to-one 
relation between the optical depth and the reionization redshift. 

The second, and more important, signature of the reionization epoch on the 
CMB power spectra, is a new feature in the large-scale part of the polarization 
power spectrum. The polarization of the CMB emerges naturally as a prediction of 
the gravitational instability paradigm, for which if the anisotropies we observe in today in the CMB
are the product of gravitational instability grow on primordial fluctuations, they would naturally polarize
the CMB. The generation of polarization has two necessary conditions: 1) the quadrupole moment of
the angular distribution of the photons temperature must be non-zero and 2) photons must undergo Thompson
scattering off free electrons. 

Therefore, the degree of linear polarization of the CMB at any scale reflects the ``local'' quadrupole anisotropy
 in the plasma when the photons last scatter at the same scales. 
The largest scale at which a quadrupole moment can form is the scale of the horizon at recombination,
$\approx 1$ degree. Thus, any polarization signature on any scale larger than the horizon scale at 
recombination is a clear signature for Thompson scattering at later epochs, when the horizon scale is larger.

After recombination, the quadrupole moment of temperature anisotropies
grows due free-streaming of photons. 
The radiation scatters off the free electrons put about by reionization
and creates a polarization signal on the scale equivalent to the horizon 
scale at the redshift of scattering. Therefore, the scale of the 
reionization feature in the polarization power spectrum is an indication
of the reionization redshift. The polarization signal will peak at a 
position $\ell \propto z_{\rm re}^{1/2}$ and it will have an amplitude 
proportional to the optical depth $\tau_{el}$.

The third modification of the CMB anisotropies due to reionization 
is in the small-scales part of the temperature power spectrum. 
In non-instantaneous models of reionization, in the ``pre-overlap'' phase 
the ionization process proceeds in a ``patchy'' way, where discrete ionizing 
sources grow and eventually coalesce at the end of the overlap phase. 
This behavior causes secondary CMB temperature anisotropy, on top 
of the general damping of the entire power spectrum. One of the secondary 
effects created by reionization is the kinetic Sunyaev-Zel'dovich effect (kSZ).

The kSZ is due to the temperature and the motion of regions of reionized electrons
along the line of sight, and therefore it occurs at small scales ($<0.1^\mathrm{o}$, $\ell >$ 2000).
This kSZ would produce a distortion in the CMB temperature power spectrum that can be expressed as
\beq
\delta_{T}\equiv\frac{\Delta T}{T}(\mathbf{\hat{{u}}})=\sigma_{T}\int dz\ c (dt/dz) e^{-{\tau_{e}}}n_{e} \mathbf{\hat{u}} \cdot \mathbf{ v} 
\eeq
where  $\mathbf{\hat{{u}}}$ is the line of sight, and $\mathbf{ v} $ is the peculiar velocity of the gas. 
The corresponding angular power spectrum is therefore 
\beq
C_{\ell}^{kSZ}\equiv T_{CMB}^2|\hat{\delta}_{T}(k)|^2,
\eeq
and it probes both the homogeneity and the efficiency of the reionization process.
The overall amplitude of the secondary spectrum appears to be model-dependent \cite{Furlanetto:2005ax}), 
especially as regards the duration of the patchy space, while the shape of the spectrum is not (see e.g. \cite{Zahn:2005fn}, \cite{Mesinger:2012ys}).

\subsubsection{{\bf X-rays Effects}}\label{xray}
In addiction to the UV-photons, X-ray photons have also to be considered if sources producing them are thought to be present during hte reionization epoch. X-rays can come from binary stars in early galaxies and/or mini-quasars, but also from thermal emission from the supernova remnants, or Compton up-scattering of CMB photons by relativistic electrons produced by supernova explosions (\cite{Venkatesan2001} - \cite{Ricotti:2004xf}). 
X-ray photons have a much larger mean free path compared to the UV photons; their escape fraction is close to unity, and so they can penetrate the IGM in a uniform way, creating a floor of semi-ionized gas and then can change the scenario of patchy reionization dramatically. They could provide the optical depth required by the WMAP polarization measurements, rapidly ending reionization at $z\sim6$. These X-ray sources would also contribute to the unresolved soft X-ray background. However, the observed X-ray background level constrains the production of X-ray by early mini-quasar to at a  negligible level (see e.g. \cite{Salvaterra05}).

There is also a known correlation between X-ray luminosity and star formation rate (SFR), which holds {for redshifts until $z\sim 3-4$ (see e.g. \cite{Persic:2004vf}-\cite{Colbert:2003jg}). The impact of X-rays can affect the reionization history in different ways. Reionization would occur earlier, and be slightly more extended; it has less small-scale structure and develops a partially-ionized ``haze'' (see Fig. \ref{fig:xH_lightcone}). X-rays also provide an additional thermal feedback, by which the reionization process would be delayed \cite{Mesinger:2012ys}.

\begin{figure}[htbp]
\centering
\includegraphics[width=1\textwidth]{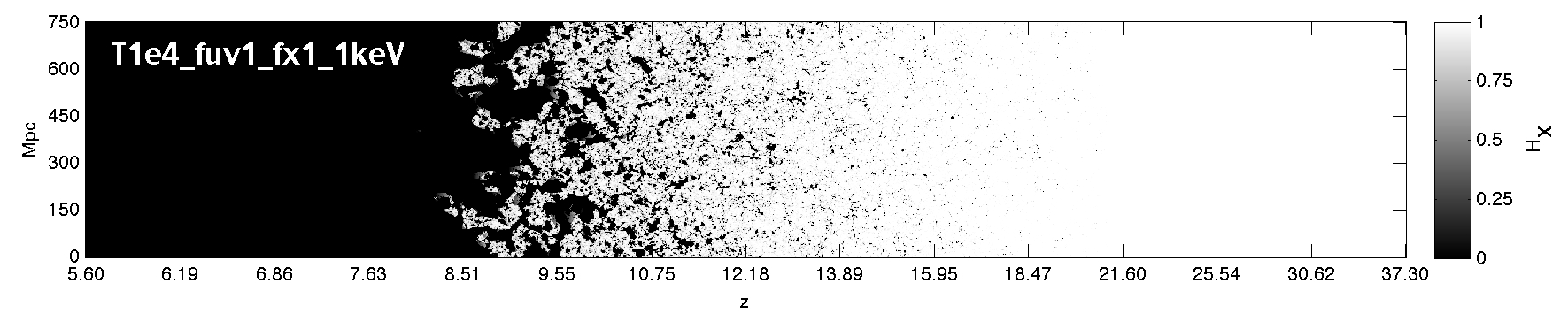}
\includegraphics[width=1\textwidth]{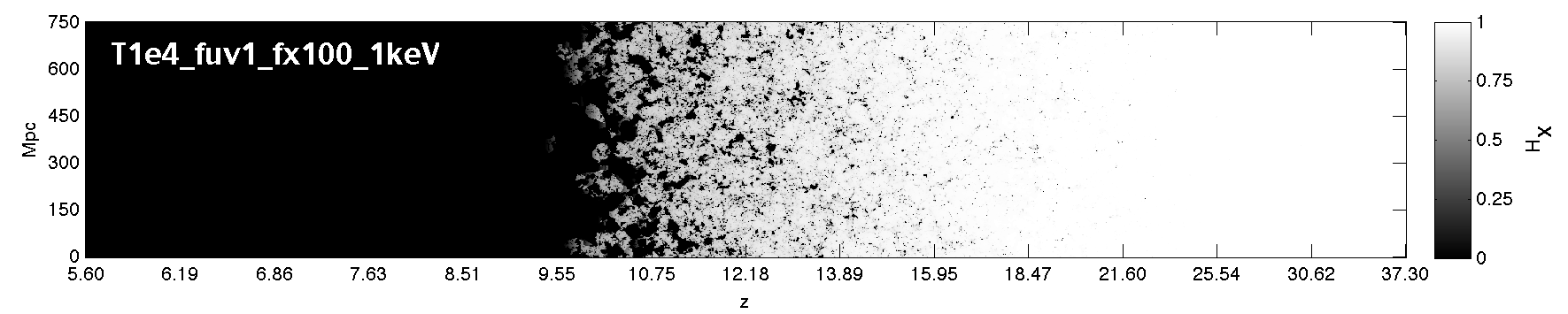}
\includegraphics[width=1\textwidth]{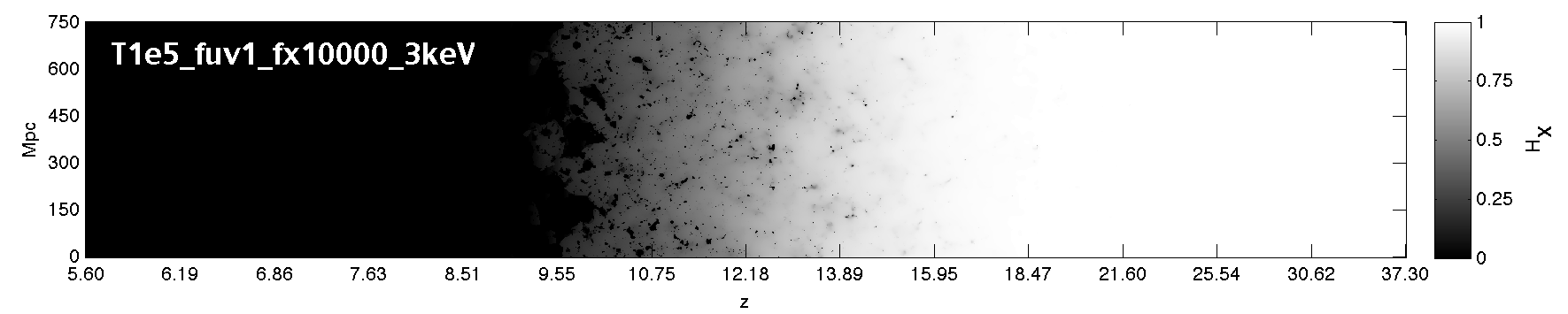}
\caption{Taken from \cite{Mesinger:2012ys}. The impact of X-rays on reionization. It occurs earlier, is slightly more extended, has less small-scale structure and develops a partially-ionized ``haze''.}
\label{fig:xH_lightcone}
\end{figure}
Another signature of the presence of the X-ray photons could be seen in the kinetic Sunyaev-Zel'dovich effetc (kSZ): at a fixed reionization history, X-rays decrease the kSZ power at $\ell=3000$ by suppressing the small-scale ionization structures. If the X-rays contribute 1/2 of the ionizations, the patchy kSZ power is decreased by $\approx0.5$ $\mu K^2$ with respect to a UV-only model with the same reionization history (see Fig. \ref{fig:kSZ}, taken from \cite{Mesinger:2012ys}).

\begin{figure}[htbp]
\centering
\includegraphics[width=8cm]{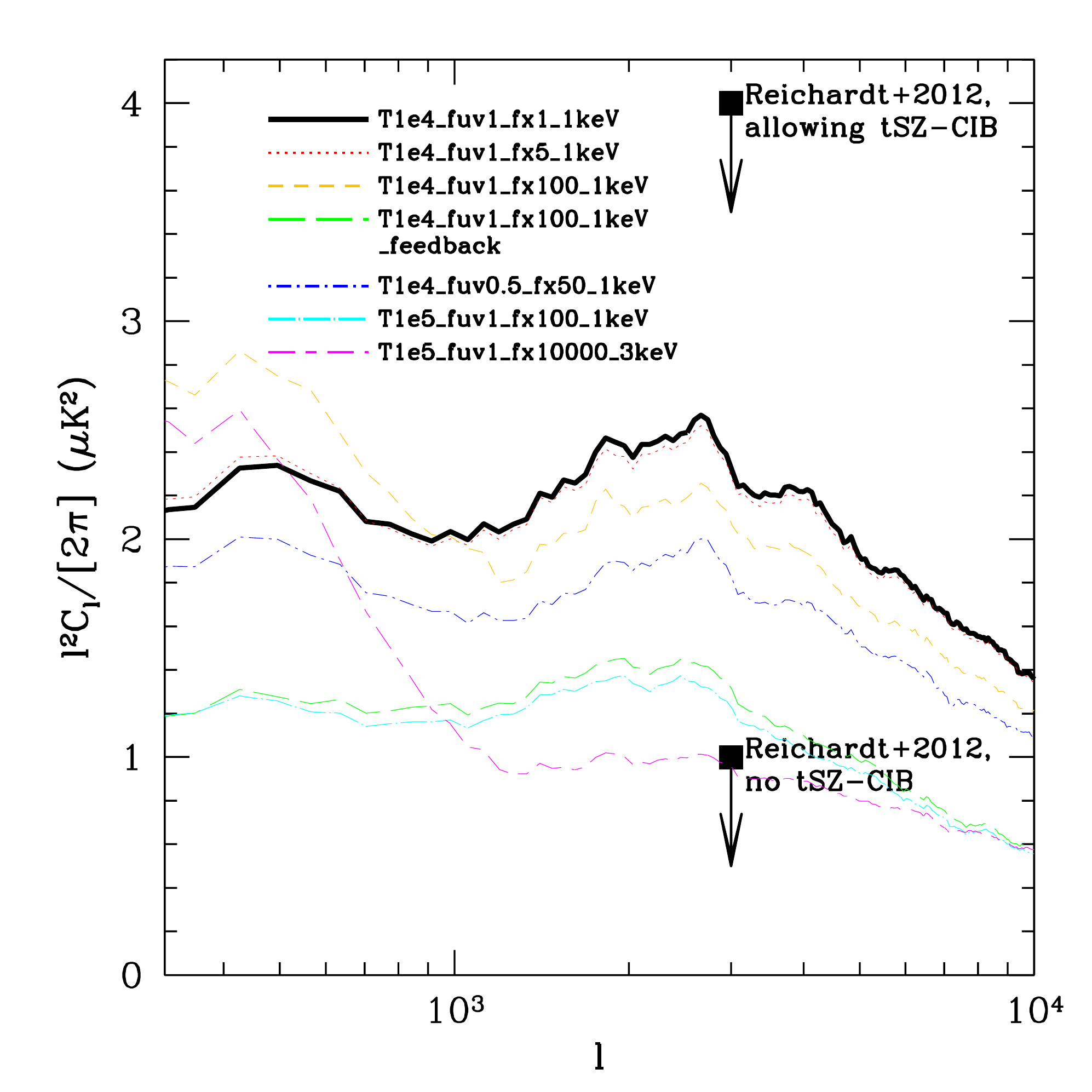}
\caption{Taken from \cite{Mesinger:2012ys}.  X-rays decrease the kSZ power at $\ell=3000$ by suppressing the small-scale ionization structures. If the X-rays contribute to a one half of the ionizations, the patchy kSZ power is decreased by $\approx0.5$ $\mu K^2$ with respect to a UV-only model with the same reionization history.}
\label{fig:kSZ}
\end{figure}

\subsubsection{{\bf Chemical feedback}}\label{metal}

As we have seen in Sect. \ref{sources}, one of the main uncertainties in modeling the reionization process is the unknown proportion between massive, zero-metallicity Pop III stars, and low-mass, metal-polluted Pop II stars. For example, the number of photons per baryon into stars can differ by more than one order of magnitude.

Several lines of though lead to the idea that the first stars might have been massive,  with masses $> 40-50 M_\odot$; unfortunately,
very little is known about their IMF. The ashes of these first supernova explosions pollute with metals the surrounding IGM and eventually lead to the creation of the subsequent generations of low-mass Pop~II/I stars, which have a ``Salpeter-like''  IMF 
when locally the metallicity approaches the critical value $Z_{cr}=10^{-5\pm1}Z_\odot$ (Schneider et al. 2002, 2006).

The uncertainty in the value of the critical metallicity is due to the role of dust cooling. Schneider et al. (2006b) pointed out that metals depleted onto dust grains ignite gas fragmentation into solar or sub-solar mass clumps already when metallicities are of the order of $Z=10^{-6} Z_\odot$; in the absence of dust grains, instead, even at $Z= 0.01 Z_\odot$ gas-phase metals can only produce fragments of mass about 100 times larger. 

Hence metal enrichment due to the first stars feeds back on the star formation mode and in this way regulates the transition from the Pop III to Pop~II/I star formation epoch, ultimately affecting the reionization evolution to a decreased ionizing photon production. 
Many aspects of chemical feedback need to be better characterized, as the number of Pop~III stars that end up their lives as supernovae, the metal injection, transport and mixing in the IGM. If the chemical feedback is efficient, Pop~III star formation
could last for a very short time and the contribution of these pristine stars to the ionizing photon budget could be negligible (see e.g. Ricotti \& Ostriker 2004a for a discussion).

It is very likely though that the transition occurred relatively smoothly because the cosmic metal distribution is observed
to be highly inhomogeneous even at epochs well beyond reionization where we can routinely detect them vis absorption line experiments. The key concept to bear in mind here is that chemical feedback is inherently a {\it local process}, with regions 
close to star formation sites rapidly becoming metal-polluted and overshooting $Z_{\rm cr}$, and others remaining essentially
metal-free. For this reason, the widely accepted scenario is that Pop~III and Pop~II stars have likely been coeval. 
Indeed, Scannapieco, Schneider \& Ferrara (2003) used an analytical model of inhomogeneous structure formation to follow the separate evolution of Pop~III/Pop~II stars as a function of star formation and wind efficiencies. They found that Pop~III stars continue to contribute appreciably to the star formation rate density at much lower redshift, rather independently from the parameter's values. Later on, these predictions have been confirmed by the authors of \cite{tornatore} (for recent discussions see Refs.~\cite{Xu:2013ata}, \cite{Latif:2013ysa} and references therein). This findings opens the intriguing possibility that detectable signatures from Pop~III stars could be found also in relatively evolved galaxies where the metallicity has become super-critical. Several experimental attempts are underway at this time and might yield the first clear detection of Pop III stars (see, e.g. \cite{Cassata:2012ix}).

\subsection{{\bf Challenges}}
Different experiments can be use to constrain the reionization process. Below is a (partial) list:
\begin{itemize}
\item \Lya and \Lyb Gunn-Peterson opacity (see Sect. \ref{GP})
\item Electron scattering optical depth (see Sect. \ref{CMB})
\item UV Background Intensity (see Sect. \ref{UV} - \ref{UV2})
\item Redshift evolution of Lyman Limit Systems
\item IGM temperature evolution (see Sect. \ref{EOS})
\item IGM Metallicity (see Sect. \ref{metal})
\item Cosmic star formation history
\item High-$z$ galaxy counts
\item Near Infrared background
\end{itemize}
\begin{figure}
\centerline{\includegraphics[width=23pc, angle=-90]{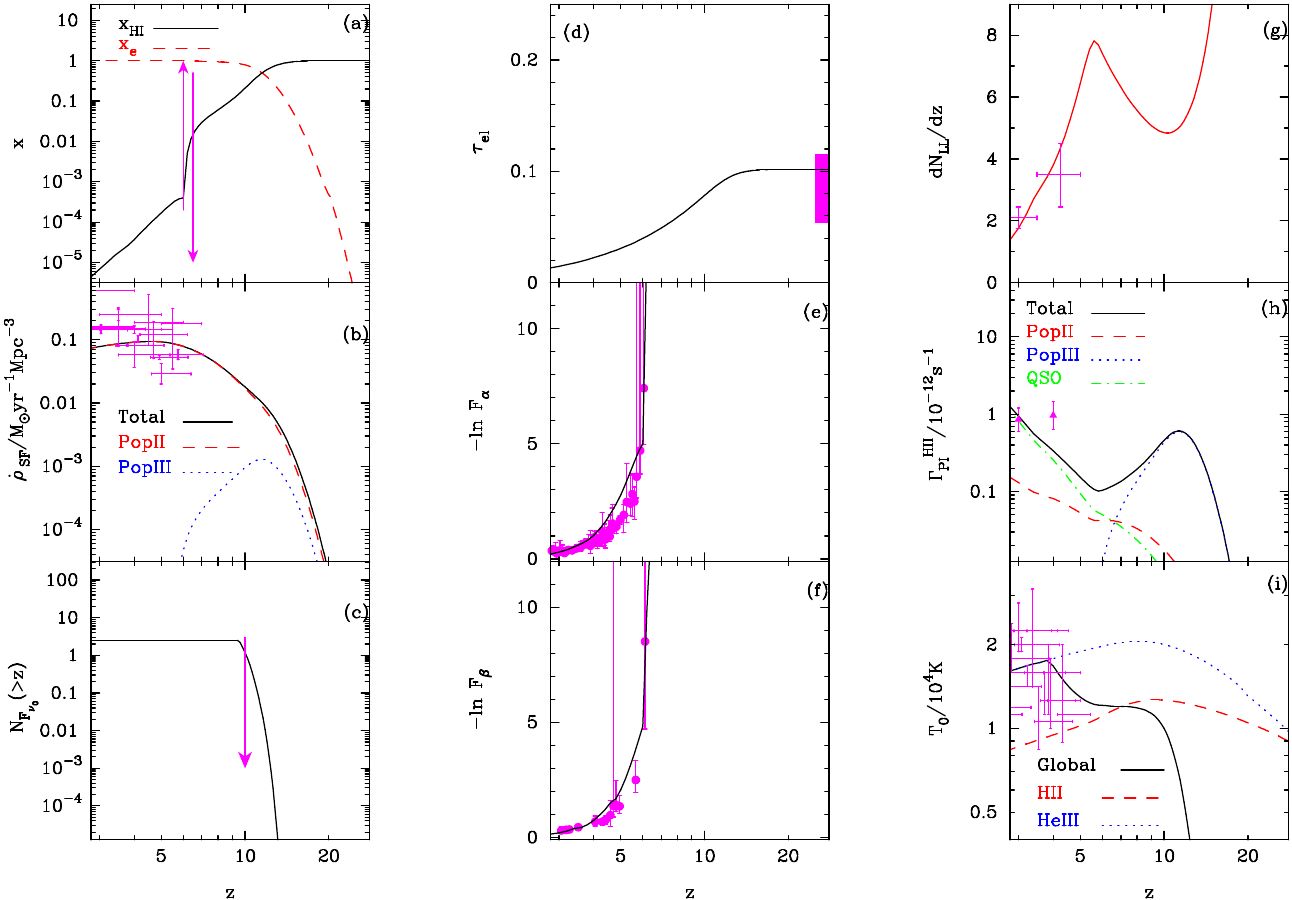}}
\caption{The best-fitting model of \cite{ChoudhuryFerrara06}, taken from \cite{Choudhury:2009kk}. The panels show
the redshift evolution of: (a) the volume-averaged electron and HI fraction. 
The arrows show an observational lower limit from QSO absorption lines at $z=6$ and upper limit from Ly$\alpha$ emitters at $z=6.5$; (b) the cosmic star formation history, with
the contribution of Pop III and Pop II stars. 
(c) the number of source counts above a given redshift, with the observational upper limit from NICMOS HUDF;
(d) the electron scattering optical depth, with observational constraints from 3-yr {\it WMAP} data; (e) Ly$\alpha$ effective optical
depth; (f) Ly$\beta$ effective optical depth; (g) the evolution of
Lyman-limit systems; (h) photoionization rates for neutral hydrogen, with estimates from numerical simulations; (i) temperature 
evolution of the mean density IGM. For tha data sources, see the original paper.
}
\label{cf06}
\end{figure}

A global model for reionization require many ingredients: a model for galaxy formation and an accurate treatment of the propagations of ionizing photons into the IGM. This, in turn requires: (i) a self consistent treatment of the thermal history of the IGM together with the evolution of the ionized regions: (ii) following the simultaneous evolution of the neutral, $\HII$ and $\HeIII$ regions (iii) resolving the sources of ionizing radiation (Pop III, Pop II stars, and QSO) (iv) a precise and complete treatment of the mechanical, radiative and chemical feedbacks (whose comprehensive and rigorous treatment lies outside the scope of these lectures, but see Sect. \ref{abinitio}, next Sect. \ref{metal} and Ref.~\cite{Ciardi:2004ru} for a review).
\subsubsection{{\bf Global reionization models}}\label{global}

Although numerical simulations of the reionization process are now in a very advanced state (see Sect. \ref{simulations}),  a great dela of our current understanding comes from analytical and semi-analytical models that offer considerable advantages: they are fast and therefore they can explore wide regions of the model parameter space; in principle, all the relevant physical processes can be implemented as numerical cost are not usually an issue. The advantages do not come for free, however, and the price is the loss of   topological information. To exemplify what can be currently achieved by semi-analytical models we will concentrate on the one developed by Choudhury and Ferrara (\cite{Choudhury:2004vs}, \cite{ChoudhuryFerrara06}).
This model implements the following main features: 
\begin{itemize}
\item It accounts for the IGM inhomogeneities ($\Delta$) by adopting a lognormal distribution for P($\Delta$); reionization is said to be complete when all the low-density regions are ionized.
\item Thermal and ionization history of neutral, $\HII$ and $\HeIII$ phases of the IGM are followed simultaneously and self-consistently, as the IGM is treated as a multi-phase medium.
\item Three types of ionization sources are assumed: Pop III, Pop II and quasars.
\item Radiative and chemical feedbacks are computed self-consistently from the evolution of the thermal properties of the IGM.
\item The predictions of the model are tested against a large range of observational tests, such as: the redshift evolution of the Lyman-limit absorption systems (LLS); the effective optical depths $\tau_{eff}$ for \Lya and \Lyb absorption; the CMB electron scattering optical depth; temperature of the mean intergalactic gas; cosmic star formation history; source number counts. 
\end{itemize}

The model has allowed to cast precise predictions. According to the model results, reionization has been a long-lasting and gradual event which was ignited by Pop III stars at $z\approx 15$, and it was $90\%$ complete at $z\approx$ 8 (see Fig. \ref{cf06}). At redshifts $z < 10$ the growth of the ionized regions slows down for the action of radiative and chemichal feedbacks, and therefore reionization is only complete at $z\approx$ 6. The process is initially driven by metal-free stars in low mass ($M < 10^8 M_\odot$) haloes and then the combined action of chemical and radiative feedback at $z < 10$ soon prevent the formation of these objects (see previous Sect. \ref{metal}). 

Fig. \ref{fig:halofrac}, taken from \cite{Choudhury:2007mr} quantitatively illustrates the properties of the sources of ionizing radiation. The plot shows the quantity $x_{\gamma}(z)$, defined as the number of ionizing photons per H-atom contributed by haloes in the mass range $[M_{\rm min}, M_{\rm max}]$ in a fraction of the Hubble time $t_H(z)$ equal to the recombination time $t_{\rm rec}(z)$. Therefore, by definition,  the IGM is reionized when $x_{\gamma} \gtrsim 1$. 

From the plot \ref{fig:halofrac} one can see how the lower mass $10^7$-$10^8 M_{\odot}$ haloes dominate the photon production rate at early redshifts providing about 0.25 photon/H-atom on
the fractional recombination timescale. Then, when the production of Pop III stars is hampered by the chemical and radiative feedback the reionization process relies on photons emitted by 
more massive haloes with $ M> 10^9 M_{\odot}$. The right-hand panel shows the fractional instantaneous contribution of 
halos above a certain mass for selected redshifts. This plot shows again that $>80$\% of the ionizing power at $z \ge 7$ is provided by haloes with masses $< 10^9 M_{\odot}$ which are predominantly harboring PopIII stars. Then,  galaxies residing in $M>10^9 M_{\odot}$ haloes, produce $\approx 60$\% of the ionizing photons at $z=6$, and this is the Pop II stars dominated
phase. 

\begin{figure}
\rotatebox{270}{\resizebox{0.4\textwidth}{!}{\includegraphics{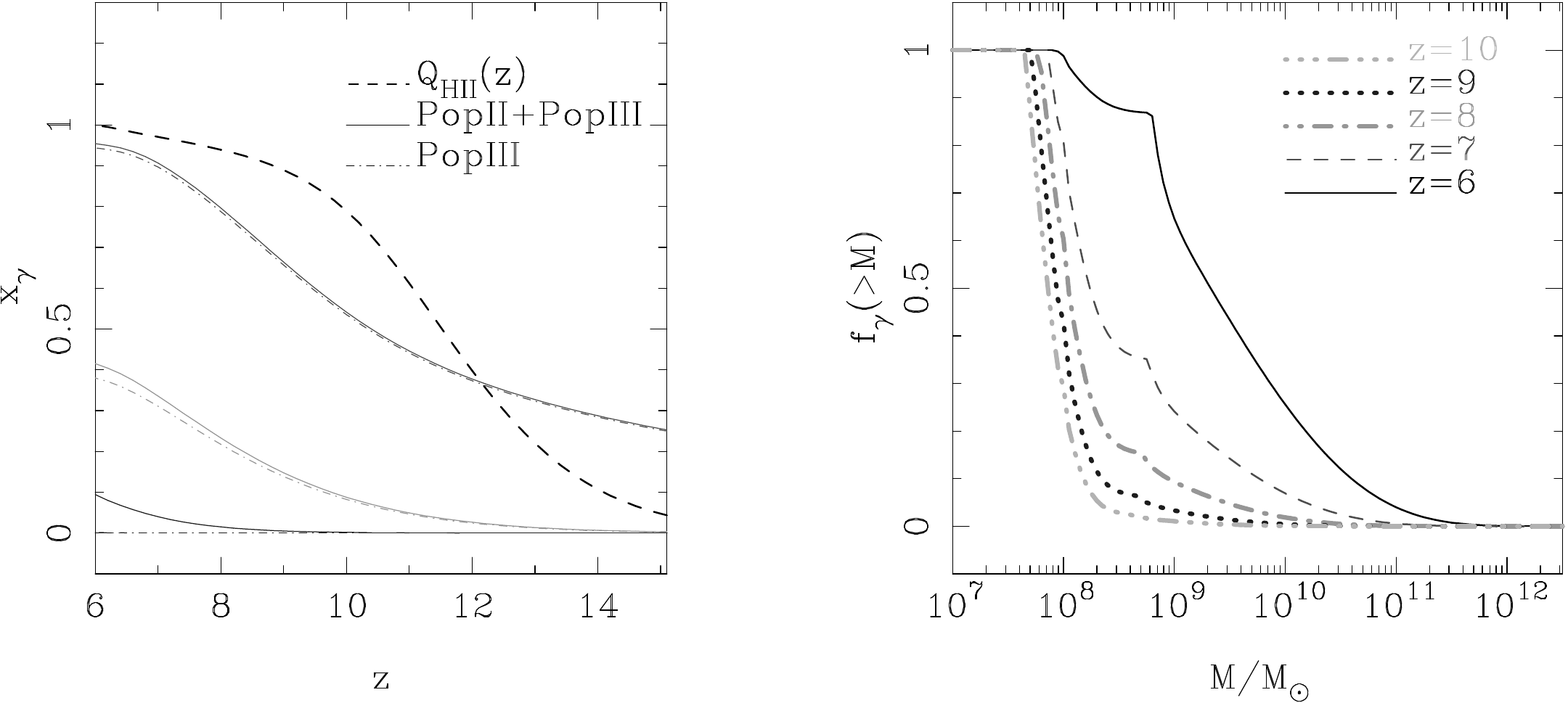}}}
\caption{(Taken from \cite{Choudhury:2007mr})  {\it Left}: Number of ionizing photons per H-atom contributed by haloes of different mass 
in a fraction of the Hubble time $t_H(z)$ equal to the recombination time $t_{\rm rec}(z)$ as a function of 
redshift $z$ for different halo masses and for different stellar populations. 
{\it Right}: Cumulative fraction of the ionizing power $f_{\gamma}$ contributed by haloes of mass $>M$. The curves from 
right to left correspond to $z = 6,7,8,9,10$ respectively.
}
\label{fig:halofrac}
\end{figure}

\subsection{{\bf Parameter uncertainties}}\label{constraining}

One of the major uncertainty in modeling reionization is to parametrize the transfer of radiation from the sources to the IGM. This  process is usually parameterized through $N_{ion}$, the number of photons entering the IGM per baryon in collapsed objects.
This parameter can be written as 
\begin{equation}\label{Nion}
N_{ion}=f_{*}f_{esc}N_{\gamma},
\end{equation} 
where $f_{*}$ is the star-forming efficiency (fraction of baryons within collapsed haloes going into stars), $f_{esc}$ is the fraction of photons escaping into the IGM, and $N_{\gamma}$ is the number photons per baryon emitted from the sources. 
The star-forming efficiency and escape fraction (and hence $N_{ion}$) are likely to be a function of redshift and halo mass,  but as the dependencies are not well understood the simplest approach is to take them as constants. A different approach has been attempted by the authors of \cite{Mitra:2010sr} and \cite{Mitra:2011uv}: they assumed $N_{ion}$ to be completely arbitrary and decompose it into its principal components (PC), a method that has  the advantage of highlighting which components of $N_{ion}$ can be more constrained by available data. 
 
Figure \ref{fig:aic_wmapclee} is taken from \cite{Mitra:2011uv}. It shows the marginalized posterior distributions of several physical quantities related to the reionization process. From the top-left plot, in a clockwise order we see the evolution of: $N_{\rm ion}(z)$; the photo-ionization rate $\Gamma_{\rm PI}(z)$ together with its observational constraints from Bolton \& Haehnelt (2007); the LLS distribution alongside with data from Songaila \& Cowie (2010); the volume filling factor of HII regions $Q_{\rm HII}(z)$; the global neutral hydrogen fraction $x_{\rm HI}(z)$ with data from Kashikawa et al. 2006, Fan et al. 2006 and Totani et al 2006; temperature, polarization and cross temperature-polarization power spectra with the data points from WMAP7 (Larson et al. 2010). The solid lines is the best fit model of their analysis and the shaded regions correspond to 2-$\sigma$ limits.

\begin{figure}
  \includegraphics[height=0.95\textwidth, angle=270]{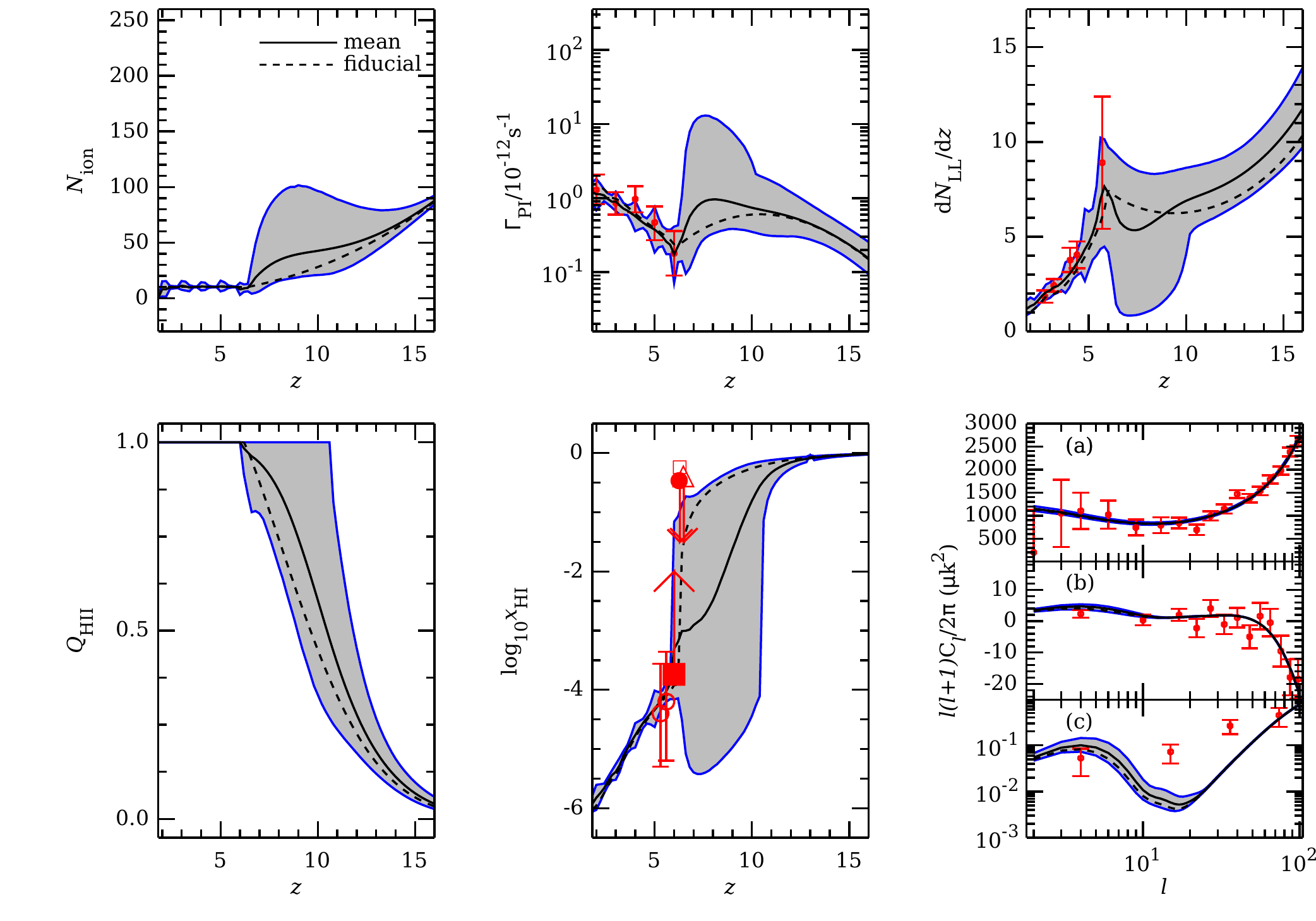}
  \caption{The two-sigma marginalized posteriori distributions of reionization-related quantities from the PC analysis of  \cite{Mitra:2011uv} with corresponding over-plotted data points.}
\label{fig:aic_wmapclee}
\end{figure}

\subsubsection{{\bf Escape fraction}}\label{escape}
In the expression Eq. \ref{Nion} for $N_{ion}$), perhaps the major uncertainty is represented by $f_{esc}$. The escape fraction, in fact, is likely to depend strongly on redshift and galaxy mass. Theoretical/numerical studies indicate a broad range of allowed values, ranging in $0.01<f_{esc}<1$ (e.g. \cite{dove_photoionization_1994}, \cite{Wise:09}; for a recent review see discussion see \cite{Ferrara:2012zy}. Many factors could induce large variation in the actual value of $f_{esc}$: among these, there are galaxy mass, morphology, redshift, gas density profiles and composition, source luminosities/spectra and distribution. Numerical studies seem to indicate that $f_{esc}$ increases with decreasing galaxy mass \cite{Shull:12}.  A similar trend is found with redshift: lower-mass galaxies and higher star formation rates at high redshifts seem to lead to higher values of $f_{esc}$ \cite{Ferrara:2012zy}. Observational constraints for $f_{esc}$ have been obtained by measuring the escaping Lyman-continuum (LyC) radiation, and the observations seem to confirm the theoretical trend (see e.g. \cite{Nestor:11} and \cite{Finkelstein:12}). The field urgently required more theoretical and numerical work to better understand the complex physics determining $f_{esc}$. For a (relatively recent) complete survey of both theoretical and observational constraints we refer the reader to \cite{fernandez_effect_2011}. 

\subsubsection{{\bf Specific ionizing photon production}}
The other uncertain factor in Eq. \ref{Nion} is the specific ionizing photon production, i.e.  the number of ionizing photons produced per baryon into stars, $N_{\gamma}$. As we saw in Sect. \ref{global}, from the studies of \cite{Choudhury:2004vs} and \cite{ChoudhuryFerrara06} emerges also that reionization is initially driven by metal-free stars in low mass ($M < 10^8M_{\odot}$) haloes but then chemical and radiative feedback at $z < 10$ gradually quench the formation of such objects.
The number of the ionizing photons depends on the emission spectrum of the source:  Fig. \ref{popstars}, taken from \cite{Tumlinson:1999iu}, shows the (log) number of ionizing photons per second for a Pop II (above) and Pop III (below) stars with a Salpeter IMF. Pop III stars are represented by the solid line with zero metallicity, $Z = 0$. As we saw in Sect. \ref{sources}, due to their metal-free composition, the first stars are hotter and have harder spectra then Pop II stars; therefore the ionizing photon production is enhanced. Metal-free stars with masses $\geq 20 M_{\odot} $ emit between $10^{47}$ and $10^{48}$ \HI and \HeI ionizing photons/s/solar mass, where the upper value is for stars with $100 M_{\odot}$ \cite{Tumlinson:1999iu}. As $N_{\gamma}$ depends on the stellar IMF, it is affected by the proportion between Pop III and Pop II. As we have seen in Sec. \ref{sources} and \ref{metal} it should also depend on the efficiency of the chemical feedback, and on the value of the critical metallicity.

\begin{figure}
\centerline{\includegraphics[width=30pc]{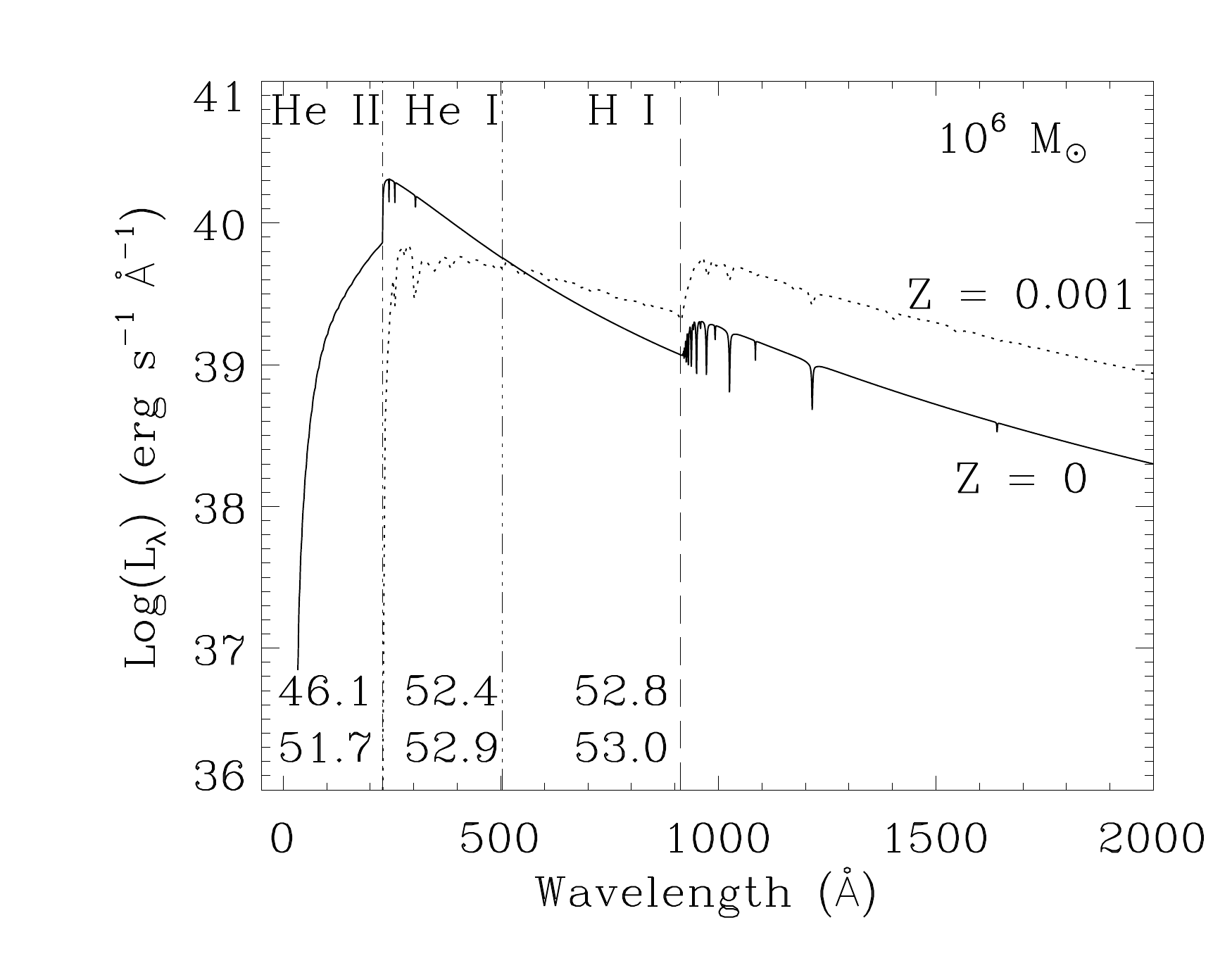}}
\caption{Synthetic spectra of Pop~II and Pop~III clusters of 10$^6$~M$_\odot$
with a Salpeter IMF. The numbers in the lower left corner, near each continuum
mark, represent the rate (in units of photons s$^{-1}$)
of ionizing photons production (log) for that continuum
(Tumlinson \& Shull 2000).}
\label{popstars}
\end{figure}

\subsubsection{{\bf Reionization and cosmology}}

A realistic and complete modeling of the reionization process is not only important per se,
but it has also important implications for cosmology at large. Most noticeably, reionization is closely
interlinked wit the study of Cosmic Microwave Background anisotropies. Inadequate assumptions on the reionization
history lead to biased reconstructed constraints on the other cosmological parameters shaping the temperature, but most importantly, the polarization power spectrum of the CMB.

In the absence of any precise knowledge, in the analysis of the CMB reionization is customarily parametrized as a ``sudden'' or instantaneous process. However, this choice of parametrization could bias the constraints of cosmological parameters, especially 
the inflationary sector's ones, which mostly regulates the shape of the polarization power spectrum
(see Ref.\cite{Pandolfi:2010mv} and reference therein).

In \cite{Pandolfi:2011kz}  the authors combined the CMB data with astrophysical datasets/results from
quasar absorption line experiments (such as the Gunn-Peterson test and the redshift evolution of Lyman Limit Systems) 
and considered the joint variation of both the cosmological and astrophysical parameters governing the evolution of the free electron fraction $x_e(z)$. They found that including a data-constrained reionization history (obtained from the \cite{ChoudhuryFerrara06} model described above) in the analysis induces appreciable changes in the cosmological parameter values deduced through a standard WMAP7 analysis. Particularly noteworthy are the variations in the baryon density parameter $\Omega_bh^2$ and the optical depth values. They conclude that the inclusion of astrophysical datasets, allowing to robustly constrain the reionization history,
in the extraction procedure of cosmological parameters leads to relatively important differences in the final determination of their values. This is an important future avenue to pursue in the era of precision cosmology.

\appendix

\acknowledgments

\end{document}